%% file: EURONUmain.tex
\documentclass{IDR}
\usepackage{url}
\usepackage[english]{babel}
\usepackage{subfigure}
\usepackage{graphicx}
\usepackage{epsfig}
\usepackage{multirow}
\usepackage{amsmath}
\usepackage{amssymb}
\usepackage{amsfonts}
\usepackage{slashed}
\usepackage{bbm}
\usepackage{ctable,longtable}
\usepackage{cancel}
\usepackage{pifont}
\newcommand\ie{\textit{i.e.}}
\newcommand\Dmq{\Delta m^2}

\newcommand\eVq{\text{eV}^2}

\newcommand\ldm{\Delta m_{31}^2}
\newcommand\sdm{\Delta m_{21}^2}
\newcommand\stheta{\sin^2 2\theta_{13}}
\newcommand{\CP}{\text{CP}}
\newcommand{\Nuc}[2][]{{\ensuremath{\ifthenelse{\equal{#1}{}}{}{\mbox{}^{#1}}\text{#2}}}}

\newcommand{\stch}{\sin^2 2\theta_{13}}
\newcommand{\deltacp}{\delta_{\mathrm{CP}}}
\newcommand{\znbbeq}{0\nu\beta\beta}
\newcommand{\znbb}{$\znbbeq$}
\def\simge{\mathrel{%
   \rlap{\raise 0.511ex \hbox{$>$}}{\lower 0.511ex \hbox{$\sim$}}}}
\def\simle{\mathrel{
   \rlap{\raise 0.511ex \hbox{$<$}}{\lower 0.511ex \hbox{$\sim$}}}}
\newcommand{\mcl}[1]{\mathcal{#1}}
\def\slc#1{\setbox0=\hbox{$#1$}           
    \dimen0=\wd0                                 
    \setbox1=\hbox{/} \dimen1=\wd1               
    \ifdim\dimen0>\dimen1                        
       \rlap{\hbox to \dimen0{\hfil/\hfil}}      
       #1                                        
    \else                                        
       \rlap{\hbox to \dimen1{\hfil$#1$\hfil}}   
       /                                         
    \fi}
\newcommand{\nn}{\nonumber}
\newcommand{\vldm}{\Delta m_{41}^2}
   

\input CONTENT/def_temp.tex

\begin{document}
%
%
\selectlanguage{english}
\makeatletter

\pagestyle{empty}
\thispagestyle{empty}
%
\input 00a-Title-pageEURONU
%
\cleardoublepage
\input 00b-AuthorListWP6

\clearpage
%
%
\parindent 10pt
\pagenumbering{roman}                   
\setcounter{page}{1}
\thispagestyle{plain}
\pagestyle{plain}
%
\tableofcontents
\clearpage
\pagenumbering{arabic}                   
\setcounter{page}{1}
%
\input WP6-Executive-summary
%
\newpage
\input CONTENT/Outline.tex

\clearpage

\newpage
\bibliographystyle{utphys}
\bibliography{./EURONUmain,%
              }
%
\end{document}

%% file: CONTENT/def_temp.tex
\usepackage{bbold}

%% file: 00a-Title-pageEURONU.tex
\begin{figure}
  \vspace{-1cm}
  \begin{flushright}
    \includegraphics[width=3cm]{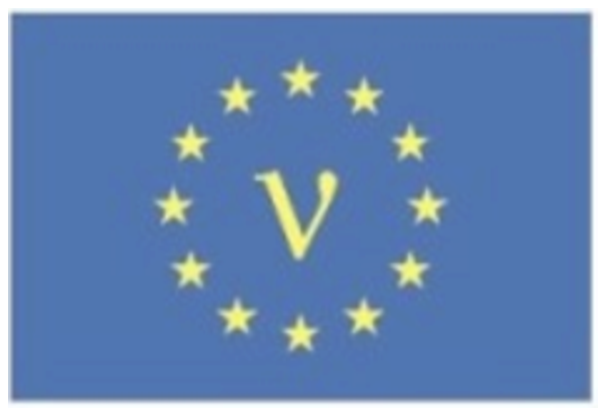}
  \end{flushright}
  \rightline{EURONU-WP6-11-34}
  \vspace{2cm}
\end{figure}
\begin{center}
  {\bf\LARGE 
    EUROnu-WP6 2010 Report \\
  }
  \vspace{0.75cm}
  {\bf\it\Large The EUROnu Working Package 6 (Physics)} \\
  \vspace{2cm}
\end{center}

\vspace{3cm}

{\setlength\parindent{0em}
S.~K.~Agarwalla$^1$, E.~Akhmedov$^2$, M.~Blennow$^3$, P.~Coloma$^{4,5}$, A.~Donini$^{1,5}$ (editor),  E.~Fern\'andez Mart\'{\i}nez$^3$, C.~Giunti$^6$, 
J.~J.~G\'omez Cadenas$^1$, M.C. Gonz\'alez Garc\'{\i}a$^{7,8,9}$, P.~Hern\'andez$^{1}$ (editor), P.~Huber$^{10}$,  M.~Laveder$^{11}$, T.~Li$^{12}$, A.~Longhin$^{13,14}$,
J.~L\'opez Pav\'on$^5$, M.~Maltoni$^5$, D.~Meloni$^{15}$, O.~Mena$^1$, J.~Men\'endez$^{5,16,17}$, M.~Mezzetto$^{11}$, P.~Migliozzi$^{18}$, T.~Ohlsson$^{19}$, 
C.~Orme$^{12}$, S.~Pascoli$^{12}$, J.~Salvado$^8$,  T.~Schwetz$^2$, L.~Scotto-Lavina$^{20,21}$,  J.~Tang$^{15}$, F. ~Terranova$^{14}$, W.~Winter$^{15}$ and
 H.~Zhang$^{19}$
}

\begin{center}
  \vspace{3cm}
  \today{}
\end{center}
\vfill

%% file: 00b-AuthorListWP6.tex
{\setlength\parindent{0em}
 {\it $^1$
    Instituto de Fisica Corpuscular (IFIC), CSIC/UVEG,
    Edificio Investigaci\'on Paterna, Apartado 22085, 46071 Valencia, 
    Spain
  } 
 \\{\it $^2$ Max-Planck-Institut f\"ur Kernphysik, PO Box 103980,  69029 Heidelberg, Germany
  } 
%
%
 \\{\it $^3$
    Max-Planck-Institut f\"ur Physik (Werner-Heisenberg-Institut), Fohringer Ring 6, D-80805 Munich, Germany
  } 
   \\{\it $^4$ Departamento de F\'{\i}sica Te\'orica, Universidad Aut\'onoma de Madrid, Cantoblanco, E-28049, Madrid, Spain
   } 
  \\{\it $^5$ Instituto de F\'{\i}sica Te\'orica, UAM/CSIC, Cantoblanco, E-28049, Madrid, Spain
   }
     \\{\it $^6$
  Istituto Nazionale di Fisica Nucleare, Sezione di Torino, Via P. Giuria 1, IÐ-10125 Torino, Italy
  }
\\ {Ê\it $^7$ 
     Instituci\'o Catalana de Recerca i Estudis Avan\c cats (ICREA) }
\\{\it $^8$
Departament d'Estructura i Constituents de la Mat\'eria and Institut de Ciencies del
Cosmos, Universitat de Barcelona, Diagonal 647, E-08028 Barcelona, Spain
}
\\ {\it $^9$ C.N. Yang Institute for Theoretical Physics, State University of New York at Stony Brook,
Stony Brook, NY 11794-3840, USA
}
 \\{\it $^{10}$
 Center for Neutrino Physics, Virginia Tech, Blacksburg, VA 24061, USA
      }
    \\{\it $^{11}$
    Dipartimento di Fisica ÒG. GalileiÓ, Universit\`a di Padova and
    Istituto Nazionale di Fisica Nucleare, Sezione di Padova,
    Via Marzolo 8, I-35131, Padova, Italy
  } 
  \\{\it $^{12}$
    Institute for Particle Physics Phenomenology, Department of
    Physics, University of Durham, Science Laboratories, South Rd,
    Durham, DH1 3LE, UK
  } 
   \\{\it $^{13}$
 Institut de Recherche sur les lois Fondamentales de l'Univers, CEA-Saclay, 91191 Gif-sur-Yvette, France
 }  
   \\{\it $^{14}$
  Istituto Nazionale di Fisica Nucleare,  Laboratori Nazionali di Frascati, Frascati, Italy
  }
    \\{\it  $^{15}$
   Institut f\"ur Theoretische Physik und Astrophysik, Universit\"at W\"urzburg,
D-97074 W\"urzburg, Germany
  } 
  \\ {\it $^{16}$
  Institut f\"ur Kernphysik, Technische Universit\"at Darmstadt, 64289 Darmstadt, Germany
  }
  \\ { \it $^{17}$
   ExtreMe Matter Institute EMMI, GSI Helmholtzzentrum f\"ur Schwerionenforschung GmbH,
64291 Darmstadt, Germany
  }
   \\{\it $^{18}$
    Istituto Nazionale di Fisica Nucleare, Sezione di Napoli, Italy
  }
\\{\it $^{19}$
Department of Theoretical Physics, School of Engineering Sciences, 
KTH Royal Institute of Technology, AlbaNova University Center,
Roslagstullsbacken 21, 106 91 Stockholm, Sweden
}
  \\{\it $^{20}$
  University of Zurich, Physik-Institut, CH-8057 Zurich, Switzerland
    }
  \\{\it $^{21}$
  Laboratoire SUBATECH, 4 rue Alfred Kastler, 44307 Nantes cedex 3, France
  }

%% file: WP6-Executive-summary.tex
\newpage
%
\section{Executive summary}

The phenomenon of neutrino oscillations, arguably the most significant
advance in particle physics over the past decade, has been established
through measurements on neutrinos and anti-neutrinos produced in the
sun, by cosmic-ray interactions in the atmosphere, nuclear reactors,
and beams produced by high-energy particle accelerators.
In consequence, we know that the Standard Model is incomplete and must
be extended to include neutrino mass, mixing among the three neutrino
flavours, and therefore lepton-flavour non conservation.
These observations have profound implications for the ultimate theory
of particle interactions and for the description of the structure and
evolution of the Universe.  

These exciting possibilities justify an energetic and far reaching
programme, an essential part of which is to make precision
measurements of the oscillation parameters. 
Assuming the three flavours and the unitary neutrino-mixing matrix
that is presently favoured, oscillation measurements can be used to
determine the three mixing angles and the critical phase parameter
that can provide a new source of CP-invariance violation. 
Neutrino oscillation measurements can also be used to determine the
two (signed) mass differences. 
This programme is similar to the long-standing investigations of quark
mixing via the CKM matrix and it would now seem to be clear that an
understanding of the flavour problem will definitely need
precision measurements in both quark and lepton sectors.  

Not all the properties of the neutrino can be determined by
oscillation experiments. 
Equally important is the determination of the Majorana or Dirac nature
of the neutrino which requires the ongoing and planned neutrinoless double beta decay experiments.
In addition, although oscillation measurements determine the mass
differences, they are insensitive to the absolute mass, $m_1$, of the 
lightest mass state. The determination of $m_1$ requires a very precise 
measurement of the end-point of the electron spectrum in beta decay. 

\subsection*{Coordination and Outreach}

The members of WP6 have held meetings during the general meeting at Strasbourg in June 2010, at   
the CERN workshop NuThemes in September 2010 and at the Rutherford lab during the IDS-NF meeting also in September 2010.
A summary of the EURONU-WP6 workshop held at from 8Ð10 June 2009 at Cosener's House, Abingdon, UK on the subject ``Flavour physics in the era of precision neutrino experiments'' has been included in this report.
 
The group has kept close contact with the IDS-NF concerning the physics of the neutrino factory, with WP4 as regards the scenarios for the beta-beam 
and with WP5 as regards detector performance and systematic errors.
During 2010, the WP6 results of the first year of the project (2009) were summarized in a report that was submitted to the  archives
to inform the wider community \cite{Bernabeu:2010rz} . 

\subsection*{Global analysis and interpretation of present data}

The members of WP6 have carried out a number of important studies for the EURONU project. The first is a revision of the global analysis of neutrino oscillation data including the latest results, such as those from MiniBOONE and MINOS, as well as new cosmological data.  The values of the atmospheric and solar parameters and the bounds 
on the angle $\theta_{13}$  are of course essential for the physics optimization of future facilities. Obviously if there is physics beyond the standard three-neutrino scenario, 
for example the presence of new sterile species, this could also change drastically the optimization of baseline and energy of the neutrino beam.

\subsection*{Evaluation of physics performance, optimization and comparison}

A number of studies of the physics potential of the future facilities as regards the standard three-neutrino oscillation scenario, as well as new physics, have been performed during the past year. The results have been published in peer-review journals and are also listed as EURONU documents, or will be presented here as internal documents. Between these, 
the report contains a new design of the SPL-Fr\'ejus Super-Beam is included (see Ref.~\cite{Longhin:2011hn} for more details), the study of
the potential of high-$\gamma$ electron-capture beta-beams in the context of the LAGUNA european project has also been reviewed \cite{Orme_LAGUNA},
a review of all beta-beam setups that have been discussed in the literature in the past \cite{Donini:2011thisrep}
 and a dedicated study of atmospheric neutrino backgrounds at the ICAL@INO  detector exposed to a high-$\gamma$ $\beta$-beam \cite{Agarwalla:2011thisrep}.

\subsection*{Tools for physics studies}

A new release of the GLoBES package including migration matrices for signal and background was made available during 2010.



%% file: CONTENT/Outline.tex
\section{The Physics case}
\label{Sect:PPEG}

The main motivation of a future neutrino physics programme is to unveil what the new physics associated to neutrino masses is. 
We know for sure that new degrees of freedom must be added to the Standard Model (e.g. right-handed neutrinos) at 
some energy scale $\Lambda$. If $\Lambda$ is much larger than the electroweak scale, there is a natural explanation of why neutrinos are so light. 
Indeed the effects of {\it any} such new physics must be generically  well described at low energies by an effective Lagrangian which contains the Standard Model, 
plus a tower of higher dimensional operators constructed with the SM fields and satisfying all the gauge symmetries:
 \begin{eqnarray}
{\mathcal L} = {\mathcal L}_{SM} + \sum_i \frac{{\alpha}_i}{\Lambda} {\mathcal O}_i^{d=5} + \sum_i \frac{\beta_i}{\Lambda^2} {\mathcal O}_i^{d=6} + ...
\label{eq:eft}
\end{eqnarray} 
The effective operators, ${\mathcal O}_i$,  are ordered by their mass dimension, since the higher the dimension, the higher the power of $\Lambda$ that suppresses them.  The dominant operator is therefore the lowest dimensional one, with $d=5$, which is precisely the Weinberg's operator:
\begin{eqnarray}
{\mathcal O}^{d=5} = {\bar L}^c \Phi \Phi L ,
\label{eq:weinberg}
\end{eqnarray}
which, as is well known, induces three new ingredients to the minimal SM:
\begin{itemize}
\item Neutrino masses
\item Lepton mixing
\item Lepton number violation
\end{itemize}
In this context, neutrino masses are very small, because they 
come from an effective operator which is  suppressed by a high energy scale. 
If we go to operators of $d=6$, that are suppressed by two powers of $\Lambda$, these will generically induce new physics in dipole moments, rare decays, etc. Beyond $d=6$ we would find operators inducing non-standard neutrino interactions (NSI).  

It is also possible that the scale  $\Lambda$  is at or below the electroweak scale, or in other words that neutrino masses are linked to light degrees of freedom, {\it i.e.} a {\it hidden} sector which we have not detected yet, because it is weakly interacting. Such scenarios do not offer an explanation of why neutrinos are light, but neutrinos are the natural messengers with such hidden sectors, since they are the only particles in the SM carrying no conserved charge. Such 
new physics could be related to other fundamental problems in particle physics such as the origin of dark matter and dark energy.

Even though it is not guaranteed that we can fully understand the new physics associated to neutrino masses by measuring them, it is quite clear that we have a good chance to learn something more about it by testing the Standard scenario of 3$\nu$ mixing with future and more precise neutrino experiments. In particular we should be able to measure all the fundamental parameters: three mass eigenstates ($m_1^2, m_2^2, m_3^2$), three angles  ($\theta_{12}, \theta_{13}, \theta_{23}$) and one or three CP-violating phases  ($\delta, \alpha_1, \alpha_2$). But, also, it will be very important  to search for new physics 
beyond neutrino masses and mixings, in particular for those effects that are generic in many models of neutrino masses, such as violations of unitarity, non-standard interactions or the presence of light sterile species. To some extent these searches can also be improved in future facilities and this should be evaluated. Typically such analyses imply dealing with a much larger parameter space, which calls for new tools to perform the fits, in particular Montecarlo methods. 

Many studies in the last ten years have shown that we can measure $\theta_{13}$, discover leptonic CP violation and determine the neutrino hierarchy in more precise neutrino oscillation experiments, searching for the subleading channel $\nu_e \leftrightarrow \nu_\mu$ or its CP-conjugate channel $\nu_\mu \leftrightarrow \nu_e$ in the atmospheric range. 
In this first section, we present the results obtained within the work of EUROnu-WP6 in 2010 concerning: the status of leptonic mixing global fits \cite{GonzalezGarcia:2010er}; 
the prospects for $\theta_{13}$ searches \cite{Mezzetto:2010zi}; and, the direct determination of solar neutrino fluxes from solar neutrino data \cite{GonzalezGarcia:2009ya}.

\input{CONTENT/01MaltoniStatus.tex}
\clearpage
\input{CONTENT/02MezzettoSchwetz.tex}
\clearpage
\input{CONTENT/03GMSresumsol.tex}

\newpage
\clearpage

\input{CONTENT/04Longhin.tex}
\label{SB}

\newpage

\section{Beta-Beams}
\label{BB}

Beta-beams are one of the two new beam technologies that have been proposed in the last decade to produce intense neutrino beams aiming at distant detectors. 
Many different beta-beams proposals have been studied in recent years. The physics performances of most of them are summarized in Sect.~\ref{sec:bbsummary}. 
The rest of the section covers: atmospheric $\nu_\mu$ background at ICAL@INO for a high-$\gamma$ beta-beam from CERN \cite{Agarwalla:2011thisrep};
minimal beta-beams that exploit at most existing european infrastructures \cite{Coloma:2010wa}; an update of the physics potential
of electron-capture beta-beams \cite{EC_Orme}; and, the physics potential of a high-$\gamma$ beta-beam within the EU LAGUNA project \cite{Orme_LAGUNA}.

\input{CONTENT/05aSummaryBB.tex}

\clearpage
\input{CONTENT/05bAtmo.tex}
\input{CONTENT/06Neapolitans.tex}
\input{CONTENT/07OrmeEC.tex}
\input{CONTENT/08OrmeLAGUNA.tex}

\newpage
\clearpage

\section{Physics at the Neutrino Factory}
\label{NF}

Neutrino factories have been studied as a mean to produce intense and collimated neutrino beams aiming at far detector(s) for the last ten years. Their impressive 
physics potential make them as the {\it ultimate} neutrino oscillation facility to improve our precision on known leptonic mixing matrix parameters and to measure
the few remaining unknowns. The main activity in the field in 2010 focused on the possibility to "stage" a Neutrino Factory, so that at each upgrade a new physics
program me is at hand \cite{Tang:2009wp}; on the physics potential of a "low-energy" Neutrino Factory that could optimize its capability to measure leptonic CP violation
in case $\theta_{13}$ were large (as we now know it is the case); and, the study of an important source of background to the "golden channel" $\nu_e \to \nu_\mu$ 
oscillation that was previously overlooked (the $\tau$-contamination problem, see \cite{Donini:2010xk}).

\input{CONTENT/09TangWinter.tex}
\input{CONTENT/10LENF.tex}

\input{CONTENT/11taus.tex}

\newpage
\clearpage

\section{Physics beyond three-family oscillations}
\label{NewPhysics}

The new facilities proposed within EUROnu have some potential to look for new physics beyond the standard three-family oscillations, such as Non-Standard Interactions, 
violation of unitarity of the PMNS matrix and light sterile neutrinos. In this section we present the WP6 results on: neutrinoless double-beta decay \cite{Blennow:2010th};
sterile neutrinos as a solution to explain 2010 MiniBooNE data \cite{Giunti:2010jt}; benefits of adding NSI's to sterile neutrinos to explain the 2010 MiniBooNE data 
\cite{Akhmedov:2010vy}; the potential of the Neutirno Factory to search for sterile neutrinos \cite{Meloni:2010zr}; the potential of the Neutrino Factory to distinguish
NSI's from violation of unitarity of the PMNS matrix \cite{Meloni:2009cg}; and, NSI's in the Zee-Babu model \cite{Ohlsson:2009vk}.

\input{CONTENT/12Neutrinoless.tex}
\input{CONTENT/13GiuntiLaveder.tex}

\input{CONTENT/14AkhmedovSchwetz.tex}

\input{CONTENT/17MeloniTangWinter.tex}

\input{CONTENT/15MeloniOhlssonWinterZhang.tex}
\input{CONTENT/16OhlssonSchwetzZhang.tex}
\clearpage
\newpage
\section{Physics potential of EUROnu facilities as of April 2011 \cite{Huber:2011thisrep}}
\label{Sect:Comparison}

\input{CONTENT/18Comparison.tex}

\clearpage
\newpage

\section{Summary of the NuFlavour workshop \cite{Pascoli:2011thisrep}}
\input{CONTENT/19NuFlavour.tex}


%% file: CONTENT/01MaltoniStatus.tex
\subsection{The leptonic flavour sector \cite{GonzalezGarcia:2010er}}
\label{sec:maltoni}

It is now an established fact that neutrinos are massive and leptonic
flavors are not symmetries of Nature~\cite{Pontecorvo:1967fh,
  Gribov:1968kq}.  In the last decade this picture has become fully
proved thanks to the upcoming of a set of precise experiments. In
particular, the results obtained with solar~\cite{Cleveland:1998nv,
  Kaether:2010ag, Abdurashitov:2009tn, Hosaka:2005um, Aharmim:2007nv,
  Aharmim:2005gt, Aharmim:2008kc, Aharmim:2009gd, Arpesella:2008mt,
  Bellini:2008mr} and atmospheric neutrinos~\cite{Ashie:2005ik,
  Wendell:2010md} have been confirmed in experiments using terrestrial
beams: neutrinos produced in nuclear reactors~\cite{Shimizu:2008zz,
  Apollonio:1999ae} and accelerators~\cite{Ahn:2006zza,
  Adamson:2008zt, Adamson:2009yc, Adamson:2010uj} facilities have been
detected at distances of the order of hundreds of
kilometers~\cite{GonzalezGarcia:2007ib}.

The minimum joint description of all the neutrino data requires mixing
among all the three known neutrinos ($\nu_e$, $\nu_\mu$, $\nu_\tau$),
which can be expressed as quantum superpositions of three massive
states $\nu_i$ ($i=1,2,3$) with masses $m_i$.  This implies the
presence of a leptonic mixing matrix in the weak charged current
interactions~\cite{Maki:1962mu, Kobayashi:1973fv} which can be
parametrized as:
\begin{equation}
  U =
  \begin{pmatrix}
    1 & 0 & 0 \\
    0 & c_{23}  & s_{23} \\
    0 & -s_{23} & c_{23}
  \end{pmatrix}
  \cdot
  \begin{pmatrix}
    c_{13} & 0 & s_{13} e^{-i\delta_\CP} \\
    0 & 1 & 0 \\
    -s_{13} e^{i\delta_\CP} & 0 & c_{13}
  \end{pmatrix}
  \cdot
  \begin{pmatrix}
    c_{21} & s_{12} & 0 \\
    -s_{12} & c_{12} & 0 \\
    0 & 0 & 1
  \end{pmatrix}
  \cdot
  \begin{pmatrix}
    e^{i \eta_1} & 0 & 0 \\
    0 & e^{i \eta_2} & 0 \\
    0 & 0 & 1
  \end{pmatrix},
  \label{eq:U3m}
\end{equation}
where $c_{ij} \equiv \cos\theta_{ij}$ and $s_{ij} \equiv
\sin\theta_{ij}$.  In addition to the Dirac-type phase $\delta_\CP$,
analogous to that of the quark sector, there are two physical phases
$\eta_i$ associated to the Majorana character of neutrinos and which
are not relevant for neutrino oscillations~\cite{Bilenky:1980cx,
  Langacker:1986jv}.

Given the observed hierarchy between the solar and atmospheric
mass-squared splittings there are two possible non-equivalent
orderings for the mass eigenvalues, which are conventionally chosen as
\begin{align}
  \label{eq:normal}
  m_1 < m_2 < m_3
  \quad\text{with}\quad
  \Dmq_{21} &\ll |\Dmq_{31} \simeq \Dmq_{32}|
  \quad\text{and}\quad \Dmq_{31} > 0 \,;
  \\
  \label{eq:inverted}
  m_3 < m_1 < m_2
  \quad\text{with}\quad
  \Dmq_{21} &\ll |\Dmq_{31} \simeq \Dmq_{32}|
  \quad\text{and}\quad
  \Dmq_{31} < 0 \,.
\end{align}
As it is customary we refer to the first option,
Eq.~\eqref{eq:normal}, as the \emph{normal} (N) scheme, and to the
second one, Eq.~\eqref{eq:inverted}, as the \emph{inverted} (I)
scheme; in this form they correspond to the two possible choices of
the sign of $\Dmq_{31}$.  In this convention the angles $\theta_{ij}$
can be taken without loss of generality to lie in the first quadrant,
$\theta_{ij} \in [0, \pi/2]$, and the phases $\delta_\CP$, $\eta_i\in
[0, 2\pi]$.

Thanks to the synergy amongst a variety of experiments involving solar
and atmospheric neutrinos, as well as man-made neutrinos at nuclear
power plants and accelerators, we have now a relatively detailed
picture of the parameters describing three--flavor neutrino
oscillations~\cite{GonzalezGarcia:2010er, Fogli:2009zza,
  Schwetz:2008er, Maltoni:2008ka}.

\subsubsection{Leading $\Dmq_{21}$ oscillations: solar and KamLAND data}
\label{sec:solar}

\begin{figure}[t]
  \includegraphics[width=0.6\textwidth]{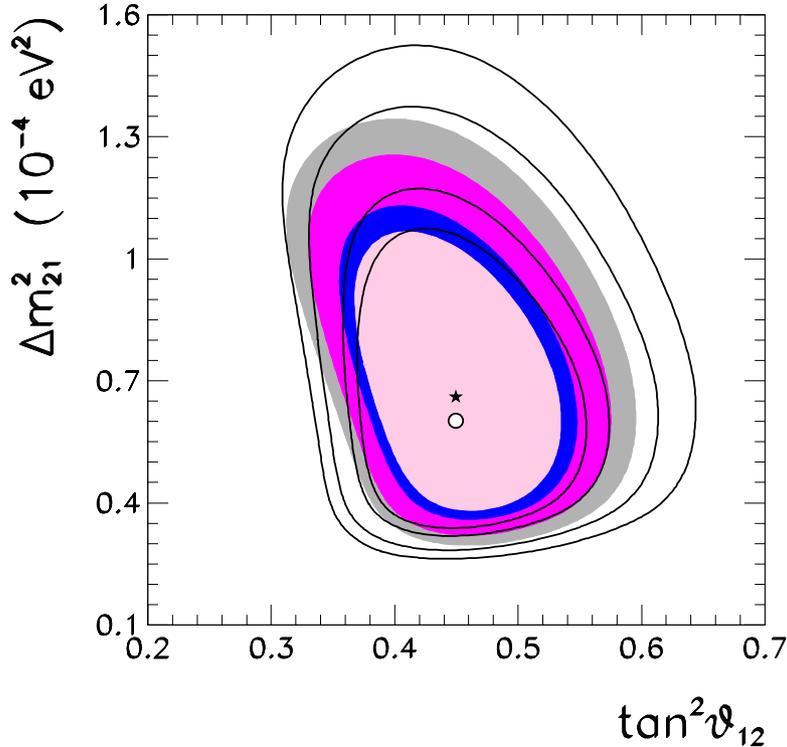}
  \caption{\label{fig:soltwo}
    Allowed parameter regions (at 90\%, 95\%, 99\% and 99.73\% CL for
    2 d.o.f.) from the combined analysis of solar data for
    $\theta_{13}=0$.  The best-fit point is marked with a star.  For
    comparison we also show as empty regions (the best-fit is marked
    by a circle) the results prior to the inclusion of the latest Ga
    capture rate of SAGE~\cite{Abdurashitov:2009tn}, the energy
    spectrum of Borexino~\cite{Arpesella:2008mt,
      Bellini:2008mr} and the low energy threshold analysis of
    the combined SNO phase I and phase
    II~\cite{Aharmim:2009gd}. In both analysis we use as
    inputs the GS98 solar model fluxes and the Gallium capture
    cross-section of Bahcall~\cite{Bahcall:1997eg}.}
\end{figure}

In the analysis of solar neutrino experiments we include the total
rates from the radiochemical experiments
Chlorine~\cite{Cleveland:1998nv}, Gallex/GNO~\cite{Kaether:2010ag} and
SAGE~\cite{Abdurashitov:2009tn}. For real-time experiments we include
the 44 data points of the electron scattering (ES) Super-Kamiokande
phase I (SK-I) energy-zenith spectrum~\cite{Hosaka:2005um} and the
data from the three phases of SNO~\cite{Aharmim:2007nv,
  Aharmim:2005gt, Aharmim:2008kc}, including the results on the low
energy threshold analysis of the combined SNO phase I and phase
II~\cite{Aharmim:2009gd} (which we label SNO-LETA). We also
include the main set of the 192 days of Borexino
data~\cite{Arpesella:2008mt} (which we label Borexino-LE) as well as
their high-energy spectrum from 246 live
days~\cite{Bellini:2008mr} (Borexino-HE).

In Fig.~\ref{fig:soltwo} we show the present determination of the
leading parameters $\Dmq_{21}$ and $\theta_{12}$ from the updated
oscillation analysis of the solar neutrino data described above in the
context of the GS98 solar model.  For comparison we also show the
results obtained prior to the inclusion of the latest Ga capture rate
of SAGE~\cite{Abdurashitov:2009tn}, the energy spectrum of
Borexino~\cite{Arpesella:2008mt, Bellini:2008mr} and the
SNO-LETA results~\cite{Aharmim:2009gd} for the same solar model.
As seen in this figure, the inclusion of these results lead to an
improvement on the determination of both $\theta_{12}$ and $\Dmq_{21}$
and for this last one the best-fit value slightly increases.
The most quantitatively relevant new information arises from the
inclusion of the SNO-LETA results. The inclusion of Borexino tends to
shift the region towards slightly lower values of $\theta_{12}$
angle. Conversely, if the analysis is done in the context of the
AGSS09 model the region is shifted towards slightly larger
$\theta_{12}$.

\subsubsection{Leading $\Dmq_{31}$ oscillations: atmospheric and accelerator data}
\label{sec:atmos}

\begin{figure}[t]
  \includegraphics[width=0.95\textwidth]{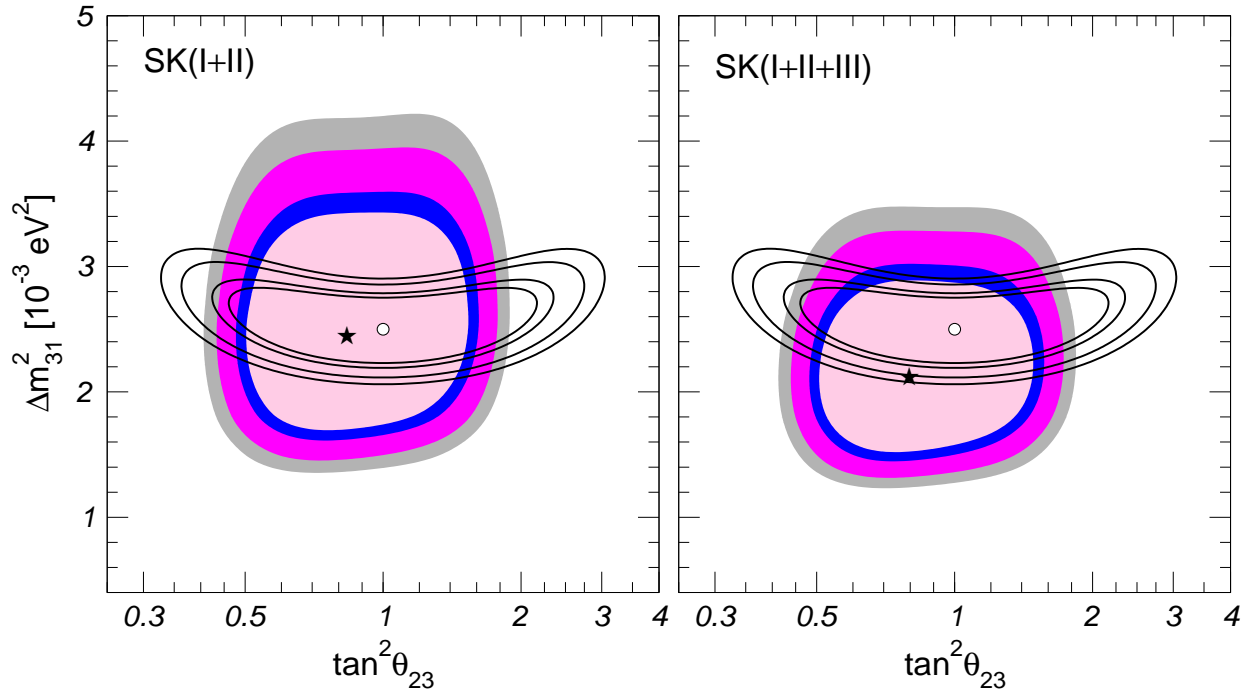}
  \caption{\label{fig:atmtwo}
    Allowed parameter regions (at 90\%, 95\%, 99\% and 99.73\% CL for
    2 d.o.f.) from the analysis of atmospheric data (full regions,
    best-fit marked with a star) and LBL data (void regions, best-fit
    marked by a circle) for $\theta_{13} = 0$ and $\Dmq_{21} = 7.6
    \times 10^{-5}~\eVq$.}
\end{figure}

In this section we present two different analyses of the atmospheric
data. The first one is very similar to the one detailed in
Ref.~\cite{GonzalezGarcia:2007ib}, and includes the results from the
first run of Super-Kamiokande, which accumulated data from May 1996 to
July 2001 (1489 day exposure) and is usually referred as
SK-I~\cite{Ashie:2005ik}, as well as the data obtained with the
partial coverage after the 2001 accident (804 day exposure), the
so-called SK-II period~\cite{atmskII}. We will refer to this analysis
as SK(I+II).
The second one is based on the new analysis recently presented by the
Super-Kamiokande collaboration including also the data taken from
December 2005 to June 2007, usually referred as
SK-III~\cite{Wendell:2010md}.  Apart from the inclusion of these new
event rates, in this data release the selection criteria and the
corresponding estimate of uncertainties for the SK-I and SK-II periods
have been changed with respect to the previous SK(I+II) analysis. We
have therefore performed a reanalysis of the new combined samples from
phases I, II and III as presented in~\cite{Wendell:2010md}. We refer
to the results of this analysis as SK(I+II+III).
It is important to point out that already since SK-II the
Super-Kamiokande collaboration has been presenting its experimental
results in terms of a large number of data samples.  The rates for
some of those samples cannot be theoretically predicted (and therefore
include in a statistical analysis) without a detailed simulation of
the detector which can only be made by the experimental collaboration
itself. Thus our results represent the most up-to-date analysis of
the atmospheric neutrino data which can be performed outside the
collaboration.  For details on our simulation of the data samples and
the statistical analysis see the Appendix of
Ref.~\cite{GonzalezGarcia:2007ib}.

For what concerns LBL accelerator experiments, we combine the results
on $\nu_\mu$ disappearance from K2K~\cite{Ahn:2006zza} with those
obtained by MINOS at a baseline of 735~km after a two-year exposure to
the Fermilab NuMI beam, corresponding to a total of $3.36 \times
10^{20}$ protons on target~\cite{Adamson:2008zt}.  We also include the
recent results on $\nu_\mu\rightarrow \nu_e$ transitions based on an
exposure of $7 \times 10^{20}$ protons on target~\cite{minapp70fnal,
  minapp70slac}.

In order to test the description of the present data in the absence of
$\theta_{13}$-induced effects we show in Fig.~\ref{fig:atmtwo} the
present determination of the leading parameters $\Dmq_{31}$ and
$\theta_{23}$ for $\theta_{13}=0$ and $\Dmq_{21}=7.6 \times
10^{-5}~\eVq$ from the two atmospheric neutrino analyses and the LBL
accelerator results. For concreteness we plot only normal ordering;
the case of inverted ordering gives practically identical results as
long as $\theta_{13} = 0$. This figure illustrates how the bounds on
the oscillation parameters $\theta_{23}$ and $\Dmq_{31}$ emerges from
a complementarity of atmospheric and accelerator neutrino data:
$|\Dmq_{31}|$ is determined by the spectral data from MINOS, whereas
the mixing angle $\theta_{23}$ is still largely dominated by
atmospheric data from Super-Kamiokande with a best-fit point close to
maximal mixing.
It is interesting to note that there is a very good agreement in the
location of the best-fit points from SK(I+II) and MINOS. This is not
the case for SK(I+II+III) for which the best-fit point in
$|\Dmq_{31}|$ is now lower than the one obtained from LBL.

\subsubsection{Status of $\theta_{13}$ from global data in 2010}

\begin{figure}\centering
  \includegraphics[width=\textwidth]{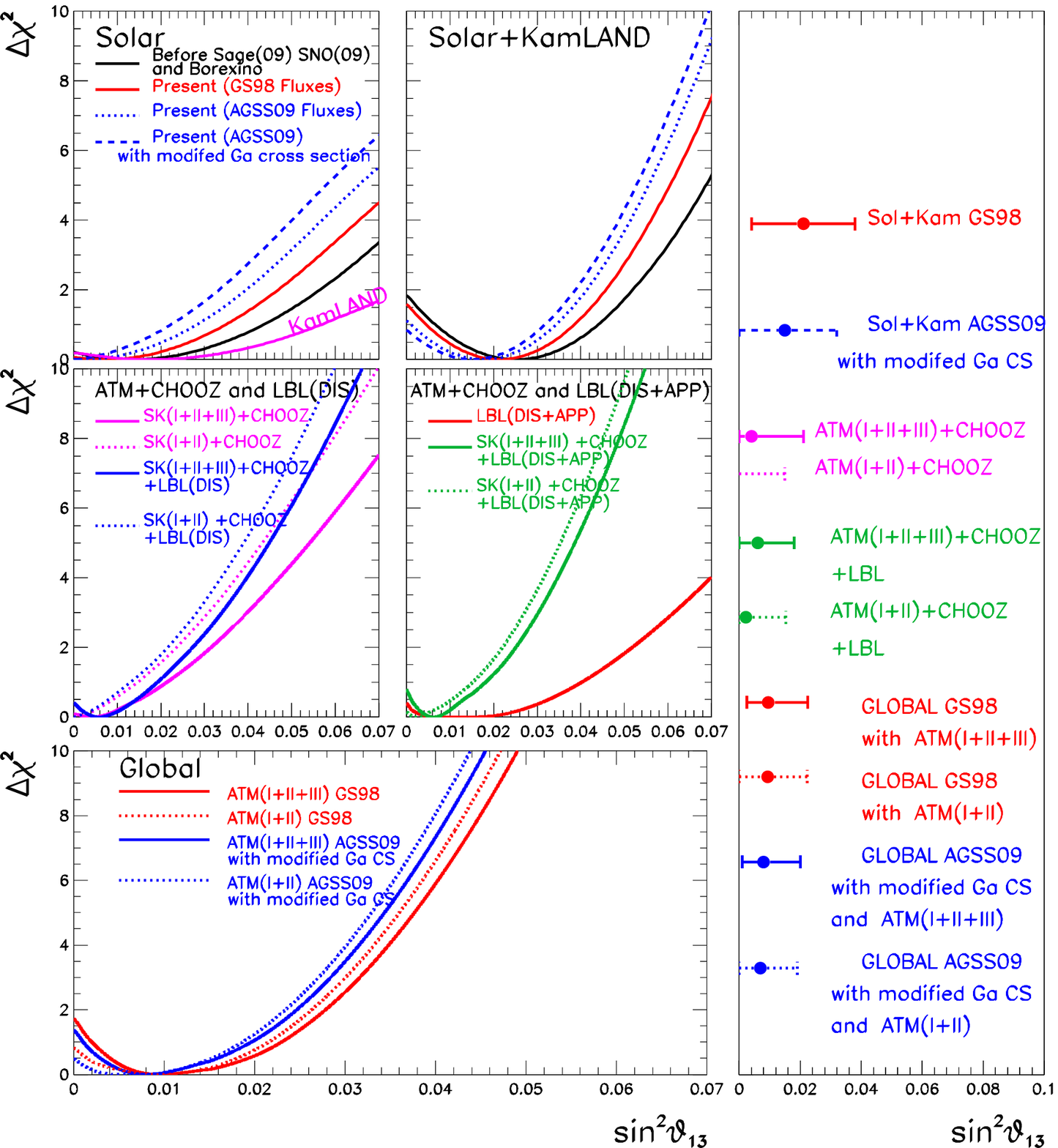}
  \caption{\label{fig:th13}
    $\Delta\chi^2$ dependence on $\sin^2\theta_{13}$ from various data
    sets as labeled in the figure. The right panel shows $1\sigma$
    ranges~\cite{GonzalezGarcia:2010er}. AGSS09 and GS98 refer to low
    and high metallicity solar models,
    respectively~\cite{Serenelli:2009yc}.}
\end{figure}

The third mixing angle $\theta_{13}$ is of crucial importance for
future oscillations experiments.
Fig.~\ref{fig:th13} summarizes the information on $\theta_{13}$
from present data, which emerges from an interplay of different data
sets. An important contribution to the bound comes, of course, from
the CHOOZ reactor experiment~\cite{Apollonio:2002gd} combined with the
determination of $|\Dmq_{31}|$ from atmospheric and long-baseline
experiments.
Using this set of data, a possible hint for a non-zero $\theta_{13}$
from atmospheric data has been found in Refs.~\cite{Fogli:2005cq,
Fogli:2008jx}. The origin of such a hint has been investigated in some
detail in Ref.~\cite{Maltoni:2008ka}, and more recently in
\cite{Fogli:2009ce, GonzalezGarcia:2010er}.
From these results one may conclude that the statistical relevance of
the hint for non-zero $\theta_{13}$ from atmospheric data depends
strongly on the details of the rate calculations and of the $\chi^2$
analysis. Furthermore, the origin of that effect might be traced back
to a small excess (at the $1\sigma$ level) in the multi-GeV $e$-like
data sample in SK-I, which however is no longer present in the
combined SK(I+II) data and is extremely weak in SK(I+II+III) data.
A very recent analysis (neglecting subleading $\Dmq_{21}$ effects)
from the Super-Kamiokande collaboration finds no evidence of such a
hint~\cite{Wendell:2010md}.

Another fragile indication of non-zero $\theta_{13}$ arises from the
results of the MINOS experiment.  In Ref.~\cite{Adamson:2009yc} the
first results on the search for $\nu_\mu\rightarrow\nu_e$ transitions
were reported, based on an exposure of $3.14 \times 10^{20}$
protons-on-target in the Fermilab NuMI beam.  The collaboration
observed 35 events in the Far Detector with a background of $27\pm 5\,
\text{(stat)} \pm 2\, \text{(syst)}$, corresponding to a $1.5\sigma$
excess which could be explained by a non-zero value of $\theta_{13}$.
Recently a new analysis with double statistics (exposure of $7 \times
10^{20}$) has been presented~\cite{minapp70fnal, minapp70slac}.  The
MINOS collaboration reported the observation of 54 events with an
expected background of $49.1\pm 7.0\, \text{(stat)} \pm 2.7\,
\text{(syst)}$, thus reducing the excess above background to
$0.7\sigma$.

An important piece of information on $\theta_{13}$ comes from solar
and KamLAND data. The relevant survival probabilities are given by
\begin{equation}\label{eq:Pee}
  P_{ee} \approx
  \begin{cases}
    \cos^4\theta_{13} \left(1- \sin^22\theta_{12}\right \langle \sin^2\phi\rangle)
    & \text{solar, low energies / KamLAND}
    \\
    \cos^4\theta_{13} \, \sin^2\theta_{12}
    & \text{solar, high energies}
  \end{cases}
\end{equation}
where $\phi = \Dmq_{21} L / 4E$ and $\langle \sin^2\phi\rangle \approx 1/2$
for solar neutrinos.  Eq.~(\ref{eq:Pee}) implies an anti-correlation
of $\sin^2\theta_{13}$ and $\sin^2\theta_{12}$ for KamLAND and low
energy solar neutrinos. In contrast, for the high energy part of the
spectrum, which undergoes the adiabatic MSW conversion inside the sun
and which is subject to the SNO CC/NC measurement, a positive
correlation of $\sin^2\theta_{13}$ and $\sin^2\theta_{12}$ emerges. As
discussed already in~\cite{Maltoni:2004ei, Goswami:2004cn}, this
complementarity leads to a non-trivial constraint on $\theta_{13}$ and
it allows to understand the hint for a non-zero value of
$\theta_{13}$, which helps to reconcile the slightly different best
fit points for $\theta_{12}$ as well as for $\Dmq_{21}$ for solar and
KamLAND separately~\cite{Balantekin:2008zm, Goswami:2004cn,
  Fogli:2008jx, Maltoni:2003da, GonzalezGarcia:2007ib}.

We found that the inclusion of the new solar data, and in particular
of the SNO-LETA results tends to lower the statistical significance of
$\theta_{13}\neq 0$ while the results from $\nu_e$ appearance from
MINOS increases it.  Within the context of the solar model with higher
metallicities (GS98) and for the original Ga capture
cross-section~\cite{Bahcall:1997eg}, we conclude that the significance
of $\theta_{13}\neq 0$ from solar+KamLAND data is 79\% ($1.26\sigma$)
which increases to 81\% ($1.31\sigma$) after inclusion of the
atmospheric, CHOOZ and LBL data. We also found that using the solar
neutrino fluxes required to fit the lower metallicity data (AGSS09)
and/or the modified (lower) cross-section for neutrino capture in Ga
lowers the best fit value of $\theta_{13}$ and its statistical
significance. So when using the AGSS09 fluxes and the lower Ga
cross-section the significance of $\theta_{13}\neq 0$ from
solar+KamLAND data is 70\% ($1.05\sigma$) and 76\% ($1.17\sigma$) for
adding atmospheric, CHOOZ and LBL data.

\subsubsection{Tritium beta decay experiments}

The neutrino mass scale is constrained in laboratory experiments
searching for its kinematic effects in Tritium $\beta$ decay which are
sensitive to the so-called effective electron neutrino
mass~\cite{Shrock:1980vy, Vissani:2000ci, Farzan:2001cj}
\begin{equation}
  \label{eq:mb}
  m^2_{\nu_e} \equiv 
  \sum_i m^2_i |U_{ei}|^2=c_{13}^2 c_{12}^2 m_1^2
  + c_{13}^2 s_{12}^2 m_2^2+s_{13}^2 m_3^2 \,,
\end{equation}
At present the most precise determination from the
Mainz~\cite{Bonn:2001tw} and Troitsk~\cite{Lobashev:2001uu}
experiments give no indication in favor of $m_{\nu_e}\neq 0$ and one
sets an upper limit
\begin{equation}
  \label{eq:nuelim}
  m_{\nu_e} < 2.2~\text{eV} \,, 
\end{equation}
at 95\% confidence level (CL).  A new experimental project,
KATRIN~\cite{Osipowicz:2001sq}, is under construction with an
estimated sensitivity limit: $m_{\nu_e} \sim 0.2$ eV.

\subsubsection{Neutrinoless double-beta decay experiments}

Direct information on neutrino masses can also be obtained from
neutrinoless double beta decay ($0\nu\beta\beta$) searches provided
they are Majorana particles. In the absence of other sources of lepton
number violation in the low energy Lagrangian, the $0\nu\beta\beta$
decay amplitude is proportional to the effective Majorana mass of
$\nu_e$, $m_{ee}$,
\begin{equation}
  \label{eq:mbb}
  m_{ee} = \left| \sum_i m_i U_{ei}^2 \right|
  = \left|
  c_{13}^2 c_{12}^2 m_1 \, {e}^{i\eta_1} +
  c_{13}^2 s_{12}^2 m_2 \, {e}^{i\eta_2} +
  s_{13}^2 m_3 \, {e}^{-i\delta_\CP} \right| \,,
\end{equation}
which, in addition to the masses and mixing parameters that affect the
tritium beta decay spectrum, depends also on the phases in the
leptonic mixing matrix. The strongest bound from $0\nu\beta\beta$
decay was imposed by the Heidelberg-Moscow group~\cite{KlapdorKleingrothaus:2000sn}
\begin{equation}
  m_{ee} < 0.26~(0.34)~\text{eV}
  \quad \text{at 68\% (90\%) CL,}
\end{equation}
which holds for a given prediction of the nuclear matrix
element. However, there are large uncertainties in those predictions
which may considerably weaken the bound~\cite{Vogel:2006sq}.  A series
of new experiments is planned with sensitivity of up to $m_{ee} \sim
0.01$ eV~\cite{Avignone:2007fu}.

\subsubsection{The impact of cosmological fits}

Neutrino oscillation data provides as unique information on the
absolute neutrino mass scale a lower bound
\begin{equation}
  \Sigma_\nu \equiv \sum_i m_i \gtrsim
  \begin{cases}
    \hphantom{1} \sqrt{|\Dmq_{31}|} 
    & \text{for Normal hierarchy,}
    \\[1mm]
    2 \sqrt{|\Dmq_{31}|}
    & \text{for Inverted hierarchy.}
  \end{cases}
\end{equation}
Furthermore, neutrinos, like any other particles, contribute to the
total energy density of the Universe.  Furthermore within what we
presently know of their masses, the three Standard Model (SM)
neutrinos are relativistic through most of the evolution of the
Universe and they are very weakly interacting which means that they
decoupled early in cosmic history.  Depending on their exact masses
they can impact the CMB spectra, in particular by altering the value
of the redshift for matter-radiation equality.  More importantly,
their free streaming suppresses the growth of structures on scales
smaller than the horizon at the time when they become non-relativistic
and therefore affects the matter power spectrum which is probed from
surveys of the LSS distribution (see~\cite{Lesgourgues:2006nd} for a
detailed review of cosmological effects of neutrino mass).
 
Within their present precision, cosmological observations are
sensitive to neutrinos only via their contribution to the energy
density in our Universe, $\Omega_\nu h^2$ (where $h$ is the Hubble
constant normalized to $H_0 = 100 ~\text{km} ~\text{s}^{-1}
~\text{Mpc}^{-1}$). $\Omega_\nu h^2$ is related to the total mass in
the form of neutrinos
\begin{equation} 
  \Omega_\nu h^2 = \Sigma_\nu \big/ (94~\text{eV}) \,.
\end{equation}
Therefore cosmological data mostly gives information on the sum of the
neutrino masses and has very little to say on their mixing structure
and on the ordering of the mass states (see Ref.~\cite{Jimenez:2010ev}
for a recent update on the sensitivity of future cosmological
observations to the mass ordering.)

\begin{figure}[t]
  \includegraphics[width=0.7\textwidth]{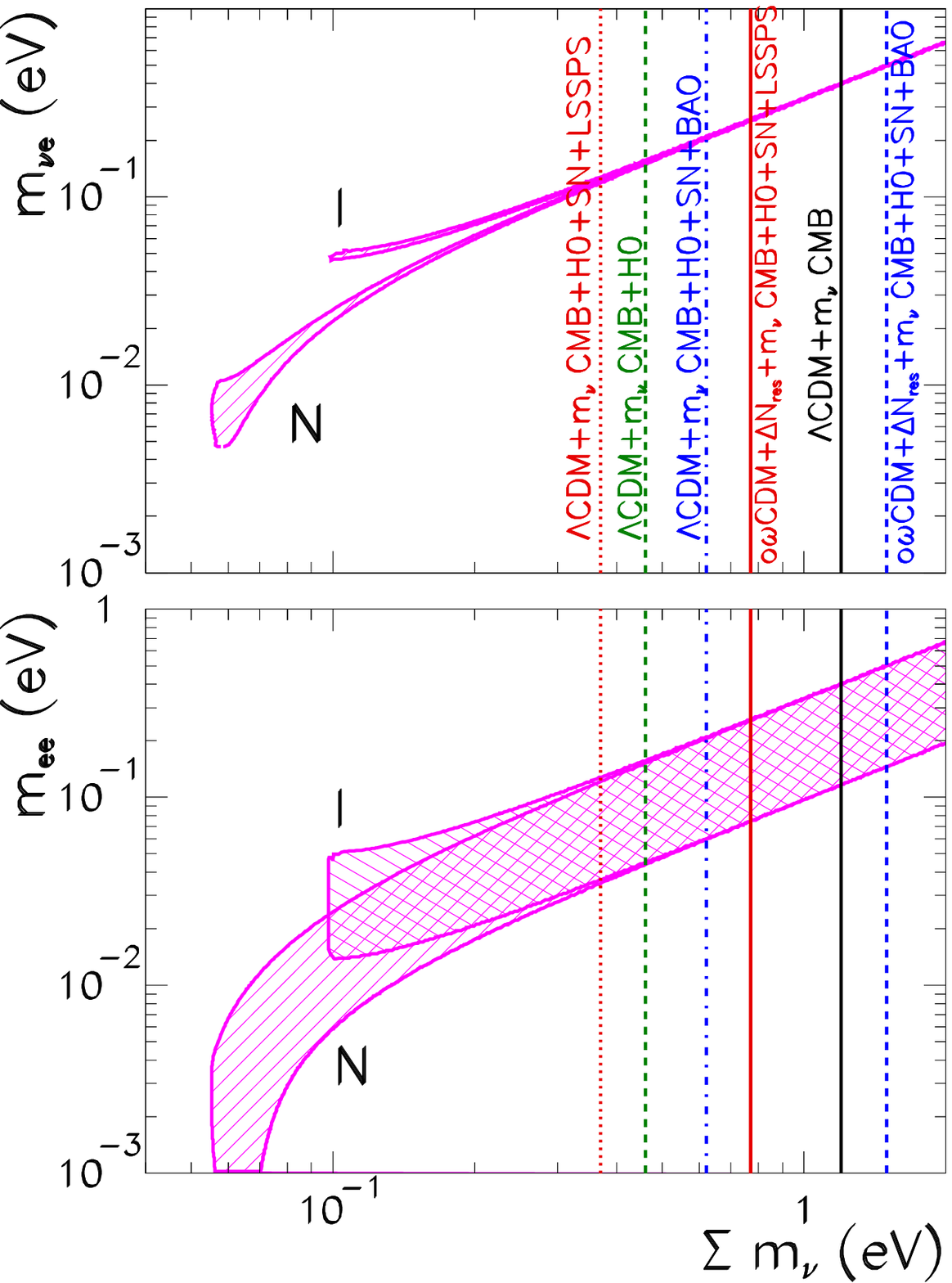}
  \caption{\label{fig:mbeta}
    95\% allowed regions (for 2 dof) in the planes ($m_{\nu_e}$,
    $\Sigma_\nu$) and ($m_{ee}$, $\Sigma_\nu$) from the global
    analysis of oscillation data (full regions). We also show
    superimposed the 95\% upper bounds on $\Sigma_\nu$ from
    cosmological constraints for the different analysis as labeled in
    the figure.}
\end{figure}

In Ref.~\cite{GonzalezGarcia:2010un} we have studied the information
on the absolute value of the neutrino mass which can be obtained from
the analysis of the cosmological data in $o\omega\text{CDM} + \Delta
N_\text{rel} + m_\nu$ cosmologies where, besides neutrino masses, one
allows for non-vanishing curvature, dark energy with equation of state
with $\omega\neq -1$ together with the presence of new particle
physics whose effect on the present cosmological observations can be
parametrized in terms of additional relativistic degrees of
freedom. To break the degeneracies in these models, at least the
information from four different cosmological probes must be combined.
Thus we have performed analysis including the data from CMB
experiments, the present day Hubble constant H0, measurement, the
high-redshift Type-I SN results and the information from large scale
LSS surveys. 

In Fig.~\ref{fig:mbeta} we plot the 95\% allowed regions (for 2 dof)
in the planes ($m_{\nu_e}$, $\Sigma_\nu$) and ($m_{ee}$,
$\Sigma_\nu$). In the figure we also show superimposed the single
parameter 95\% bounds on $\Sigma_\nu$ from different cosmological
analysis.  The figure illustrates the well-known fact that currently
for either mass ordering the results from neutrino oscillation
experiments imply a lower bound on $m_{\nu_e}$. On the contrary
$m_{ee}$ is only bounded from below for the case of the normal
ordering while full cancellation due to the unknown Majorana phases is
still allowed for the inverted ordering.
These results show that, even for the most restrictive analysis
including LSSPS, part of the allowed ranges for $m_{\nu_e}$ in the
context of the $o\omega\text{CDM} + \Delta N_\text{rel} + m_\nu$
cosmologies are within the reach of the KATRIN experiment. This is not
the case for $\Lambda\text{CDM} + m_\nu$ models unless only the
information of CMB and BAO (or SN) is included.  We also find that
near future neutrinoless double beta decay can test some of the
allowed ranges in all these scenarios.  This will be complementary to
the improvement on the expected sensitivity from upcoming cosmological
probes such as the Planck mission~\cite{Planck:2006uk}.

%% file: CONTENT/02MezzettoSchwetz.tex
\subsection{$\theta_{13}$: phenomenology, present status and prospect \cite{Mezzetto:2010zi}}
\label{sec:mezzettoschwetz}

The leptonic mixing angle $\theta_{13}$ is currently a high-priority topic in the field
of neutrino physics, with five experiments under way, searching for neutrino oscillations
induced by this angle: the reactor neutrino experiments Daya Bay~\cite{Guo:2007ug},
Double Chooz~\cite{Ardellier:2006mn}, RENO~\cite{Ahn:2010vy} and the accelerator experiments NO$\nu$A~\cite{Ambats:2004js} and
T2K~\cite{Itow:2001ee}. The results of these experiments will be essential for the planning towards
a possible next generation of long-baseline neutrino experiments able to
address leptonic CP violation and the neutrino mass hierarchy.

On the theoretical side, the determination of $\theta_{13}$ will provide important information
on the mechanism of neutrino mass generation and the flavour structure
in the lepton sector. Considering neutrino mass models without any flavour
structure, so-called anarchical models, one does expect a value of $\theta_{13}$ close to the
present bound~\cite{deGouvea:2003xe}. If on the contrary experiments would indicate a very tiny value
for $\theta_{13}$ one might wish to have a symmetry reason as an explanation. For example,
rather symmetric patterns for the mixing matrix are the tri-bimaximal~\cite{Harrison:2002er} or
the bimaximal~\cite{Barger:1998ta} mixing matrices.

The present situation obtained in the global fit of all relevant oscillation data,
can be summarized according to the updated analysis1 of~\cite{Schwetz:2008er}. 
The following bounds at 90\% (3) CL are obtained:

\begin{equation}
\sin^2 \theta_{13} \leq 
\left \{
\begin{array}{ll}
0.053 (0.078) & {\rm solar+KamLAND} \\
0.033 (0.058) & {\rm CHOOZ+atm+K2K+MINOS} \\
0.031 (0.047) & {\rm global \, data}
\end{array}
\right .
\end{equation}

The ÒhintÓ for $\theta_{13} > 0$ coming from the different data sets can be quantified by considering the $\Delta \chi^2$ for $\theta_{13} = 0$:\begin{equation}
\sin^2 \theta_{13} \leq 
\left \{
\begin{array}{ll}
2.2 (1.5) & {\rm solar+KamLAND} \\
0.8 (0.9) & {\rm CHOOZ+atm+K2K+MINOS} \\
0.6 (0.7) & {\rm MINOS} \, \nu_e \, {\rm appearance} \\
1.8 (1.3) & {\rm global \, data}
\end{array}
\right .
\end{equation}

In table \ref{tab:fit-comparison} are compared the best-fit values for $\sin^2 \theta_{13}$ and the significance of
the hint for $\theta_{13} > 0$ from the global fits to neutrino oscillation data from three
different groups.
All groups find a non-zero best-fit point in the range $\sin^2 \theta_{13} = 0.01- 0.02$.
While it is premature to draw strong conclusions from these results, upcoming
experiments will answer very soon the question whether $\theta_{13}$ is indeed in the range
indicated by present global analyses.
Reactor experiments see a large signal of $\bar \nu_e$ events, and search for a small
deviation from the non-oscillation prediction due to $\theta_{13}$-induced $\bar \nu_e$ disappearance.
These are a precision experiment, whose success relies on statistical as well
as systematical errors below the percent level. Table \ref{tab:exp-summary} summarises a few key
parameters of reactor experiments.
Accelerator experiments look for the appearance of the $\nu_e$ flavour in an almost
pure $\nu_\mu$ beam, due to oscillations.

\begin{table}
  \centering
  \begin{tabular}{lcc}
  \hline
    reference & best-fit and $1\sigma$ errors & significance \\
  \hline
    Fogli et al.~\cite{Fogli:2009ce} &
    $\sin^2\theta_{13} = 0.02\pm 0.01$ &
    $2\sigma$ \\
    Gonzalez-Garcia et al.~\cite{GonzalezGarcia:2010er} (GS98) &
    $\sin^2\theta_{13} = 0.0095^{+0.013}_{-0.007}$ &
    $1.3\sigma$ \\
    Gonzalez-Garcia et al.~\cite{GonzalezGarcia:2010er} (AGSS09) &
    $\sin^2\theta_{13} = 0.008^{+0.012}_{-0.007}$ &
    $1.1\sigma$ \\
    Schwetz et al.~\cite{Schwetz:2008er} (GS98) &
    $\sin^2\theta_{13} = 0.013^{+0.013}_{-0.010}$ &
    $1.5\sigma$\\
    Schwetz et al.~\cite{Schwetz:2008er} (AGSS09) &
    $\sin^2\theta_{13} = 0.010^{+0.013}_{-0.008}$ &
    $1.3\sigma$\\
  \hline
  \end{tabular}
  \caption{Comparison of the best-fit values for $\sin^2\theta_{13}$ and the
  significance of the hint for $\theta_{13} > 0$ from different global fits
  to neutrino oscillation data. The numbers from
  \cite{GonzalezGarcia:2010er} and \cite{Schwetz:2008er} include $7\times
  10^{20}$~pot $\nu_e$ appearance data from MINOS,
  whereas~\cite{Fogli:2009ce} is based on $3.14 \times 10^{20}$~pot. AGSS09
  and GS98 refer to low and high metallicity solar models,
  respectively~\cite{Serenelli:2009yc}. \label{tab:fit-comparison}}
\end{table}

\begin{table}
\begin{tabular}{lcccrcc}
\hline
Setup &  $P_{\mathrm{Th}}$ [GW] & $L$ [m] & $m_{\mathrm{Det}}$ [t] & Events/year & Backgrounds/day\\
\hline
 Daya Bay~\cite{Guo:2007ug} &  17.4  & 1700  &   80\hspace{1ex}  & $10 \hspace{1.7ex}\cdot 10^4$ & 0.4 \\
 Double Chooz~\cite{Ardellier:2006mn} &  \hspace{1ex}8.6  & 1050 &   \hspace{1ex}8.3 & $  1.5 \cdot 10^4$ & 3.6 \\
 RENO~\cite{Ahn:2010vy}    &  16.4  & 1400  &  15.4  & $3\hspace{1.7ex} \cdot 10^4$ & 2.6 \\
\hline
\end{tabular}
\caption{\label{tab:exp-summary} Summary of experimental key parameters of
upcoming reactor neutrino experiments. We give the thermal reactor power,
the approximate distance between reactors and far detector, and detector
mass, neutrino events per year, and background events per day, all for the
far detector. RENO backgrounds are the sum of correlated backgrounds as
computed in \protect \cite{Ahn:2010vy} and uncorrelated backgrounds as
estimated in \protect \cite{Mention:2007um}.}
\end{table}

The T2K (TokaiÐtoÐKamioka) experiment~\cite{Itow:2001ee} will use a high intensity off-axis
($2.5^\circ$) neutrino beam, with a peak energy of 700 MeV, generated by a 30 GeV
proton beam at J-PARC (Japan Proton Accelerator Research Complex) fired to
the Super Kamiokande detector, located 295 km from the proton beam target.

The NO$\nu$A experiment~\cite{Ambats:2004js}  will run at an upgraded NuMI neutrino beam
expected to deliver $6.5\times 10^{20}$ pot/year, corresponding to a beam power of 700
kW, generating a neutrino beam with an average energy $E_\nu \sim 2$ GeV and a $\nu_e$
contamination less than 0:5\%. The far detector, placed at baseline of 810 km,
14 mrad ($0.8^\circ$) off-axis, will be a Òtotally activeÓ tracking liquid scintillator,
constructed from liquid scintillator contained inside extruded PVC cells.

\begin{figure}
\centering \includegraphics[width=0.6\textwidth]{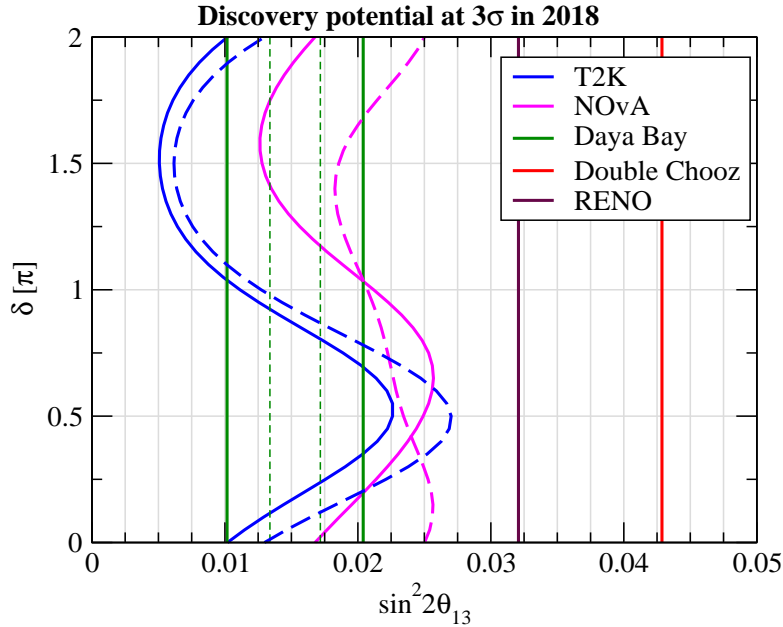}
   \caption{\label{fig:th-del} Discovery potential of the five upcoming
   experiments in the plane of $\stheta$ and $\delta$ expected in 2018, see
   section 5.3 of Ref.~\cite{Mezzetto:2010zi} for our assumptions on exposure. To the right of
   the curves a non-zero value of $\theta_{13}$ can be established at
   $3\sigma$. For the beam experiments we show normal (solid) and inverted
   (dashed) hierarchies, while reactor experiments are independent of the
   hierarchy. The four lines for Daya Bay correspond to different
   assumptions on the achieved systematic uncertainty, from weakest to
   strongest sensitivity: 0.6\% correlated among detector modules at one
   site, 0.38\% correlated, 0.38\% uncorrelated among modules, 0.18\%
   uncorrelated.}
\end{figure}

Fig. \ref{fig:th-del} shows the $\theta_{13}$ discovery reach of the five upcoming experiments expected
in 2018. It is clear from the figure that the discovery potential of the appearance
experiments strongly depends on the CP-phase as well as on the neutrino mass
hierarchy. We observe that the inverted hierarchy gives a weaker sensitivity.
Hence, in case no appearance signal is found the final $\theta_{13}$ limit will be set by the
IH. The different shape of the IH curve for NO$\nu$A results from the anti-neutrino
running included in the NO$\nu$A run plan. As evident from the figure, reactor
experiments are neither sensitive to the value of $\delta$ nor to the mass hierarchy.

\begin{figure}[t]
\begin{center}
\includegraphics[width=0.6\textwidth]{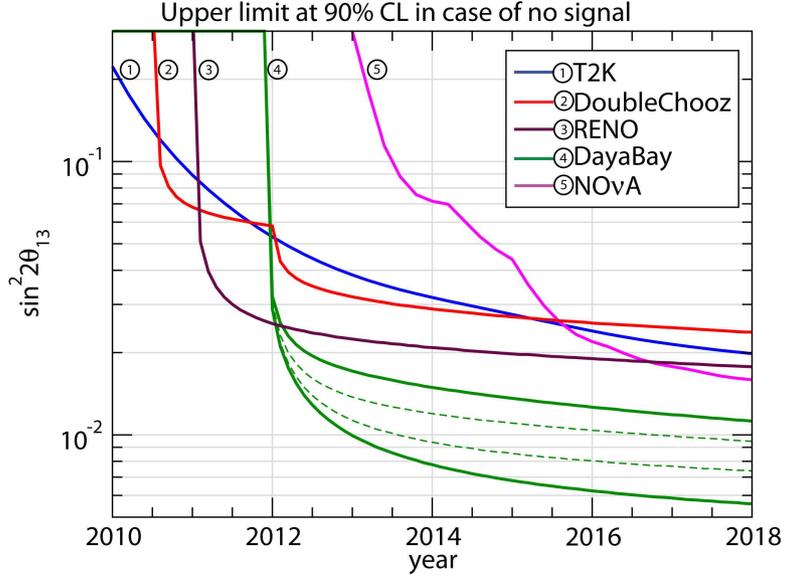}
\end{center}
\caption{\label{fig:evolsens} Evolution of the $\theta_{13}$ sensitivity
  limit as a function of time (90\% CL), i.e., the 90\%~CL limit which will
  be obtained if the true $\theta_{13}$ is zero. The four curves for Daya
  Bay correspond to different assumptions on the achieved systematic
  uncertainty, from weakest to strongest sensitivity: 0.6\% correlated among
  detector modules at one site, 0.38\% correlated, 0.38\% uncorrelated among
  modules, 0.18\% uncorrelated.}
\end{figure}

The sensitivity of the different experiments to $\theta_{13}$ can be discussed using two
different performance indicators: the $\theta_{13}$ sensitivity limit and the $\theta_{13}$ discovery
potential. The $\theta_{13}$ sensitivity limit describes the ability of an experiment to
constrain $\theta_{13}$ if no signal is seen. It is basically determined by the worst case
parameter combination which may fake the simulated $\theta_{13} = 0$.
The $\theta_{13}$ sensitivity limit time evolution is shown in Fig. \ref{fig:evolsens}. We observe that
the global sensitivity limit will be dominated by reactor experiments.

\begin{figure}
  \centering
  \includegraphics[width=0.6\textwidth]{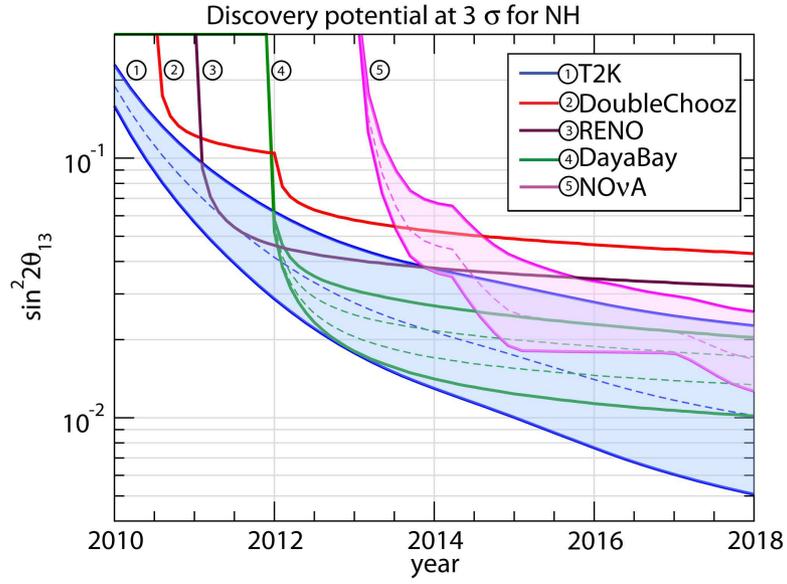}
  \caption{\label{fig:evoldisc} 
  Evolution of the $\theta_{13}$ discovery potential as a function of time (3 $\sigma$ CL)
for NH, showing the global sensitivity reach. The bands for the beams and the
global reach reflect the (unknown) true value of $\delta$. For Daya Bay it is assumed
a systematical uncertainty of 0.38\% correlated among detector modules at one
site.}
\end{figure}

In case of no signal, the $\theta_{13}$ limit from beam experiments suffers from the
marginalization over the CP phase and the mass hierarchy. This situation is very
different in case of the discovery potential, since there a favourable value of  $\delta$ can
greatly enhance the sensitivity of the appearance experiments. The $\theta_{13}$ discovery
potentials are shown in Fig. \ref{fig:evoldisc} as a function of time. For the beam experiments,
the dependence on the true value of $\delta$ is reflected by the interval between the solid
curves for a given time (shaded regions). The dashed curves for T2K and NO$\nu$A
correspond to a fixed value for the CP phase of $\delta = 0$.\footnote{Evolution of sensitivities under this condition have been shown recently in \cite{Mezzetto:2009cr, Suzuki:2010zz}.} 
The reactor experiments are not affected by the true $\delta$; the various curves for Daya Bay again correspond
to the different assumptions concerning systematics as described above.
The comparison of Figs. \ref{fig:evoldisc} and \ref{fig:evolsens} shows that suitable values of $\delta$ may significantly
improve the discovery potential of beams compared to their sensitivity
limit. Indeed, T2K may discover $\theta_{13}$ for smaller $\theta_{13}$ than Daya Bay in a significant
fraction of the parameter space, depending on the achieved systematics in
Daya Bay. The NO$\nu$A band becomes more narrow due to the complementary
information from the anti-neutrino running, with the clear disadvantage of being
somewhat late.

In figure \ref{fig:evoldisc} is also illustrated how the world sensitivity to $\theta_{13}$ could look like
under the assumptions of the above schedules and that at each point in time a
combined analysis of all available data is performed. The discovery reach will be
set roughly by the optimal sensitivity of T2K, where the reactor experiments play
an important role in providing sensitivity for the values of $\delta$ unfavourable for T2K.
This plot nicely illustrates the interplay between reactor and beam experiments
and shows that the global reach can be enhanced significantly if experiments of
both types are available simultaneously with comparable sensitivities.

%% file: CONTENT/03GMSresumsol.tex
\subsection{Direct Determination of the Solar Neutrino Fluxes from Neutrino Data \cite{GonzalezGarcia:2009ya}}

In 1939, Hans Bethe described in an
epochal paper~\cite{Bethe:1939bt} two nuclear fusion mechanisms by
which main sequence stars like the Sun could produce the energy
necessary to power their observed luminosities.  The two mechanisms
have become known as the pp-chain and the
CNO-cycle~\cite{Bahcall:1989ks}.  
In order to precisely determine the rates of the different reactions
in the two chains, which are responsible for the final neutrino fluxes
and their energy spectrum, a detailed knowledge of the Sun and its
evolution is needed. Standard Solar Models 
(SSM's)~\cite{Bahcall:1987jc, TurckChieze:1988tj, Bahcall:1992hn, Bahcall:1995bt, Bahcall:2000nu, Bahcall:2004pz,Serenelli:2009yc}
describe the properties of the Sun and its evolution after entering
the main sequence. 

Till recently SSM's have had notable successes in predicting other
observations. In particular, quantities measured by helioseismology
such as the radial distributions of sound speed and
density~\cite{Bahcall:1992hn, Bahcall:1995bt, Bahcall:2000nu,
  Bahcall:2004pz} showed good agreement with the predictions of the
SSM calculations and provided accurate information on the solar
interior.  A key element to this agreement is the input value of the
abundances of heavy elements on the surface of the
Sun~\cite{Grevesse:1998bj}. However, recent determinations of these
abundances point towards substantially lower values than previously
expected~\cite{Asplund:2004eu, Asplund:2009fu}. A SSM which
incorporates such lower metallicities fails at explaining the
helioseismological observations~\cite{Bahcall:2004yr}, and changes in
the Sun modeling (in particular of the less known convective zone) are
not able to account for this discrepancy~\cite{Chaplin:2007uh, Basu:2006vh}.

So far there has not been a successful solution of this puzzle. Thus
the situation is that, at present, there is no fully consistent SSM.
This led to the construction of two different sets of SSM's, one
(labeled ``GS'') based on the older solar
abundances~\cite{Grevesse:1998bj} implying high metallicity, and one
(labeled ``AGS'') assuming lower metallicity as inferred from more
recent determinations of the solar abundances~\cite{Asplund:2004eu,
Asplund:2009fu}. In Ref.~\cite{Serenelli:2009yc} the solar fluxes
corresponding to such two models were detailed, based on updated
versions of the solar model calculations presented in
Ref.~\cite{Bahcall:2004pz}.

Alternatively one may attempt to directly determine the solar neutrino
fluxes from the solar neutrino data itself. 
In here we summarize the results of the most up-to-day extraction 
of the solar neutrino fluxes  directly from the solar neutrino data
from Ref ~\cite{GonzalezGarcia:2009ya} 
in the framework of three-neutrino oscillations.
The data included comprises  the total
rates from the radiochemical experiments
Chlorine~\cite{Cleveland:1998nv}, Gallex/GNO~\cite{Hahn:2008zz} and
SAGE~\cite{Hahn:2008zz, Abdurashitov:2009tn}. For real-time
experiments in the energy range of \Nuc[8]{B} neutrinos we include the
44 data points of the electron scattering (ES) Super-Kamiokande phase
I (SK-I) energy-zenith spectrum~\cite{Hosaka:2005um}, the 34 data
points of the day-night spectrum from SNO-I~\cite{Aharmim:2007nv}, the
separate day and night rates for neutral current (NC) and ES events
and the day-night energy-spectrum for charge current (CC) events from
SNO-II (a total of 38 data points)~\cite{Aharmim:2005gt}, the three
rates for CC, ES and NC from SNO-III~\cite{Aharmim:2008kc}, and the 6
points of the high-energy spectrum from the 246 live days of
Borexino~\cite{Bellini:2008mr}.
Finally, we include the main set of the 192 days of
Borexino data~\cite{Arpesella:2008mt}.
Besides solar experiments, we also include the latest results from the
long baseline reactor experiment KamLAND~\cite{Shimizu:2008zz,
  GonzalezGarcia:2007ib}, which in the framework of three neutrino
mixing also yield information on the parameters $\Dmq_{21}$,
$\theta_{12}$, and $\theta_{13}$.
In addition, we include the information on $\theta_{13}$ obtained
after marginalizing over $\Dmq_{31}$, $\theta_{23}$ and
$\delta_\textsc{cp}$ the results from the complete SK-I and SK-II
atmospheric neutrino data sets (see the Appendix of
Ref.~\cite{GonzalezGarcia:2007ib} for full details on our analysis),
the CHOOZ reactor experiment~\cite{Apollonio:1999ae},
K2K~\cite{Ahn:2006zza}, the latest MINOS $\nu_\mu$ disappearance data
corresponding to an exposure of $3.4\times 10^{20}$
p.o.t.~\cite{Adamson:2008zt}, and the first MINOS $\nu_\mu \to \nu_e$
appearance data presented in Ref.~\cite{Collaboration:2009yc}.

We do a Bayesian analysis in order to produce the posterior
porbability distribution for the parameters $(\Dmq_{21}, \theta_{12},
\theta_{13}, f_{\Nuc{pp}}, f_{\Nuc[7]{Be}}, f_{\Nuc{pep}},
f_{\Nuc[13]{N}}, f_{\Nuc[15]{O}}, f_{\Nuc[17]{F}},
f_{\Nuc[8]{B}},f_{\Nuc{hep}})$.
In this model independent analysis we assume a
uniform prior probability over which we impose a set of
constraints, such as the luminosity constraint which relates the 
number of neutrinos produced with the total Sun luminosity
~\cite{Bahcall:2001pf}, as well as those needed 
to ensure consistency in the pp-chain and CNO-cycle, and some 
relations from nuclear physics. For details on the normalization
of the fluxes and the nuclear constraints see~\cite{GonzalezGarcia:2009ya}. 
An important  arises from the 

\begin{figure}
  \includegraphics[width=0.95\textwidth]{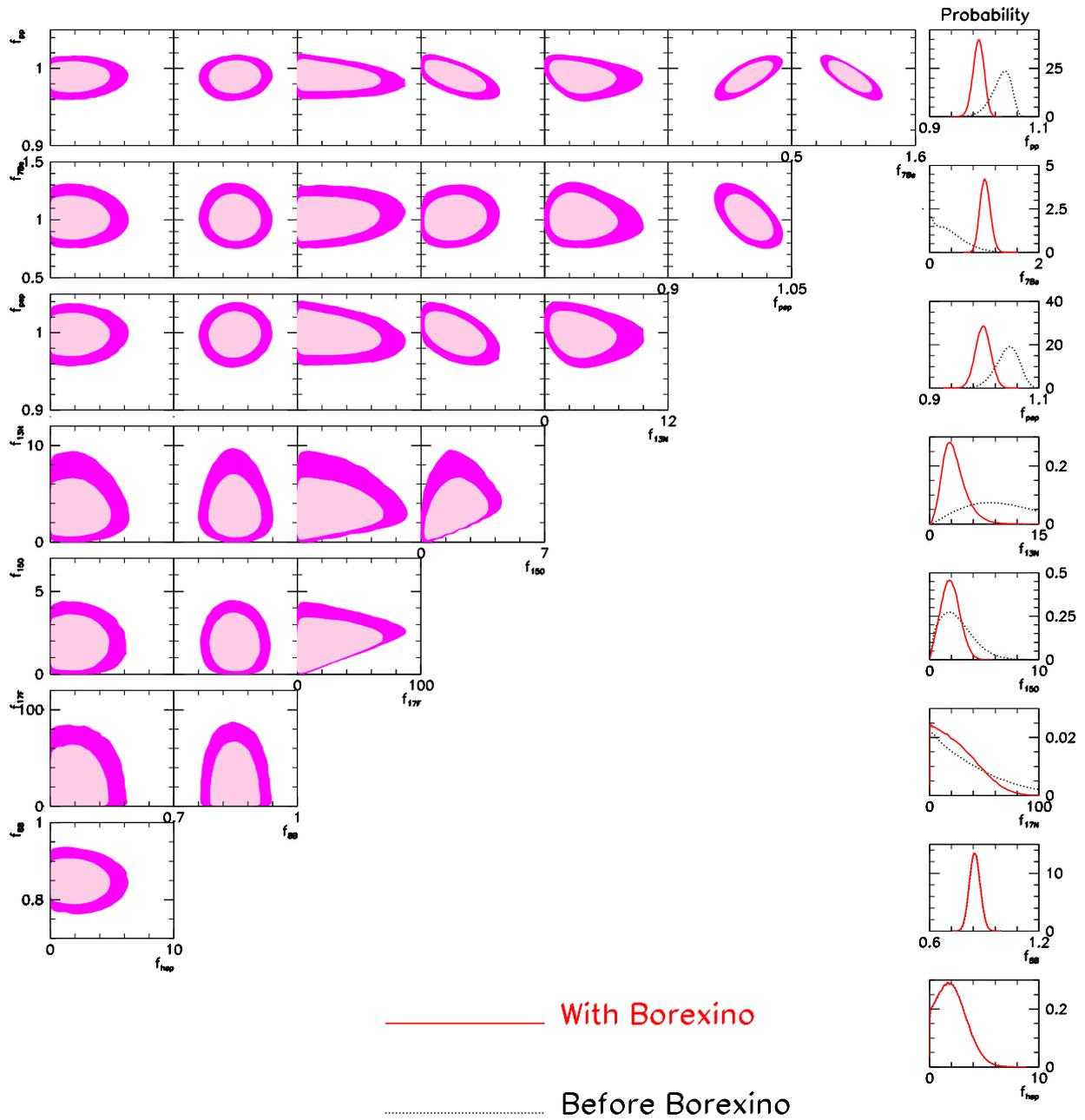}
  \caption{\label{fig:fitlc}
    Constraints from our global analysis on the solar neutrino
    fluxes. The curves in the rightmost panels show the marginalized
    one-dimensional probability distributions, before and after the
    inclusion of the Borexino spectral data. The rest of the panels
    show the 90\% and 99\% CL two-dimensional credibility regions (see
    text for details).}
\end{figure}

Our results for the analysis with luminosity constraint are displayed
in Fig.~\ref{fig:fitlc}, where we show the marginalized
one-dimensional probability distributions for the eight solar neutrino 
fluxes as well as the 90\% and 99\% CL two-dimensional allowed regions. 
The corresponding ranges at  $1\sigma$ (and at the 99\% CL in square brackets) 
on the oscillation parameters are:

\begin{equation}
  \label{eq:bestosc}
  \begin{aligned}
    \Dmq_{21}
    &= 7.6 \pm 0.2 \, [\pm 0.5]  \times 10^{-5}~\eVq \,,
    \\
    \sin^2\theta_{12}
    &= 0.33 \pm 0.02 \, [\pm 0.05] \,,
    \\
    \sin^2\theta_{13}
    &= 0.02 \pm 0.012 \, [^{+0.03}_{-0.02}] \,,
  \end{aligned}
\end{equation}
while for the solar neutrino fluxes are (in units of 
$~\text{cm}^{-2}~\text{s}^{-1}$):
\begin{equation}
  \label{eq:bestlc}
  \begin{aligned}
    & \Phi_{\Nuc{pp}}
    & = 5.910 ^{+0.057}_{-0.063} [^{+0.14}_{-0.16}]
    \times 10^{10} \,,
&&
    & \Phi_{\Nuc[7]{Be}}
    & = 5.08^{+0.52}_{-0.43} \, [^{+1.3}_{-1.0}]
    \times 10^{9} \,,
    \\
    & \Phi_{\Nuc{pep}}
    & = 1.407 ^{+0.019}_{-0.020} \, [^{+0.054}_{-0.057}]
    \times 10^{8} \,,
&&    
    & \Phi_{\Nuc[13]{N}}
    & = 7.8^{+5.0}_{- 3.4} \, [^{+16}_{-7.0}]
    \times 10^{8} \,,
    \\
    & \Phi_{\Nuc[15]{O}}
    & = 4.0^{+1.8}_{-1.9} \, [^{+4.8}_{-3.8}]
    \times 10^{8} \,,
&&
    & \Phi_{\Nuc[17]{F}}
    & \leq 5.9 \, [43]
    \times 10^{7} \,,
    \\
    & \Phi_{\Nuc[8]{B}}
    & = 5.02^{+0.18}_{-0.17} \, [^{+0.45}_{-0.42}]
    \times 10^{6} \,,
&&    
    & \Phi_{\Nuc{hep}}
    & = 1.3 \pm 1.0 \, [^{+3.0}_{-1.3}]
    \times 10^{4} \,.
  \end{aligned}
\end{equation}

All these results imply the following share of the energy production
between the pp-chain and the CNO-cycle
\begin{equation}
  \label{eq:ppcnolum1}
  \frac{L_\text{pp-chain}}{L_\odot} =
  0.986 ^{+0.005}_{-0.006} \, [^{+0.011}_{-0.014}]
  \quad\Longleftrightarrow\quad
  \frac{L_\text{CNO}}{L_\odot} =
  0.014^{+0.006}_{-0.005} \, [^{+0.014}_{-0.011}] \,,
\end{equation}
in perfect agreement with the SSM's which predict $L_\text{CNO} /
L_\odot \leq 1$\% at the $3\sigma$ level.

As seen in Figs.~\ref{fig:fitlc} the inclusion
of Borexino has a very important impact on the determination of the
\Nuc[7]{Be}, \Nuc{pep} and CNO fluxes, and indirectly on the \Nuc{pp}
flux. 

In order to statistically compare our results with the SSM's
predictions we perform two diferent significance tests. 
First we do the analisis without asuming gausianity constructing an
statistical estimator $t$ from the likelihood, where we found 
that the GS model has a lower $t$, $t_\text{GS} = 8.5$, while
$t_\text{AGS} = 11.0$ which corresponds to $P^\text{agr}_\text{GS} =
43$\% and $P^\text{agr}_\text{AGS} = 20$\%.

In the second case we use an estimation of the covariance matrix from 
the posterior probability distribution and we do a chi-square test, 
we found $\chi^2_\text{GS} = 5.2$($P^\text{agr}_\text{GS} = 74\%$) 
versus $\chi^2_\text{AGS} = 7.4$ ($P^\text{agr}_\text{AGS} = 50\%$).

From these results we conclude that, while the fit shows a slightly
better agreement with the GS model corresponding to higher
metallicities, the difference between the two is not statistically
significant.  This is partly due to the lack of precision of present
data. But we also notice that, while the measurements of SNO and SK
favor a lower \Nuc[8]{B} flux as predicted by the low metallicity
models, the determination of the \Nuc[7]{Be} flux in Borexino and the
corresponding determination of the \Nuc{pp} flux from the luminosity
constraint show better agreement with the GS predictions.

Finally in order to check the consistency of our results we have 
performed the same analysis without imposing the luminosity constraint.
This allow us to test the relation between 
the luminosity of the Sun as directly measured with the one infered
from the determination of the solar fluxes. 
\begin{equation}
    \label{eq:lnutot}
    \frac{L_\odot \text{(neutrino-inferred)}}{L_\odot}
    = 1.00 \pm 0.14 [^{+0.37}_{-0.34} ] \,.
\end{equation}
Thus  at present, the neutrino inferred luminosity
perfectly agrees with the one directly determined and this agreement
is known with a $1\sigma$ uncertainty of 15\%.

%% file: CONTENT/04Longhin.tex
\section{A new design for the SPL-Fr\'ejus Super-Beam \cite{Longhin:2011hn}}

In this study we consider a graphite target: this choice constitutes a proven 
technology in existing neutrino beams (i.e. T2K and CNGS).
We assume a cylindrical shape with $r=1.5$~cm, $L=78$~cm
and $\rho=1.85$ g/cm${}^3$.
A granular Ti target with the same geometry will also be discussed.

\begin{figure}[hbpt!]
\begin{center}
\vskip -0.4cm
\includegraphics[width=7cm]{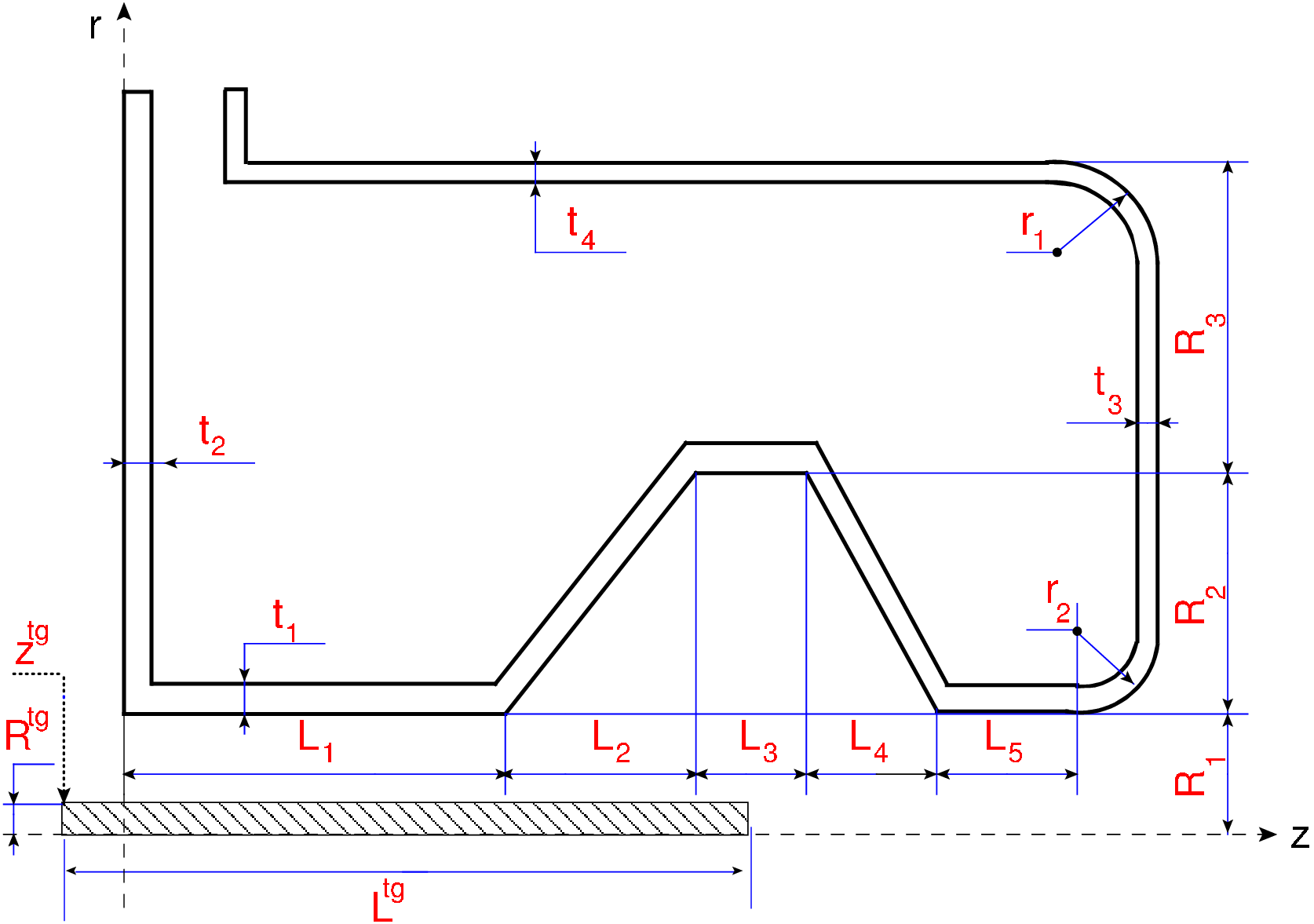}
\includegraphics[width=8cm]{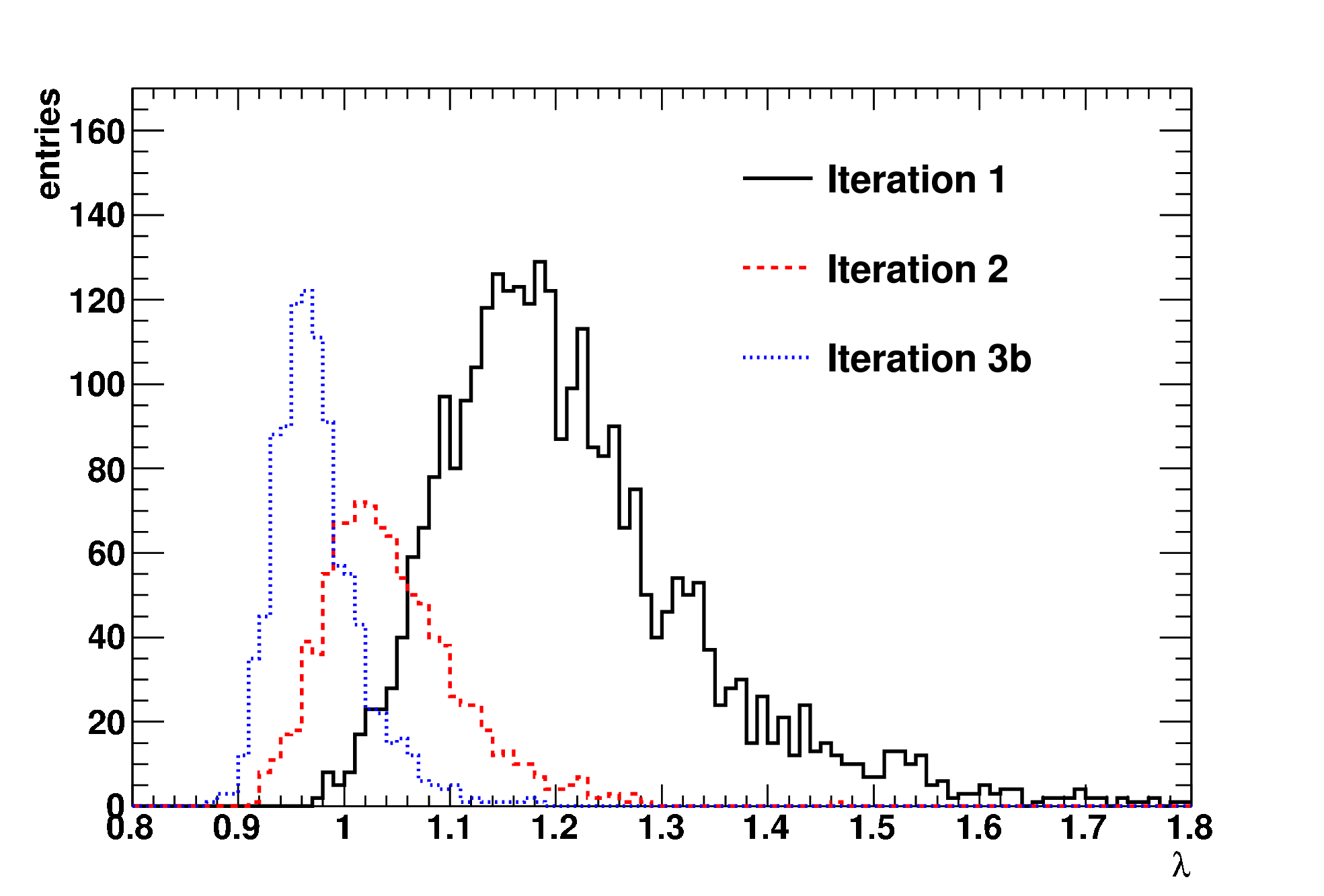}
\caption{ \label{fig1} Left: parametrization of the forward--closed horn. 
Right: distribution of the figure of merit $\lambda$.
  See the text for the definition of the samples.}
\end{center}
\end{figure}

A new horn model
inspired by the one used for the MiniBooNE beam,
having a large acceptance for forward produced pions, has been adopted giving
a reduced contamination from wrong--charge pions.
The generic layout of the horn is shown in Fig.~\ref{fig1} (Left). 
Taking advantage of the small transversal dimensions,
the idea of using a battery of four horns in parallel has been
proposed. This arrangement allows reduced stress on the targets thanks to
the lower frequency pulsing (12.5 Hz) which brings 
the average beam power on each target to a level
which is currently considered as a viable upper limit for
solid targets operations ($\sim$ 1 MW). 
We have verified that placing the horns as central as
possible (i.e. in mutual contact) causes a minor loss
of $\nu_\mu$ of the order of 1-2\% with a mild loss
as a function of the radial displacement.

The approach which was followed in the optimization of the forward--closed horn and 
the decay tunnel uses the final $\sin^2 2\theta_{13}$ sensitivity,
as a guiding principle in the ranking of the system.
Given the well known dependence on the $\sin^22\theta_{13}$ limit on the $\delta_{CP}$ phase, we introduced the figure of merit of the focusing system $\lambda$ defined as the $\delta_{CP}$-averaged 99~\% C.L. sensitivity limit on $\sin^2 2\theta_{13}$ 
in units of $10^{-3}$.
In a first stage the parameters of the forward--closed horn and of 
the tunnel were sampled with uniform probability distributions within
large ranges.
The maximal length and radius of the horn were limited
to 2.5~m and 80~cm in order to maintain a compact design 
suitable for the operation of four horns in parallel.
Moreover, the inner radius $R_1$ was limited in [1.2, 4] cm, the lower 
limit corresponding to the ``integrated target'' limit. 
At this level the target geometry
and the current were not varied ($I$ = 300 kA).
The obtained distribution for $\lambda$ is shown in the continuous histogram of 
Fig.~\ref{fig1} (Right). 
A second scan was performed
after fixing the horn inner radius at 1.2 cm and restricting the
ranges of variation of a set of relevant parameters\cite{Longhin:2011hn}.
The distribution of $\lambda$ for this sample is shown by the dashed histogram of Fig.~\ref{fig1}. 
Finally the horn shape was fixed 
and a further tuning of the tunnel length and radius was performed. 
The dependence of $\lambda$ on the tunnel variables can be reasonably fitted
with a quadratic function:
{\small{
$\lambda=0.94+2.1\cdot 10^{-4}(L^{\rm{tun}}[{\rm{m}}]-31.8)^2+2.4\cdot 10^{-2}(R^{\rm{tun}}[\rm{m}]-2.9)^2$.
}}
The distributions of $\lambda$ for $L^{\rm{tun}}$  and $R^{\rm{tun}}$ 
in the neighborhood of the minimum, is shown by the dotted 
histogram of Fig.~\ref{fig1}: an improvement 
of 25-30~\% is obtained with respect to the initial sample.
Since the minimum is relatively broad we chose  
$L^{\rm tun} = 25$~m  and $R^{\rm tun} = 2$~m as central values based on practical considerations related to the excavation and shielding of large volumes. This compares to the previous values of $40$ m of length and 2 m of radius.

We have also observed that an increase in the current (between 300 and 400 kA)
tends to systematically produce better sensitivity limits. 
Data are well fitted by a linear function in ($I$, $R_1$):
{\small{
$\lambda=(9.2-0.81\cdot I [100~\rm{kA}])/(7.3-0.37\cdot R_1[\rm{cm}])$.
}}
The effect of increasing the current, i.e. a stronger magnetic field in the vicinity of the target, is physically equivalent to decreasing the minimum horn inner radius. In this way, using a constant $I/R_1(\propto B)$, allows to roughly work at fixed sensitivity. 

The fractions of $\nu_\mu$, $\bar{\nu}_\mu$, $\nu_e$ and $\bar{\nu}_e$ with respect to the total
are (98.0\%, 1.6\%, 0.42\%, 0.015\%) and (95.3\%, 4.4\%, 0.28\%, 0.05\%) for the positive and negative focusing 
modes respectively.
In positive (negative) focusing mode the $\nu_e$ ($\bar{\nu}_e$) 
fluxes are dominated by muon decays: 82\% (90\%). 
The c.c. fluxes receive instead a large contribution from $K$ 3-body decays 
(81~\% and 75~\% in "+" and "-" focusing respectively)
with $\mu$ decays from the decay chain of ``wronge charge'' $\pi$ at low energy
contributing for the rest. These fluxes are available on the internet \cite{Longhin:2011hn}.

The discovery potential for $\theta_{13}\neq 0$ and CP violation
improves with respect to the previous design.
The uncertainty on hadro-production has also been addressed, for the graphite target, at the level of sensitivities  by exploiting the data of the HARP experiment and different models (FLUKA and GEANT4-QGSP). More detailed information on the subject can be found in \cite{Longhin:2011hn}.

As undergoing studies in the context of the EUROnu design study
have shown technical challenges for a solution with an integrated horn-target system, 
we studied the performances of two additional configurations assuming:
1) a graphite target separated from the horn (ST); 
2) a granular target composed of titanium spheres with diameters of $\mathcal{O}$(mm) (PB);   while keeping the target geometry unchanged.
In both cases we set for the inner radius of the horn ($R_1$) a value of 3 cm and
for the current a value of 350 kA. The 1.5 cm gap between the target and the horn 
is intended to accomodate the cooling infrastructure.
Thanks to the favorable surface to volume
 ratio and the possibility to flow transversely the coolant within the interstices of the spheres (i.e. a high pressure flow of He gas) the granulat target is expected to have a good behaviour even under extreme irradiation conditions.

\begin{figure}[hbpt!]
\vskip -0.2cm
\centering
\includegraphics[width=7cm]{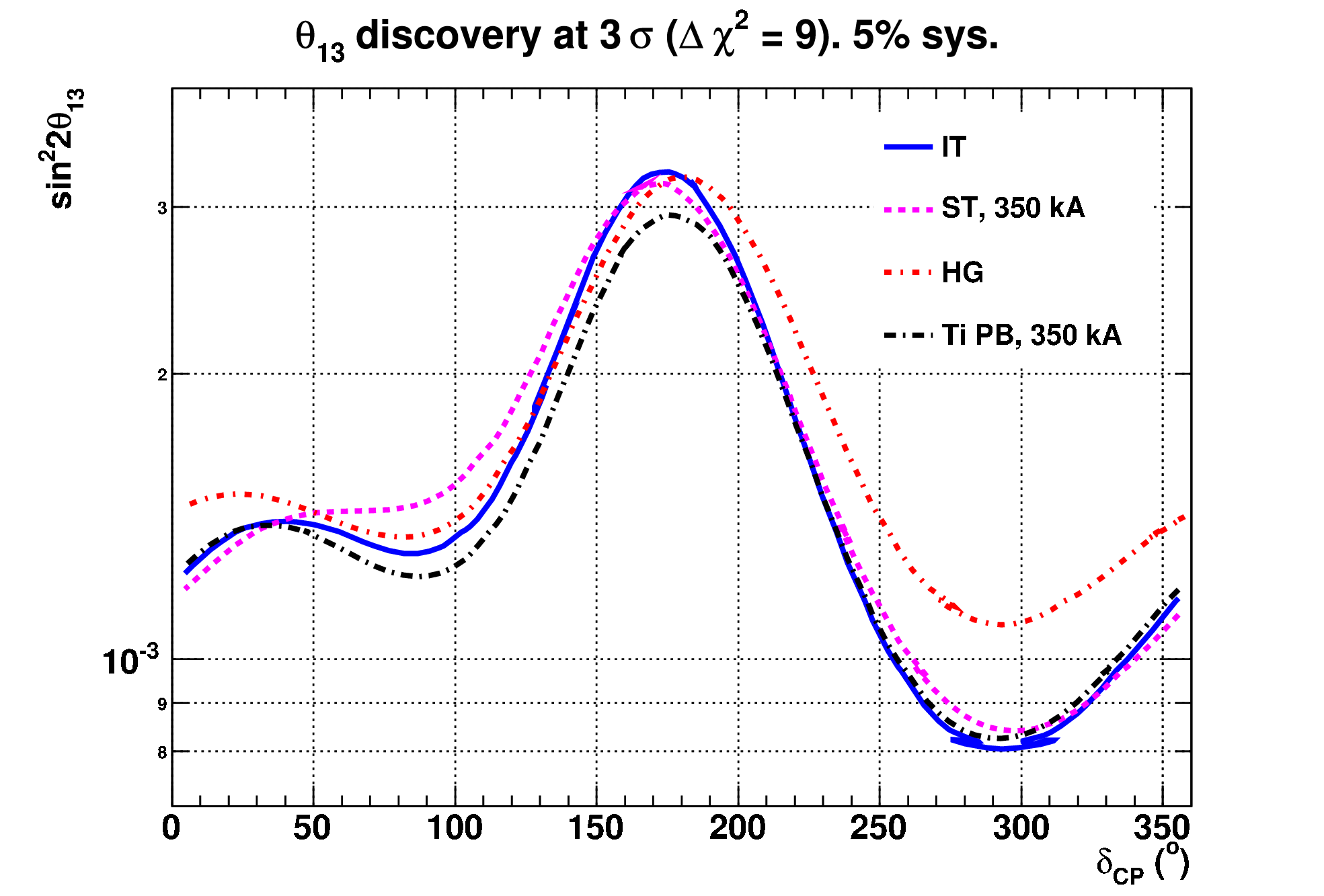}
\includegraphics[width=7cm]{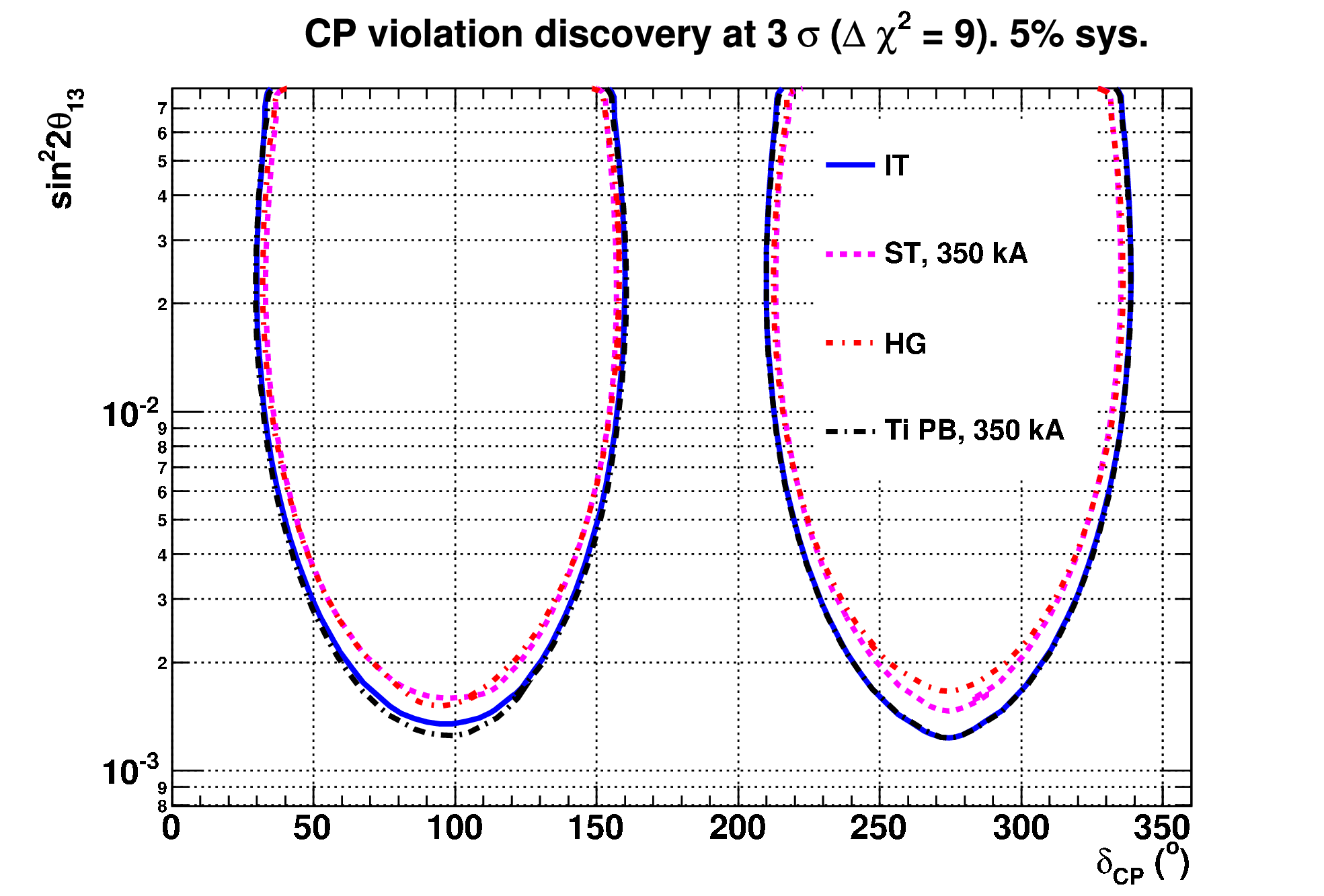}
\caption{\label{fig2} The $\theta_{13}$ (Left) and CP violation discovery potential (Right) at 
3~$\sigma$.  
Known parameters were included in the fit assuming a prior knowledge with an accuracy of 10\% for $\theta_{12}$, $\theta_{23}$, 5\% for $\Delta m^2_{31}$ and 3\% for $\Delta m^2_{12}$ at 1 $\sigma$ level. The running time is (2$\nu$+8$\bar{\nu}$) years.
Curves are calculated with GLoBES 3.0.14.
The parametrization of the MEMPHYS Water Cherenkov detector
is implemented in the publicly distributed AEDL file SPL.glb.
}
\vskip -0.5cm
\end{figure}

The discovery potentials for  $\theta_{13}\neq 0$
and CP violation for these two configurations are compared to the ones obtained with the former design based on a mercury target (HG) and 
to the performance with the integrated target (IT) in Fig. \ref{fig2}.  With respect to the IT design the 
solution with a separated monolithic graphite target and increased current gives limits which are only slightly worsened; the PB solution gives practically unchanged
performance for $\delta>\pi$ and some improvement for $\delta<\pi$. 
The granular Ti-target in association with the optimized horn represents 
then possibly the most appealing solution in terms of
both physics performance and engineering.
%
%

%% file: CONTENT/05aSummaryBB.tex
\subsection{Performances of Beta-Beam setups as of January 2011 \cite{Donini:2011thisrep}}

In this short EUROnu-WP6 internal note, we review the physics potential of several beta-beam setups that have been proposed in recent years, comparing 
their performances in the sensitivity to $\sin^2 2 \theta_{13}$, the CP discovery potential and the sensitivity to the neutrino mass hierarchy. 
Combination of these facilities with other facilities (as the proposed synergy of the SPL super-beam and the $\gamma = 100$ beta-beam aiming at the Fr\'ejus 
underground laboratory) and/or with atmospheric neutrino oscillation data collected at the same detector of the beta-beam setup under study
 \cite{Campagne:2006yx} are not considered in this review. 

\subsubsection{``Low''-$\gamma$: $\gamma = 100$}
Belong to this category all the setups that use existing CERN facilities to boost the ions up to the desired energy. In particular, the reference setup is the $\gamma = 100$
$^6$He/$^{18}$Ne beam aiming at a 1 Mton class water \v Cerenkov detector located in the Fr\'ejus underground laboratory \cite{Zucchelli:2002sa,Bouchez:2003fy}, 
with a baseline $L = 130$ Km. Three options have been considered for this setup, depending on the achievable ion rates.
A second possibility using the same infrastructures at CERN (i.e., the PS and the SPS) is to change the stored ions, going from low-$Q$ ones such as $^6$He/$^{18}$Ne 
to high-$Q$ ones, such as $^8$Li/$^8$B. This possibility was advanced in Refs.~\cite{Rubbia:2006pi,Mori:2006zn}. Also in this case, 
several options depending on the achievable ion rates have been considered. 
In addition to that, it is conceivable to store a mixture of low- and high-$Q$ ions aiming at the same
baseline, in order to use the first- and second-peak of the oscillation probability, as in Ref.~\cite{Donini:2006dx}. 
The characteristics of the low-$\gamma$ setups are summarized in Tab.~\ref{tab:setupslow}.

\begin{table}
  \centering
  \begin{tabular}{|c|c|c|c|c|c|c|}
  \hline
   Setup & $\gamma$ & Ions & Fluxes $[10^{18}]$ & Years & Baseline & Detector Technology \\
  \hline
   CERN-Fr\'ejus, 1 & 100 & $\begin{array}{c} ^6{\rm He} \\ ^{18}{\rm Ne} \end{array}$ & $\begin{array}{c} 2.9 \\  1.1  \end{array}$ & 
   $\begin{array}{c} 5 \\ 5 \end{array}$ & 130 km & 440 Kton WC \\
     \hline
   CERN-Fr\'ejus, 2 & 100 & $\begin{array}{c} ^6{\rm He} \\ ^{18}{\rm Ne} \end{array}$ & $\begin{array}{c} 2.9 \times 2 \\ 1.1 / 2  \end{array}$ & 
   $\begin{array}{c} 2 \\ 8 \end{array}$ & 130 km & 440 Kton WC  \\
  \hline
   CERN-Fr\'ejus, 3 & 100 & $\begin{array}{c} ^6{\rm He} \\ ^{18}{\rm Ne} \end{array}$ & $\begin{array}{c} 2.9 \times 2 \\ 1.1 / 5  \end{array}$ & 
   $\begin{array}{c} 2 \\ 8 \end{array}$ & 130 km & 440 Kton WC  \\
    \hline
    \hline
   CERN-Canfranc, 4 & 100 & $\begin{array}{c} ^8{\rm Li} \\ ^{8}{\rm B} \end{array}$ & $\begin{array}{c} 2.9  \\  1.1   \end{array}$ & 
   $\begin{array}{c} 5 \\ 5 \end{array}$ & 650 km & 440 Kton WC \\
   \hline
   CERN-Canfranc, 5 & 100 & $\begin{array}{c} ^8{\rm Li} \\ ^{8}{\rm B} \end{array}$ & $\begin{array}{c} 2.9 \times 2  \\  1.1  \times 2 \end{array}$ & 
   $\begin{array}{c} 5 \\ 5 \end{array}$ & 650 km & 440 Kton WC \\
\hline
   CERN-Canfranc, 6 & 100 & $\begin{array}{c} ^8{\rm Li} \\ ^{8}{\rm B} \end{array}$ & $\begin{array}{c} 2.9  \times 5 \\  1.1 \times 5  \end{array}$ & 
   $\begin{array}{c} 5 \\ 5 \end{array}$ & 650 km & 440 Kton WC \\
   \hline
\hline
  CERN-Canfranc, 4 & 100 & $\begin{array}{c} ^8{\rm Li} \\ ^{8}{\rm B} \\ ^{6}{\rm He} \end{array}$ & $\begin{array}{c} 2.9  \\  1.1  \\ 2.9 \end{array}$ & 
   $\begin{array}{c} 3 \\ 5 \\ 2 \end{array}$ & 650 km & 440 Kton WC \\
   \hline
        CERN-Canfranc, 5 & 100 & $\begin{array}{c} ^8{\rm Li} \\ ^{8}{\rm B} \\ ^{6}{\rm He} \end{array}$ & $\begin{array}{c} 2.9 \times 2  \\  1.1  \times 2 \\ 2.9 \times 2 \end{array}$ & $\begin{array}{c} 3 \\ 5 \\ 2 \end{array}$ & 650 km & 440 Kton WC \\
        \hline
CERN-Canfranc, 6 & 100 & $\begin{array}{c} ^8{\rm Li} \\ ^{8}{\rm B} \\ ^{6}{\rm He} \end{array}$ & $\begin{array}{c} 2.9  \times 5 \\  1.1  \times 5 \\ 2.9 \times 5 \end{array}$ &   $\begin{array}{c} 3 \\ 5 \\ 2 \end{array}$ & 650 km & 440 Kton WC \\
   \hline
  \end{tabular}
  \caption{
  Summary of the characteristics of the $\gamma = 100$ beta-beam setups that have been shown in the literature (for a review, see Ref.~\cite{FernandezMartinez:2009hb}).
   \label{tab:setupslow}}
\end{table}

In Tab.~\ref{tab:lowresults} we compare the performances of the $\gamma = 100$ beta-beam setups defined above in terms of three observables: (1) the minimum value
of $\sin^2 2 \theta_{13}$ that can be excluded at 3$\sigma$ in case of a null result of a given experiment (sensitivity to $\theta_{13}$); (2) the fraction of the $\delta$-parameter
space (known as the CP-fraction) for which a non-vanishing $\delta$ can be distinguished by $\delta = 0, \pi$ at 3$\sigma$ (CP discovery potential), computed for\footnote{
This value is approximately consistent with the recent measurement of $\theta_{13}$ by reactor experiments \cite{Abe:2011fz,An:2012eh,Ahn:2012nd}.
} 
$\sin^2 2 \theta_{13} = 0.1$; and, (3) the fraction of the $\delta$-parameter space for which a true positive $\Delta m^2_{13}$ can be distinguished by a negative $\Delta m^2_{13}$
at 3$\sigma$ (sensitivity to the neutrino mass hierarchy), computed for $\sin^2 2 \theta_{13} = 0.1$.
Notice that in all  cases we have considered no atmospheric neutrino background. This means that we have always assumed that the duty cycle at which the beta-beam
is operated is tight enough to make this background negligible. The duty cycle for which this approximation is valid differs depending on the setup. How the performance
are deteriorated when the duty cycle is relaxed is not shown in this table. The impact of the atmospheric neutrino background on the beta-beam performance has
been studied in Ref.~\cite{FernandezMartinez:2009hb} for the low-$\gamma$ case. See Sect.~\ref{sec:atmosanjib} in this report for an estimate of the effect of atmospheric 
neutrino background for high-$\gamma$ setups.

\begin{table}
  \centering
  \begin{tabular}{|c|c|c|c|}
  \hline
   Setup & $\left ( \sin^2 2 \theta_{13} \right )_{\rm min} $ & $\begin{array}{c} {\rm  CP \; discovery\;  potential} \\
                                                                                                         {\rm CP-fraction \; for\; } \sin^2 2 \theta_{13} = 0.1 \end{array}$ & 
                                                                                                         $\begin{array}{c} {\rm Sensitivity \; to \; sign}(\Delta m^2_{13}) \\
                                                                                                         {\rm CP-fraction \; for\; } \sin^2 2 \theta_{13} = 0.1 \end{array}$  \\
  \hline
   CERN-Fr\'ejus, 1 & $5 \times 10^{-4}$ &  70\% &   NO \\
      \hline
   CERN-Fr\'ejus, 2 & $6 \times 10^{-4}$ &  70\% & NO  \\
      \hline
  CERN-Fr\'ejus, 3 & $1 \times 10^{-3}$ &  61\% &  NO \\
      \hline
    \hline
  CERN-Canfranc, 1 & $1.5 \times 10^{-3}$ & 58\%  &  51\% \\
      \hline
   CERN-Canfranc, 2 & $7 \times 10^{-4}$ &  72\% &  61\% \\
      \hline
  CERN-Canfranc, 3 & $2 \times 10^{-4}$ &  78\% &  100\% \\
      \hline
CERN-Canfranc, 4 & $1.7 \times 10^{-3}$ &  66\% &  100\% \\
      \hline
   CERN-Canfranc, 5 & $7 \times 10^{-4}$ &  71\% &  100\% \\
      \hline
  CERN-Canfranc, 6 & $3 \times 10^{-4}$ &  79\% &  100\% \\
      \hline
     \hline
  \end{tabular}
  \caption{
  Summary of the performances of the $\gamma = 100$ beta-beam setups at 3$\sigma$ in terms of:
   sensitivity to $\theta_{13}$; CP discovery potential; sensitivity to the neutrino mass hierarchy. From Ref.~\cite{FernandezMartinez:2009hb}.
   \label{tab:lowresults}}
\end{table}

\subsubsection{``High''-$\gamma$:  $\gamma \geq 350$}

In Ref.~\cite{Burguet-Castell:2003vv,Burguet-Castell:2005pa}, the possibility of using an upgrade of the SPS\footnote{The SPS+, proposed at that time within the 
of maintenance and upgrade programme  of the LHC.} to boost radioactive ions up to higher $\gamma$ values. In particular, it was shown that the sensitivity to $\theta_{13}$ and 
the CP discovery potential of a beta-beam with $^6$He/$^{18}$Ne ions boosted at $\gamma = 350$ was extremely good and competitive with the Neutrino Factory in some part of the parameter space. As for the $\gamma = 100$ option, the beam was directed towards a 1 Mton class water \v Cerenkov detector, located this 
time at the underground laboratory of Canfranc ($L = 650$ Km from CERN). This option was later adapted for the Gran Sasso underground laboratory, where such 
a big detector cannot be hosted, by replacing it with a 100 Kton iron detector (that can be magnetized, to reduce backgrounds) \cite{Donini:2005qg,Donini:2006tt,Donini:2007qt} 
or with a 50 Kton TASD \cite{Huber:2005jk}.
It was later noticed that, when using high-$Q$ ions with high-$\gamma$, the neutrino flux is peaked around the resonant energy for $\nu_e \to \nu_\mu$ conversion in 
Earth matter \cite{Agarwalla:2006vf,Agarwalla:2007ai}. This makes a high-$Q$ high-$\gamma$ beta-beam aiming to a very far 50 Kton iron detector (at $L \sim 7000$ Km) an extremely good experiment to measure the neutrino mass hierarchy. As a consequence, several two-baseline beta-beam setups have been proposed, using a resonant beam aiming to $L = 7000$ Km and a second beam aiming to a moderate distance detector ($L \sim 2000$ Km when using Li/B \cite{Coloma:2007nn}, $L = 650$ Km when using He/Ne \cite{Choubey:2009ks}; see also Refs.~\cite{Agarwalla:2008gf,Agarwalla:2008ti}). 
These beams was shown to be competitive with the Neutrino Factory in most part of the parameter space.
The characteristics of the high-$\gamma$ setups are summarized in Tab.~\ref{tab:setupshigh}.
In Tab.~\ref{tab:highresults} we compare the performances of the high-$\gamma$ beta-beam setups defined above.

\begin{table}
  \centering
  \begin{tabular}{|c|c|c|c|c|c|c|}
  \hline
   Setup & $\gamma$ & Ions & Fluxes $[10^{18}]$ & Years & Baseline & Detector Technology \\
  \hline
   High-$\gamma$, 1 \cite{Burguet-Castell:2003vv} & 350 & $\begin{array}{c} ^6{\rm He} \\ ^{18}{\rm Ne} \end{array}$ & $\begin{array}{c} 2.9  \\  1.1   \end{array}$ & 
   $\begin{array}{c} 5 \\ 5 \end{array}$ & {\rm 700 km} & {\rm 500 Kton WC} \\
     \hline
   High-$\gamma$, 2 \cite{Huber:2005jk} & 350 & $\begin{array}{c} ^6{\rm He} \\ ^{18}{\rm Ne} \end{array}$ & $\begin{array}{c} 2.9  \\ 1.1   \end{array}$ & 
   $\begin{array}{c} 2 \\ 8 \end{array}$ & {\rm 700 km} & {\rm 50 Kton TASD}  \\
  \hline
  \hline
   Two baselines, 1 \cite{Coloma:2007nn} & 350 & $\begin{array}{c} ^8{\rm Li} \\ ^{8}{\rm B} \end{array}$ & $\begin{array}{c}  3  \\  3  \end{array}$ & 
   $\begin{array}{c} 5 \\ 5 \end{array}$ & $\begin{array}{c} {\rm 2000 \; km} \\ {\rm 7000 \; km} \end{array}$ & $\begin{array}{c} {\rm 50 \; Kton \; MIND} \\ {\rm 50 \; Kton \; MIND} \end{array}$ \\
   \hline
      Two baselines, 2 \cite{Coloma:2007nn} & 350 & $\begin{array}{c} ^8{\rm Li} \\ ^{8}{\rm B} \end{array}$ & $\begin{array}{c}  5  \\  5  \end{array}$ & 
   $\begin{array}{c} 5 \\ 5 \end{array}$ & $\begin{array}{c}  {\rm 2000 \; km} \\ {\rm 7000 \; km} \end{array}$ & $\begin{array}{c} {\rm 50 \; Kton \; MIND} \\ {\rm 50 \; Kton \; MIND}  \end{array}$ \\
   \hline
   Two baselines, 3 \cite{Coloma:2007nn} & 350 & $\begin{array}{c} ^8{\rm Li} \\ ^{8}{\rm B} \end{array}$ & $\begin{array}{c}  10  \\  10  \end{array}$ & 
   $\begin{array}{c} 5 \\ 5 \end{array}$ & $\begin{array}{c} {\rm 2000 \; km} \\ {\rm 7000 \; km} \end{array}$ & $\begin{array}{c} {\rm 50 \; Kton \; MIND} \\ {\rm 50 \; Kton \; MIND}  \end{array}$ \\
   \hline
 \hline
  Cocktail, 1 \cite{Choubey:2009ks} & $\begin{array}{c} 390 \\ 656 \\ 350 \\ 350 \end{array}$ & $\begin{array}{c} ^8{\rm Li} \\ ^8{\rm B} \\ ^6{\rm He} \\ ^{18}{\rm Ne}   \end{array} $&
   $\begin{array}{c} 0.6 \times 3 \\ 0.6 \times 3 \\ 3 \\ 3 \end{array}$ & $\begin{array}{c} 2.5 \\ 2.5 \\ 2.5 \\ 2.5 \end{array}$ & $\begin{array}{c} {\rm 7000 \; km} \\ \\ {\rm 650 \; km} \end{array}$ &
    $\begin{array}{c} {\rm 50 \; Kton \; MIND} \\ \\ { \rm 500 \; Kton \; WC} \end{array}$ \\
\hline 
  Cocktail, 2 \cite{Agarwalla:2008ti}& $\begin{array}{c} 390 \\ 656 \\ 575 \\ 575 \end{array}$ & $\begin{array}{c} ^8{\rm Li} \\ ^8{\rm B} \\ ^6{\rm He} \\ ^{18}{\rm Ne}   \end{array}$ &
   $\begin{array}{c} 3 \\ 3 \\ 3 \\ 3 \end{array}$ & $\begin{array}{c} 2.5 \\ 2.5 \\ 2.5 \\ 2.5 \end{array}$ & $\begin{array}{c} {\rm 7000 \; km} \\ \\ {\rm 650 \; km}  \end{array}$ &
    $\begin{array}{c} {\rm 50 \; Kton \; MIND} \\ \\ {\rm 50 \; Kton \; TASD} \end{array}$ \\
\hline 
\hline
     \end{tabular}
  \caption{
  Summary of the characteristics of the high-$\gamma$ beta-beam setups that have been shown in the literature.
   \label{tab:setupshigh}}
\end{table}

\begin{table}
  \centering
  \begin{tabular}{|c|c|c|c|}
  \hline
   Setup & $\left ( \sin^2 2 \theta_{13} \right )_{\rm min} $ & $\begin{array}{c} {\rm  CP \; discovery\;  potential} \\
                                                                                                         {\rm CP-fraction \; for\; } \sin^2 2 \theta_{13} = 0.1 \end{array}$ & 
                                                                                                         $\begin{array}{c} {\rm Sensitivity \; to \; sign}(\Delta m^2_{13}) \\
                                                                                                         {\rm CP-fraction \; for\; } \sin^2 2 \theta_{13} = 0.1 \end{array}$  \\
  \hline
   High-$\gamma$, 1 & $1.6 \times 10^{-4}$ & 93\% & 100\% \\
      \hline
   High-$\gamma$, 2 & $5 \times 10^{-4}$ & 75\% &  85\% \\
      \hline
    \hline
  Two-baselines, 1 & $1 \times 10^{-3}$ & 44\% & 100\% \\
      \hline
   Two-baselines, 2 & $3 \times 10^{-4}$ & 62\% &  100\% \\
      \hline
  Two.baselines, 3 & $1.5 \times 10^{-4}$ & 74\% &  100\% \\
      \hline
      \hline
    Cocktail, 1 & $1.8 \times 10^{-4}$ &  81\% &  100\% \\
      \hline
      Cocktail, 2 & $5 \times 10^{-4}$ &  73\% &  100\% \\
      \hline
     \hline
  \end{tabular}
  \caption{
  Summary of the performances of the $\gamma \geq 350$ beta-beam setups at 3$\sigma$ in terms of:
   sensitivity to $\theta_{13}$; CP discovery potential; sensitivity to the neutrino mass hierarchy.
   \label{tab:highresults}}
\end{table}

%% file: CONTENT/05bAtmo.tex
\subsection{Atmospheric neutrino events at ICAL@INO and high Q $\beta$-beam \cite{Agarwalla:2011thisrep}}
\label{sec:atmosanjib}

\begin{enumerate}
\item {\underline{\em \bf Atmospheric events:}}
 
\begin{itemize}

\item
First of all, we compute the expected number of atmospheric events in 50 kton ICAL@INO 
detector in 5 years. We have calculated separately the neutrino and anti-neutrino events.
For neutrinos, we have considered ($\nu_{\mu} \rightarrow \nu_{\mu}$ + $\nu_e \rightarrow \nu_{\mu}$)
oscillation channels. In case of anti-neutrinos, the considered channels are
($\bar\nu_{\mu} \rightarrow \bar\nu_{\mu}$ + $\bar\nu_e \rightarrow \bar\nu_{\mu}$).
Fig. 1 shows the expected number of events as a function of neutrino energy at ICAL@INO detector 
in 5 years. Here we have done the integration over $\theta$ and $\phi$ in their entire range.
All other details of the simulation have been mentioned on the body of the figure itself.
Next, we will present these number of events in a tabular form (Table. 1 and 2).  

\begin{figure}[t]
\begin{center}
\includegraphics[width=10.0cm]{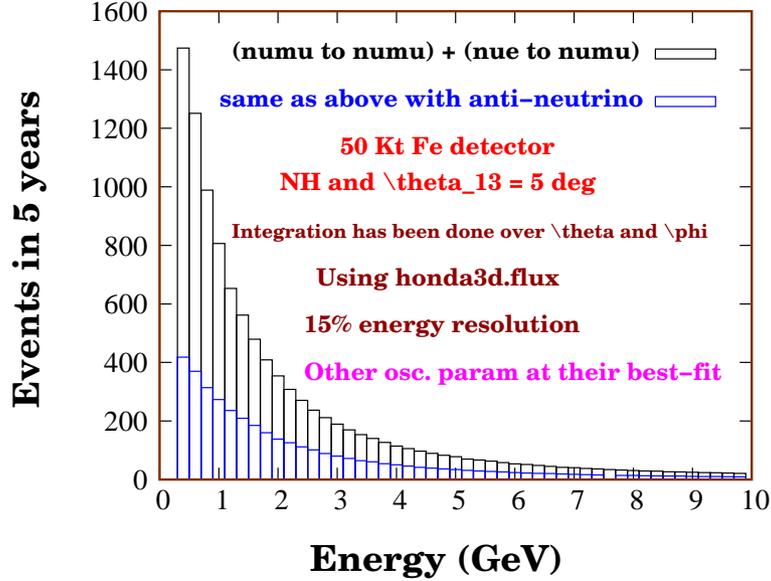}
\caption{\label{atm-events}
Atmospheric neutrino events in 50 Kton ICAL@INO detector in 5 years.}
\end{center}
\end{figure}

\begin{table}[t!] 
\begin{center}
 \begin{tabular}{||l|r|r||}
\hline\hline
Energy Bins & $\nu$ events & $\bar\nu$ events \\ 
GeV & ($\nu_{\mu} \rightarrow \nu_{\mu}$ + $\nu_e \rightarrow \nu_{\mu}$) &
($\bar\nu_{\mu} \rightarrow \bar\nu_{\mu}$ + $\bar\nu_e \rightarrow \bar\nu_{\mu}$) \\
\hline
0.4 - 0.6 & 1474 & 418 \\
0.6 - 0.8 & 1251 & 370 \\
0.8 - 1   &  989 & 314 \\
1   - 1.2 &  807 & 273 \\
1.2 - 1.4 &  653 & 236 \\
1.4 - 1.6 &  562 & 209 \\
1.6 - 1.8 &  479 & 185 \\
1.8 - 2   &  409 & 160 \\
2   - 2.2 &  354 & 139 \\
2.2 - 2.4 &  308 & 126 \\
2.4 - 2.6 &  271 & 112 \\
2.6 - 2.8 &  237 & 102 \\
2.8 - 3   &  212 &  89 \\
3   - 3.2 &  189 &  80 \\
3.2 - 3.4 &  170 &  72 \\
3.4 - 3.6 &  153 &  64 \\
3.6 - 3.8 &  140 &  61 \\
3.8 - 4   &  127 &  54 \\
4   - 4.2 &  114 &  51 \\
4.2 - 4.4 &  106 &  46 \\
4.4 - 4.6 &   97 &  42 \\
4.6 - 4.8 &   89 &  39 \\
4.8 - 5   &   83 &  36 \\
\hline
\end{tabular}
\end{center}
\caption{\label{atm-events}
Atmospheric neutrino events in 0.4 to 5 GeV range at 50 Kton ICAL@INO detector in 5 years.}
\end{table}
\begin{table}[t!]
\begin{center}
 \begin{tabular}{||l|r|r||}
\hline\hline
Energy Bins & $\nu$ events & $\bar\nu$ events \\
GeV & ($\nu_{\mu} \rightarrow \nu_{\mu}$ + $\nu_e \rightarrow \nu_{\mu}$) &
($\bar\nu_{\mu} \rightarrow \bar\nu_{\mu}$ + $\bar\nu_e \rightarrow \bar\nu_{\mu}$) \\
\hline
5   - 5.2 &   78 &  34 \\
5.2 - 5.4 &   70 &  31 \\
5.4 - 5.6 &   67 &  29 \\
5.6 - 5.8 &   63 &  28 \\
5.8 - 6   &   58 &  26 \\
6   - 6.2 &   54 &  24 \\
6.2 - 6.4 &   51 &  22 \\
6.4 - 6.6 &   48 &  21 \\
6.6 - 6.8 &   45 &  20 \\
6.8 - 7   &   42 &  19 \\
7   - 7.2 &   41 &  18 \\
7.2 - 7.4 &   38 &  17 \\
7.4 - 7.6 &   36 &  16 \\
7.6 - 7.8 &   34 &  15 \\
7.8 - 8   &   33 &  14 \\
8   - 8.2 &   31 &  14 \\
8.2 - 8.4 &   29 &  13 \\
8.4 - 8.6 &   29 &  13 \\
8.6 - 8.8 &   27 &  12 \\
8.8 - 9   &   26 &  12 \\
9   - 9.2 &   25 &  11 \\
9.2 - 9.4 &   24 &  11 \\
9.4 - 9.6 &   23 &  10 \\
9.6 - 9.8 &   22 &  10 \\
9.8 - 10  &   21 &   9 \\
\hline
\end{tabular}
\end{center}
\caption{\label{atm-events}
Atmospheric neutrino events in 5 to 10 GeV range at 50 Kton ICAL@INO detector in 5 years.}
\end{table}

\vskip0.2cm

\item
For beam studies, the atmospheric events which will occur along the beam direction and arount it
will serve as background. The zenith angle for CERN-INO baseline is $124^{\circ}$ 
and in ICAL@INO detector, the angular resolution will be around $15^{\circ}$ at most and with 
higher energies, the angular resolution improves a lot. Therefore, in our next study, 
we have considered a zenith angle range of $109^{\circ}$ to $139^{\circ}$.
In this zenith angle range, the atmospheric events that you expect
at 50 kton ICAL@INO detector with 5 years of data taking is given in Table. 3.

\begin{table}[t!]
\begin{center}
 \begin{tabular}{||l|r|r||}
\hline\hline
Energy range (GeV) & Total $\nu$ events & Total $\bar\nu$ events \\
\hline
1 - 12 & 214 & 94 \\
2 - 12 & 155 & 69 \\
3 - 12 & 114 & 48 \\
4 - 12 & 90  & 39 \\  
\hline
\end{tabular}
\end{center}
\caption{\label{atm-events}
Atmospheric neutrino events at 50 Kton ICAL@INO detector in 5 years in the zenith angle range 
of $109^{\circ}$ to $139^{\circ}$. Here full integration has been done over $\phi$.}
\end{table}

\vskip0.2cm

\item
One can see from Table. 3 that the atmospheric neutrino flux falls steeply with energy 
and is expected to produce much fewer events for the energy range that we are interested in
for CERN-INO beam study. Therefore, we need to see that how sensitivity will be affected
in CERN-INO $\beta$-beam set-up with the increase in threshold. 

\begin{figure}[t]
\begin{center}
\includegraphics[width=10.0cm]{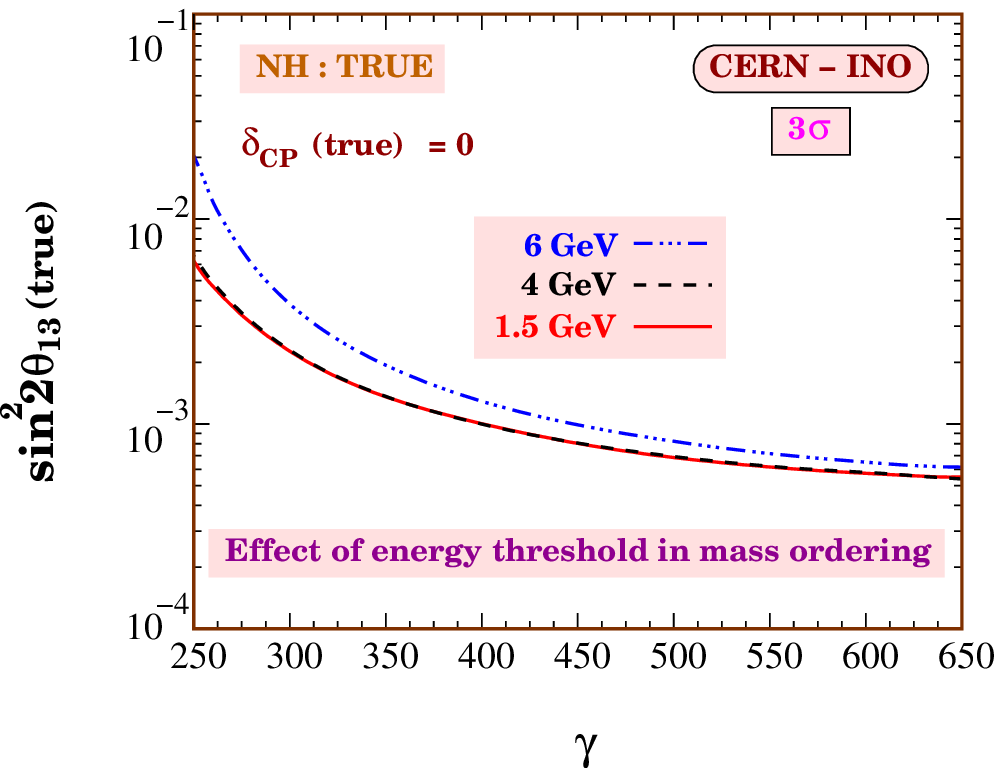}
\caption{\label{threshold}
It shows the effect of changing the threshold energy of the detector. 
The sensitivity of the experiment remains almost stable against the variation of the threshold 
energy upto 4 GeV. Here we have used $^8$B and $^8$Li ions with 5 years of running 
and we have assumed $1.1\times 10^{18}$ ($2.9\times 10^{18}$) useful ions
decays per year. This figure has been taken from~\cite{Agarwalla:2007ai}.
}
\end{center}
\end{figure}

Fig. 2 shows that the sensitivity of the experiment remains almost stable against the 
variation of the threshold energy upto 4 GeV. It means that we can work with a threshold
of 4 GeV or so and in that way, we can reduce the atmospheric background a lot as can be
seen from Table. 3.

\vskip0.2cm

\item
The fact that INO has charge identification capability further reduces the
atmospheric background. The most important handle on the
reduction of this background comes from the timing information
of the ion bunches inside the storage ring.
For 5T magnetic field and $\gamma=650$ for $^8$B ions, the
total length of the storage ring turns out to be
19564 m. We have checked that with
eight bunches inside this ring at any given time, a
bunch size of about 40 ns would give an
atmospheric background to {\it signal} ratio of about $10^{-2}$,
even for a very low $\stch$ of $10^{-3}$. For a smaller
bunch span, this will go down even further. In addition,
atmospheric neutrinos will be measured in INO during deadtime and
this can also be used to subtract them out.
   
\end{itemize}
\end{enumerate}

%% file: CONTENT/06Neapolitans.tex
\subsection{A minimal Beta Beam with high-Q ions to address leptonic CP violation \cite{Coloma:2010wa}}
\label{sec:intro}

The Beta-Beam concept and its different energy configurations have been discussed in details in~\cite{Zucchelli:2002sa,Lindroos:2010zza} and References therein.
Here we focus on a Beta Beam designed with the aim of leveraging at most existing facilities and, in particular, the CERN acceleration complex. Beta Beams that are able to accelerate radioactive ions to high energies and produce multi-GeV $\nu_e$ and $\bar \nu_e$ allow for the use of high-density detectors, which, in turn, might be hosted in moderate-size underground laboratories.
For a CERN-based Beta-Beam, the natural option to host the far detector is a laboratory located at a distance $O(600-700)$ Km from the neutrino source.
The facility that {\it exploits at most existing European infrastructures} is a multi-GeV Beta-Beam based on the CERN-SPS accelerator pointing to a massive, high-density detector located in one of the experimental halls of the Gran Sasso laboratories \cite{Coloma:2010wa}. The next cheapest alternative could be represented by the Canfranc Underground Laboratories in Spain,  where some engineering would be however needed (albeit not so impressive as for a Mton class Water \v Cerenkov detector).

Beside the huge practical interest of exploiting in an optimal manner all European facilities without additional infrastructure investment~\cite{Battiston:2009ux}, this detailed assessment is particularly relevant at present times: since 2009, machine studies for the Beta-Beam are concentrated on facilities that accelerate ions with Q-values larger than originally envisioned ($Q \sim 13$~MeV for $^8$Li and $^8$B, to be compared with $Q \sim 3$~MeV for the ions considered in the original design, $^6$He and $^{18}$Ne) 
using the existing SPS machine~\cite{Wildner:2010zz,Bernabeu:2010rz}. This option ideally fits the ``minimal'' scheme mentioned above provided that neutrinos are pointed toward the underground halls of LNGS. Other options either based on low density detectors and/or on new terminal boosters at larger energies than the CERN-SPS have also been studied in literature: for details, we refer the reader to Ref.~\cite{Lindroos:2010zza} and, in particular, to Refs.~\cite{Bouchez:2003fy,Donini:2004hu,Donini:2004iv} for low-Q ions
accelerated by the SPS, Refs.~\cite{Burguet-Castell:2003vv,Burguet-Castell:2005pa,Terranova:2004hu,Agarwalla:2005we,Huber:2005jk,Donini:2006tt,Donini:2007qt} for high-$\gamma$ Beta Beams (using facilities different from the SPS to accelerate ions) and Refs.~\cite{Rubbia:2006pi,Rubbia:2006zv,Donini:2006dx,Agarwalla:2006vf,
Agarwalla:2007ai,Coloma:2007nn,Jansson:2007nm,Winter:2008cn,Agarwalla:2008ti, Agarwalla:2008gf,Meloni:2008it,Choubey:2009ks,FernandezMartinez:2009hb} for high-Q
Beta Beams (either at low-$\gamma$ and high-$\gamma$).

The facility that we consider here does not differ from the baseline EURO$\nu$~\cite{euronu,FernandezMartinez:2009hb} design, but for 
a high density far detector located in a pre-existing hall at LNGS (for details on the facility and on the detector set-up and simulation we refer to \cite{Coloma:2010wa,Donini:2006tt}.

The physics performance of the proposed setup have been studied in terms of two observables, defined as follows:
\begin{description}
\item[the CP discovery potential] for a given point in the parameter space, we will say that CP violation can be discovered if we can rule out the no CP violation hypothesis ($\delta=0^\circ$ and $180^\circ$) at $3\sigma$ 1 d.o.f., after marginalizing over all the remaining parameters for both possible hierarchies.

\item[the $\textrm{sgn}(\Delta m_{23}^2)$ reach] this is defined as the region of the ($\sin^2 2 \theta_{13},\delta$) plane  for which the wrong hierarchy can be eliminated at $3\sigma$. 
Below this value of $\sin^2(2\theta_{13}) $, the predictions for the wrong hierarchy cannot be distinguished from the data corresponding to the right hierarchy, at a statistical significance of $3\sigma$.

\end{description}

For a far detector of 100~kton mass, a $\beta^+$-emitters ($^8$B) flux of approximately $6 \times 10^{18}$ useful decays per year\footnote{This is about three times the flux proposed for $^{18}$Ne, where $F_0 \sim 2 \times 10^{18}$. } is needed to observe CP violation in a large fraction of the parameter space (60\%) for any value of $\theta_{13}$ that gives a positive signal at T2K ($\theta_{13} \simge 3^\circ$).
This sensitivity to $\delta$ is deteriorated for $\delta<0$ due to the occurrence of the $\pi$-transit, as observed in other facilities.
The $^8$B flux  must be accompanied by a $^8$Li flux of $\sim 3\times 10^{19}$ decays per year. 
Present studies on the ionization cooling technique or on ISOL-type targets indicate that such a large $^8$Li flux could be feasible.
Moreover, the former technique should produce $\beta^+$ and $\beta^-$ emitters at a similar rate  although $^8$B ions interact stronger than $^8$Li ions with  materials in the target and in the recollection region.  To achieve the fluxes above clearly represents the most challenging task for accelerator R\&D but it is a viable option with respect to  $^{18}$Ne, where  ISOL-type targets fall almost two orders of magnitude short of the goal. 

In the same configuration, we find a non-negligible sensitivity to the neutrino mass hierarchy that extends
up to $\theta_{13} \simeq 4^\circ$ for positive (negative) values of $\delta$ for normal (inverted) hierarchy. 
In the opposite parameter area, i.e. for negative (positive) values of $\delta$ and inverted
(normal) hierarchy, the combination with atmospheric data collected
during the Beta Beam run by the same magnetized detector further
improves such sensitivity at large $\theta_{13}$ ($\simeq 6^\circ$).
Combination of atmospheric data with Beta Beam--driven ones should also be able to solve part of the $\pi$-transit deterioration 
discussed above.

%% file: CONTENT/07OrmeEC.tex
\subsection{Update on the physics of Electron Capture neutrino beams \cite{EC_Orme}}

Electron capture is a decay channel available to proton-rich nuclei that competes 
with positron decay in a manner that depends on the decay Q-value. Such decay channels 
are sources of mono-enegetic neutrinos since there is no electron in the final state with 
which to share energy. It has been 
suggested~\cite{EC_original, EC_energy_dep, EC_Sato, Rolinec_Sato} that ions with 
dominant electron capture channels can be used as an alternative to the standard Beta 
Beam ions to instead source an intense and collimated beam of mono-energetic neutrinos. 
A neutrino beam of this type could be used to accurately probe the energy 
dependence of the $\nu_{e}\rightarrow\nu_{\mu}$ appearance probability: a means to 
observe CP-violation without the need for an anti-neutrino channel. However, since a 
single neutrino energy is insufficient to determine the unknown neutrino mixing parameters 
and resolve any degeneracy, it is necessary to include at least two ion boosts in any physics 
strategy. In a recent study~\cite{EC_Orme}, the previous works~\cite{EC_original,EC_energy_dep} were
extended to examine the physics reach of a $^{150}$Dy electron capture beam sourced 
from CERN and directed towards a 440 kton fiducial mass Water \v{C}erenkov
detector 650 km distant to include the study of the choice of boost, the relative run times of 
each boost, the number of useful ion decays, and the number of atmospheric neutrino events.

\begin{figure}
\begin{center}
\includegraphics[width=14cm]{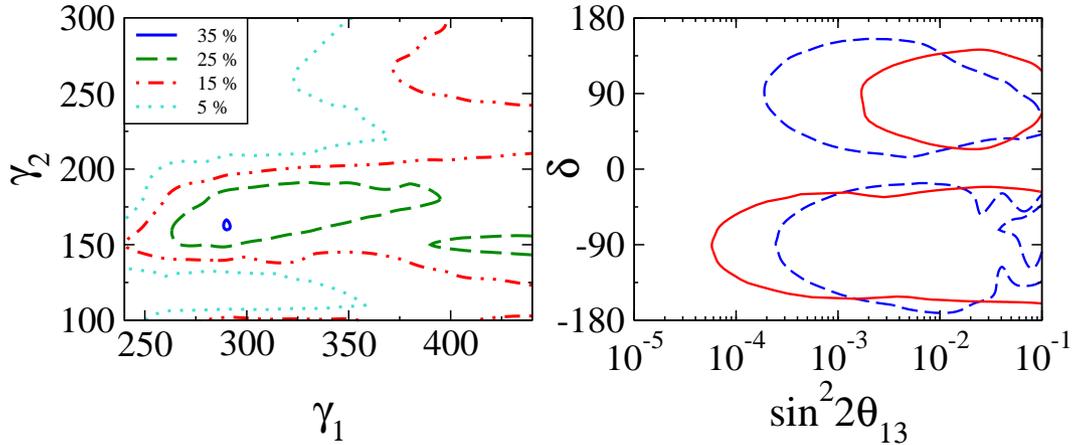}
\end{center}
\caption{(Left panel) A discrete approximation to the integrated CP-fraction for a 
mono-energetic neutrino beam, sourced at CERN and directed at a 440 kton Water \v{C}erenkov
detector located at the Canfranc laboratory, as a function of the choice of boost pairings chosen. 
The plot shows the results for a run time ratio of low boost : high boost = $7 : 3$. 
(Right panel) CP-violation discovery plots for the boost choices $(\gamma_{1},\gamma_{2})=(450,150)$ 
(red solid lines) and $(280,160)$ (blue dashed lines).}
\label{Fi:gam_scan_CP}
\end{figure}

In the first instance, the number of useful ion decays was fixed to $10^{18}$ per year, and 
atmospheric neutrino backgrounds were neglected so that the effect of altering the ion
boosts and run times could be investigated. The results of this initial analysis is presented 
in the left panel of Fig.~\ref{Fi:gam_scan_CP}: for a choice of ion boosts 
$(\gamma_{1},\gamma_{2})$, the percentage of the parameter space 
$\sin^{2}2\theta_{13}\in (10^{-5},10^{-1})$, $\delta \in (-180,180)$ for which the existence of 
CP-violation can be demonstrated (referred to hereafter as CP-coverage) at 99~\% confidence 
level is shown. (The procedure is outlined in detail in~\cite{EC_Orme}.) The results presented 
are for a low boost run taking 70~\% of the whole experimental run. All results in this section use 
2 degrees of freedom statistics. Little variation was found for the relative run times; however, two
regions of the $(\gamma_{1},\gamma_{2})$~ plane returned large CP-coverage. The naive 
pairing of (almost) first and second oscillation maximum had a CP-coverage of approximately 
35~\% but was asymmetric in $\delta\rightarrow -\delta$ (right panel, red solid lines). 
For $\delta < 0$, this choice could rule out CP-conservation down to $\sin^{2}2\theta_{13}= 10^{-4}$, 
but only $\sin^{2}2\theta_{13}\sim 2\cdot 10^{-3}$ for $\delta > 0$. There exists a slightly
larger coverage for the pair (280, 160). This choice returned a (roughly) symmetrical
CP-sensitivity region with CP-conservation ruled out down to 
$\sin^{2}2\theta_{13}\sim 4\cdot 10^{-4}$ (right panel, blue dashed lines).

\begin{figure}
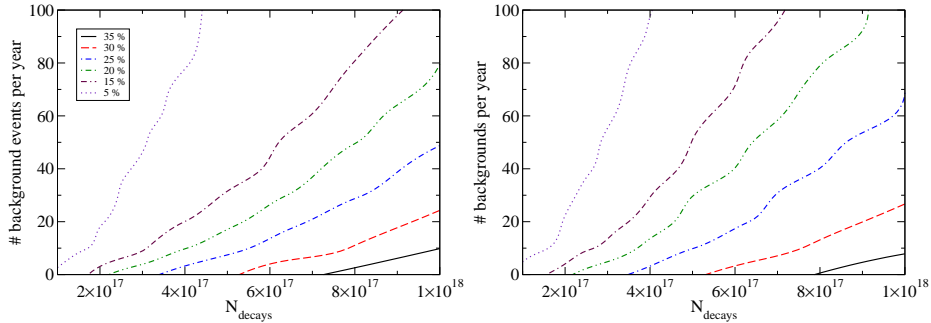

\begin{center}
\includegraphics[width=6cm]{CONTENT/FiguresReport/FiguresOrme/fig_dec_atm_280_160.eps}
\includegraphics[width=6cm]{CONTENT/FiguresReport/FiguresOrme/fig_dec_atm_440_150.eps}\\
\end{center}
\caption{CP-violation coverage for $(\gamma_{1},\gamma_{2})=(280,160)$ (left) and $(\gamma_{1},\gamma_{2})=(440,150)$ (right) as a function of the number of useful decays and number of atmospheric backgrounds per year; both with $\gamma_{1}$ for 70~\% of the experimental run time.}\label{Fi:Nvar_atmvar}
\end{figure}

However, no R\&D has been carried out on the feasibility of sourcing $10^{18}$ useful decays 
per year for high proton number ions. A brief study adapting EURISOL Beta Beam 
codes~\cite{Fraser_ions} indicated that the decays are too slow and large tune-shifts would need to
be accommodated. With existing technology, the number of useful decays is two orders of 
magnitude too small. At best, therefore, $10^{18}$ useful decays per year appears to be overly 
optimistic and should be considered as the hard limit on the yearly rate. To take this experimental 
fact into account, the ability to uncover CP-violation was investigated, for the two boost pairings 
identified above, as a function of the number of useful ion decays and the level of atmospheric 
neutrino background. The results are summarised in Fig.~\ref{Fi:Nvar_atmvar}, again using the 
CP-coverage as an indicator.  In both cases, the results were more volatile to small changes in 
the useful decay rate than the atmospheric background. It was found that drops in the total event 
rate can lead to degenerate solutions dropping below the required statistical significance of the 
test thus reducing the sensitivity in a manner beyond simply a scaling of the $\chi^{2}$~\cite{EC_Orme}. 
The background merely reduces the $\chi^{2}$ without interfering with its ability to rule out degenerate solutions.
Its manifestation is simply to push the sensitivity contours inwards.

\begin{figure}
\begin{center}
\includegraphics[width=6cm]{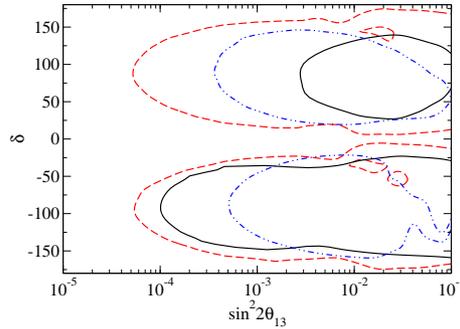}
\end{center}
\caption{CP-violation sensitivity at 99~\% confidence level for $N_{\rm atm}=20$ per year for 
$(\gamma_{1},\gamma_{2})=(440,150)$ (black solid) and (280,160) (blue dot-dashed). The red
dashed line displays the physics return for a standard Beta Beam directed along the 
CERN-Canfranc baseline with a boost $\gamma=350$ for both neutrinos and anti-neutrinos.}\label{Fi:comparison}
\end{figure}

The feasibility of an electron capture beam remains open question; however, the most 
optimistic parameterisation considered in~\cite{EC_Orme}, and shown in Fig.~\ref{Fi:gam_scan_CP}, 
is likely to be beyond the hard limits imposed by available technology. It must be conceded
that if  $\theta_{13}$ were to be very small, the electron capture will not be a competitive facility, 
unless a fast decaying ion that can be produced in large quantities is found. This can be seen 
explicitly in Fig.~\ref{Fi:comparison} where the CP-discovery for the boost pairs (280,160) and 
(440,150), and 20 atmospheric events per year, are compared to the $\gamma=$ 350,350 
Beta Beam introduced in~\cite{Burguet-Castell:2005pa} with the baseline shortened to CERN-Canfranc, 
as in this study. The simulation used $1.1 \times 10^{18}$ useful $^{18}$Ne decays per year 
and $2.9\times 10^{18}$ useful $^{6}$He decays year with an equal split between neutrino and 
anti-neutrino running for the Beta Beam.

We see that for a (440,150) electron capture facility, the physics return is poorer but 
relatively competitive for $\delta < 0^{o}$. For these $\delta$, the (280,160) facility is only
 competitive for $\sin^{2}2\theta_{13}\in [10^{-3},10^{-2}]$. For $\delta > 0^{o}$, the 
(440,150) facility is outperformed by the standard Beta Beam by 2 orders of magnitude 
whilst, again the (280,160) facility is only competitive for $\sin^{2}2\theta_{13}\in [10^{-3},10^{-2}]$. 
The physics reach of the electron capture machine is ultimately limited by the need to run for 
substantial periods at low energies where the appearance event rate is small. The availability of a 
mono-energetic anti-neutrino beam could resolve this issue. Ions that decay through bound beta decay 
could source such a beam; however, this is not possible from a practical point of view with 
available technology~\cite{EC_Orme,BBD}. 

%% file: CONTENT/08OrmeLAGUNA.tex
\subsection{High-$\gamma$ Beta Beams within the LAGUNA design study \cite{Orme_LAGUNA}}

Within the LAGUNA design study~\cite{LAGUNA}, seven candidate sites are being assessed for 
their feasibility to host a next-generation, very large neutrino observatory.
Such a detector will be expected to feature within a future European accelerator
neutrino programme: Superbeam, Beta Beam or, if the detector is magnetised, a low energy Neutrino Factory. 
In~\cite{Orme_LAGUNA}, a high boost $^{18}$Ne and $^{6}$He 
Beta Beam sourced at CERN and directed towards a 50 kton Liquid Argon detector~\cite{Rubbia_LAr} 
located at the LAGUNA sites: Slanic ($L=1570$ km) 
and Pyh\"{a}salmi ($L=2300$ km) was investigated. The study assumed the availability of an 1 TeV SPS, as envisaged in
some LHC upgrade options~\cite{LHC_upgrade}. Using boosts of $\gamma=350$ for both neutrinos and anti-neutrinos
run for 5 years each, the ability to distinguish $\theta_{13}$ from zero, rule out CP-conservation, and determine the correct
neutrino mass ordering was tested for both baselines. The simulations were then repeated but with the boost for 
$^{18}$Ne increased to $\gamma =570$. At present, no detector response data is available for a Beta Beam neutrino flux
incident on a large Liquid Argon detector; however, they are expected to possess excellent energy resolution for 
$\nu_{\mu}$ appearance events and background reduction capabilities whist maintaining a high detection efficiency. The 
results presented here take this optimism at face value and therefore should be considered near the limit of the possible 
physics reach of the two setups.

\begin{figure}[t]
\begin{center}
\includegraphics[width=15cm]{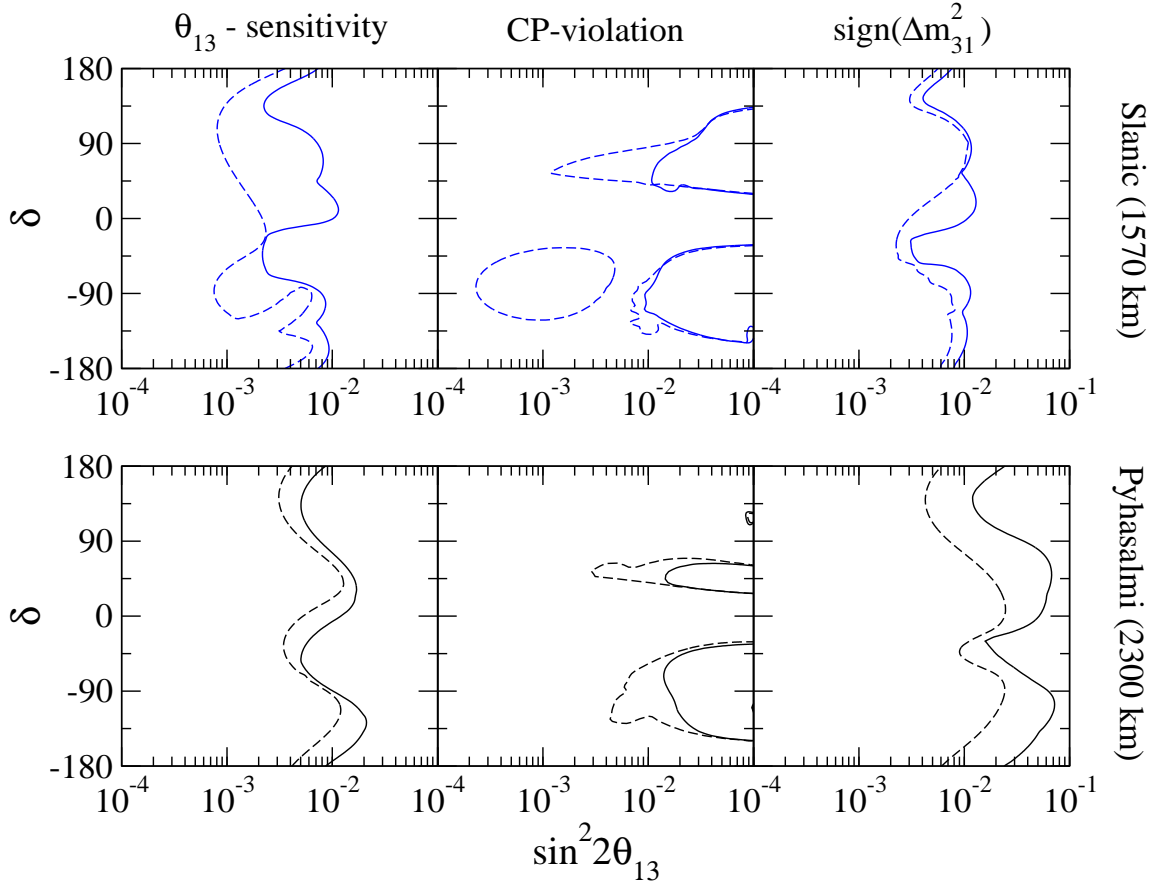}

\end{center}
\caption{$3\sigma$ C.L. contours for discovery for non-zero $\theta_{13}$ (left),
 CP-violation (centre), and sign($\Delta m_{31}^{2}$) determination (right). 
In each plot, the solid line corresponds to a $\gamma=350,350$ $^{18}$Ne and 
$^{6}$He Beta Beam for a 50 kton Liquid Argon detector located at 
Slanic ($L=1570$) (top line); and 
Pyh\"{a}salmi ($L=2300$ km) (bottom line). The dashed lines correspond to a 
$\gamma=570,350$ $^{18}$Ne and $^{6}$He Beta Beam.}\label{Fi:single}
\end{figure}

The physics reach of the two baselines considered in summarised is 
Fig.~\ref{Fi:single}. The results presented here have been simulated assuming $3\times 10^{18}$ useful decays per year
for each ion. An energy threshold of 0.4 GeV is taken with energies up to 4 GeV considered with 12 energy bins and a 
$dE/E =15~\%$. $\nu_{\mu}$-appearance detection efficiency is flat at 80~\% and neutral current backgrounds are taken at 0.1~\% of the 
unoscillated flux. The LAGUNA study assumes
a Liquid Argon detector of mass up to 100 kton. 50 kton is taken here to bring the simulations in line with other studies in 
the literature and to introduce some redundancy in the event that the parametrisation used here is too optimistic (e.g
$1.5\times10^{18}$ useful decays per year instead of $3\times 10^{18}$).   
   
The best sensitivity to non-zero $\theta_{13}$ and CP-violation is found for the CERN-Slanic
baseline. This is to be expected since, with the same source, the flux is larger for this baseline. The weaker matter
effect means that the sign($\Delta m_{31}^{2}$) degeneracy interferes less with these measurements.
For the $\gamma=570,350$ boost pair, non-zero $\theta_{13}$ can be seen down to $\sin^{2}2\theta\sim 10^{-3}$, and, for both 
boost pairs, at all values of $\delta$
for $\sin^{2}2\theta_{13}>10^{-2}$. However, there is a marked difference between the two boost pairings for sensitivity to CP-
violation. For the
$\gamma=350,350$ pair, the ability to rule out $\delta =0^{\circ}$ or $180^{\circ}$ 
is restricted to $\sin^{2}2\theta_{13}>10^{-2}$, but increasing the boost of the $^{18}$Ne ions
returns a large region of parameter space for $\delta <0^{\circ}$ and centred on $\sin^{2}2\theta_{13} =10^{-3}$. There is little 
enhancement on the region for 
low boost pairing. This is suggestive that degeneracy is a problem for the $\gamma=350,350$ boost pairing; especially 
for $\sin^{2}2\theta_{13}\sim 5\cdot10^{-3} - 1\cdot 10^{-2}$ where there is a gap in CP-violation sensitivity.  
The lower event rates for the longer 
baselines mean that the data is insufficient to reduce the significance of some 
degenerate 
solutions. Although, the ability to rule out the incorrect mass ordering is poor relative to Neutrino Factories, for a Beta Beam it is not 
intrinsically bad.
For the high boost run, the correct ordering can be indentified down to $\sin^{2}2\theta_{13}=2\cdot 10^{-3}$, with determination for 
all values of $\delta$
for $\sin^{2}2\theta_{13}>10^{-2}$ in both cases. Although the increase in the $^{18}$Ne boost improves the reach, it does not do so 
significantly. Increasing 
the boost makes data from higher enegies available without significantly altering the event rate and composition at lower energies. 
Since
European baselines make use of low and high energies in combination, improving the event rate in one region without the improving 
other need not, and has not, 
dramatically 
improved the physics return. The low event rate at low energies is still insufficient to break the degeneracy for small values of $
\theta_{13}$.

The physics return for the CERN-Pyh\"{a}salmi baseline is weaker for each of the physics indicators, with little sensitivity for $
\sin^{2}2\theta_{13}<10^{-2}$.
Principally, this is due to the $L^{-2}$ dependence of the un-oscillated neutrino flux. In particular, the ability to determine the correct 
mass ordering is
very poor even given the large matter effect at this baseline. The true and incorrect mass ordering solutions will be sufficiently 
separated in $(\theta_{13},\delta)$
space; however, the low
event rate means that the solution regions at $3\sigma$ will be large and possibly merged together. When combined with a large 
solution region from the 
low energy bins (also owing to low event rates), the data is insufficient to break the degeneracy for small $\theta_{13}$.

%% file: CONTENT/09TangWinter.tex
\subsection{Neutrino Factory in stages \cite{Tang:2009wp}}

Here we report on neutrino oscillation physics with a neutrino factory in stages, Ref.~\cite{Tang:2009wp}. We include the possibility of upgrading the muon energy, adding another baseline, and increasing the detector mass within the same program. Two aspects are taken into account depending on the possible $\theta_{13}$ discovery by the next generation of experiments.
For the large $\theta_{13}$ case ($\sin^22\theta_{13}\gtrsim 10^{-2}$), on the one hand, a low energy neutrino factory (LENF) with $E_\mu \simeq 4 \, \mathrm{GeV}$ to $5 \, \mathrm{GeV}$, discussed in Refs.~\cite{Geer:2007kn,Bross:2007ts,Huber:2007uj,Bross:2009gk}, could be sufficient to discover CP violation (CPV) and the mass hierarchy (MH). We perform an optimization study of the physics reach, and we identify the ``minimal'' neutrino factory. 
On the other hand, we propose a staging scenario to chase $\theta_{13}$ if $\theta_{13}$ is smaller, i.e., $\sin^22\theta_{13}\lesssim 10^{-2}$. For details, see Ref.~\cite{Tang:2009wp}.

\textbf{Minimal neutrino factory for the large $\boldsymbol{\theta_{13}}$.} We investigate the minimal requirements for a neutrino factory suggested in Ref.~\cite{Winter:2008cn}:
\begin{enumerate}
\item
 $5\sigma$ independent confirmation of $\stheta>0$ (for any $\deltacp$).
\item
 $3\sigma$ determination of MH for {\em any} (true) $\deltacp$.
\item
 $3\sigma$ establishment of CPV for a certain fraction (such as 80\%) of all (true) $\deltacp$.
\end{enumerate}
\begin{figure}[t]
\begin{center}
\includegraphics[width=0.4\textwidth]{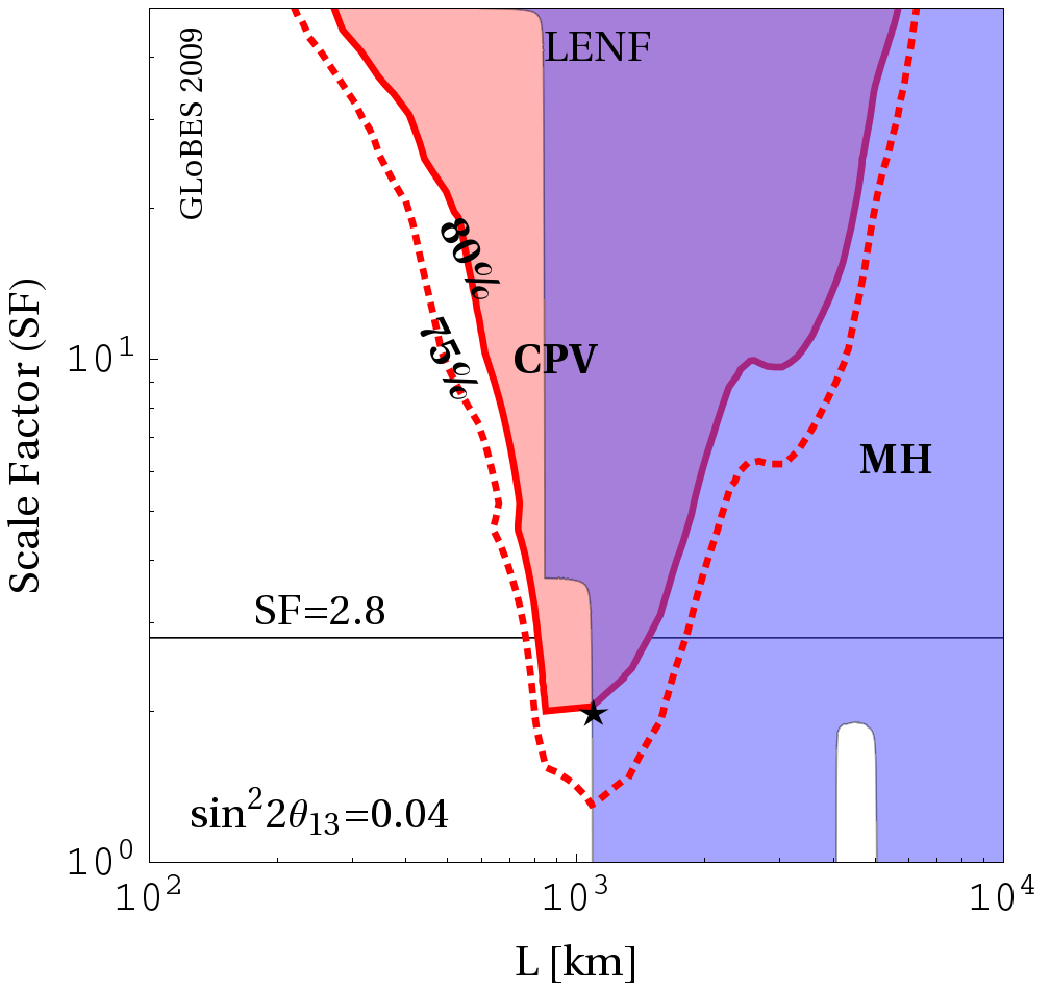} \hspace*{0.05\textwidth} \includegraphics[width=0.4\textwidth]{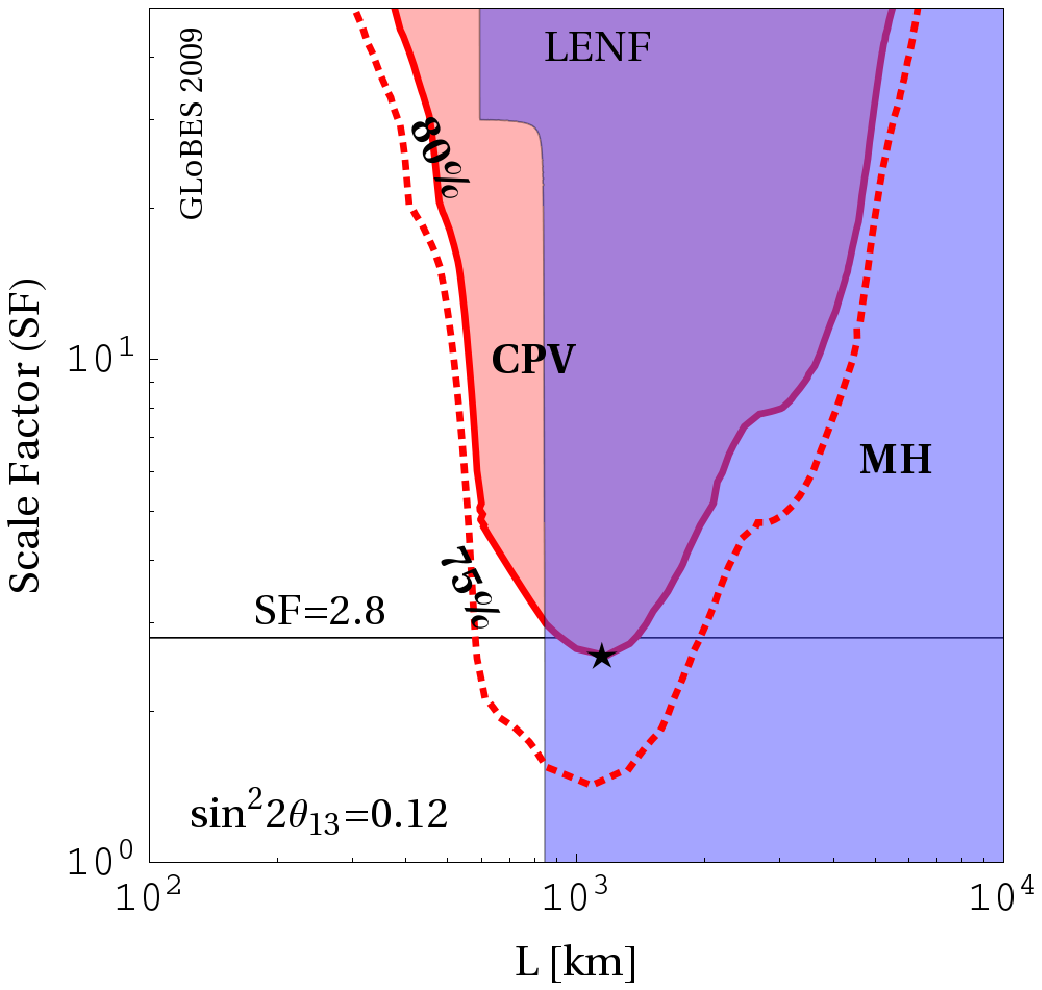}
\end{center}
\caption{\label{fig:mLENF}Discovery of  CPV (dark/red) and MH (medium gray/light blue) for the one baseline (minimal) LENF as a function of baseline and luminosity scale factor SF.  Discovery reach is given within the shaded regions at the $3\sigma$ CL, where for CPV a fraction of $\deltacp$ of 75\% or 80\% is required (as indicated), and for the MH a fraction of $\deltacp$ of 100\%. The stars show the baseline with the minimal SF: in the left panel $(1100 \, \mathrm{km},2.0)$ and in the right panel $(1150 \, \mathrm{km},2.6)$. The nominal luminosity of the LENF is given by SF=2.8. Here the true value of $\sin^22\theta_{13}$ is chosen as given in the plot panels, and a normal hierarchy is assumed. The matter density uncertainty is assumed to be 2\%. Figure taken from Ref.~\cite{Tang:2009wp}.}
\end{figure}
We determine the minimal necessary luminosity as a function of baseline with respect to the above performance indicators in Fig.~\ref{fig:mLENF}, where the requirement for $\theta_{13}$ can typically be easily met. The luminosity is expressed by a scale factor (SF) where SF=1 corresponds to $2.5 \cdot 10^{20}$ useful muon decays per year, which is the IDS-NF standard per baseline and polarity.
For the LENF, the minimal baseline is determined by the MH reach, and the minimal SF by the CPV reach. The nominal luminosity SF=2.8, coming from the injection of all muons in the same storage ring and the re-optimization of the frontend, is sufficient for the CPV measurement for 80\% of all true $\deltacp$ and for the MH measurement for all $\deltacp$ in the baseline window $1100 \, \mathrm{km} \lesssim L \lesssim 1400 \, \mathrm{km}$ for both values of $\stheta$ (left and right panel). One can read off these figures that luminosity is clearly an issue for large $\stheta$. If, for instance, only a lower SF can be achieved, the CPV discovery reach decreases accordingly. The ``minimal'' (optimal) LENF, \ie, the one with the lowest SF, is in both panels at about $L \simeq 1100 \, \mathrm{km}$, which is close to the FNAL-Homestake baseline $L = 1290 \, \mathrm{km}$.

\label{sec:smalltheta13}
\begin{figure}[t]
\begin{center}
\includegraphics[width=\textwidth]{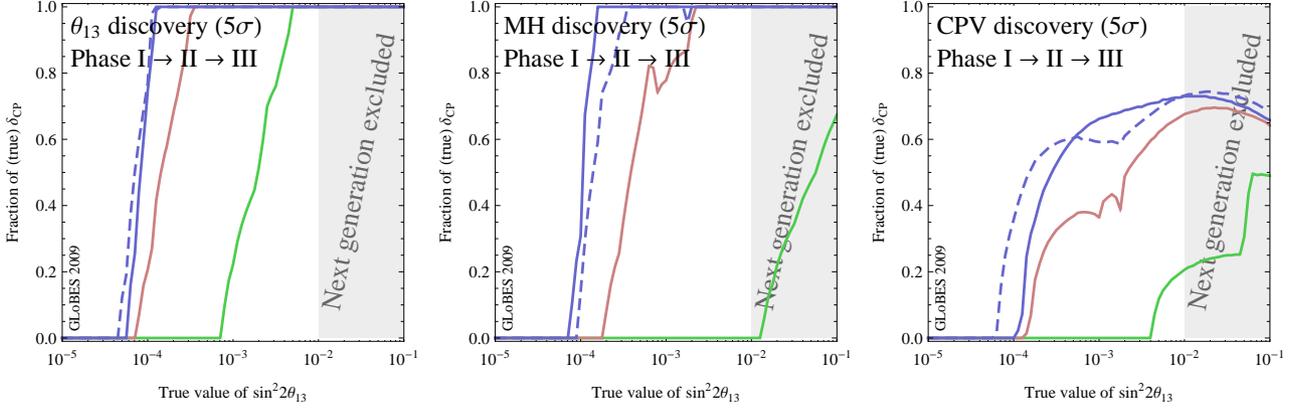}
\end{center}
\caption{\label{fig:phases} $\theta_{13}$, MH, and CP discovery reaches for a neutrino factory in three phases: Phase~I (light gray/green) is a low energy neutrino factory with a magnetized TASD, phase~II (medium gray/red) adds an energy upgrade with a MIND at the 4000~km baseline, and phase~III (dark gray/blue) includes another (magic) baseline (solid curves) or a detector upgrade at the 4000~km baseline (dashed curves). $5\sigma$ confidence level. Figure taken from Ref.~\cite{Tang:2009wp}.}
\end{figure}

\textbf{Staging for the small $\boldsymbol{\theta_{13}}$.} After several years of data taking from Daya Bay and the other next generation experiments, we will know whether $\stheta \lesssim 0.01$~\cite{Huber:2009cw}. What has to be done if $\theta_{13}$ has not been found by these experiments? Here we propose a plausible staging scenario, starting with a low energy neutrino factory. Meanwhile, it is open to take three upgrade options increasing the muon energy, adding another baseline and increasing the detector mass. We proceed in three phases of data taking, five years each. As we show in Fig.~\ref{fig:phases} ($5\sigma$), phase~I represents a low energy neutrino factory, phase~II includes the energy upgrade to a high energy neutrino factory, and phase~III considers additional upgrades, such as a larger detector or an additional baseline. In any phase, A combination of the data with the preceding phase is considered by default.
We see the neutrino factory in stages for small $\stheta$, where a second baseline is used in phase~III (solid curves) or the detector upgrade (dashed curves). At the $5\sigma$ confidence level, the strength of the magic baseline to resolve degeneracies becomes more important, especially for the mass hierarchy discovery reach. The largest increase in the discovery reaches just comes from the energy upgrade.

In summary, a realistic program may include components of LENF and HENF, even if $\stheta$ is small.

%% file: CONTENT/10LENF.tex
\subsection{LENF Overview \cite{FernandezMartinez:2010zza}}
\label{sec:lenf}

The low-energy neutrino factory (LENF), which uses muons of energy $\sim5$ GeV and a baseline of $\sim1000$km, was first proposed in Ref.~\cite{Geer:2007kn,Bross:2007ts}. Since then, developments to the accelerator and detector designs have enabled the experimental simulations to be refined and detailed optimisation 
studies to be performed \cite{FernandezMartinez:2010zza}. The key finding was that given sufficiently high statistics, an optimised LENF can have excellent sensitivity to the standard oscillation parameters, competitive with, and even better than, the high energy neutrino factory for large $\theta_{13}$ ($\sin^{2}2\theta_{13}\gtrsim10^{-3}$). The possibility of observing the \emph{platinum} channels ($\nu_{\mu}\rightarrow\nu_{e}$ and $\bar{\nu}_{\mu}\rightarrow\bar{\nu}_{e}$) in addition to the golden channels ($\nu_{e}\rightarrow\nu_{\mu}$ and $\bar{\nu}_{e}\rightarrow\bar{\nu}_{\mu}$) \cite{golden} was considered. It is found that the complementarity between these channels can be of great benefit if statistics are limited. There are also indications that the addition of these platinum channels is vital in order to resolve the degeneracy between standard oscillation parameters and non-standard interactions. Full details of the experiment can be found in Ref.~\cite{FernandezMartinez:2010zza}. In brief, we studied a setup having a baseline of 1300 km with a beam capable of delivering $1.4\times10^{21}$ useful muon decays per year \cite{lownfdesign} for 5 years per polarity. For the detector we considered either a 20 kton totally active scintillating detector (TASD) \cite{Abe:2007bi} and a 100 kton liquid argon (LAr) detector \cite{LAr}, both of which would be magnetised. These detectors would be capable of detecting and identifying the charges of both electrons and muons, providing access to multiple oscillation channels - the $\nu_{\mu}$ ($\bar{\nu}_{\mu}$) disappearance channels, as well as the golden and platinum channels.

\subsubsection{Sensitivity to standard oscillation parameters} 

\begin{figure}[htp]
     \centering
     \subfigure[~CP discovery potential]{
          \includegraphics[scale=0.4]{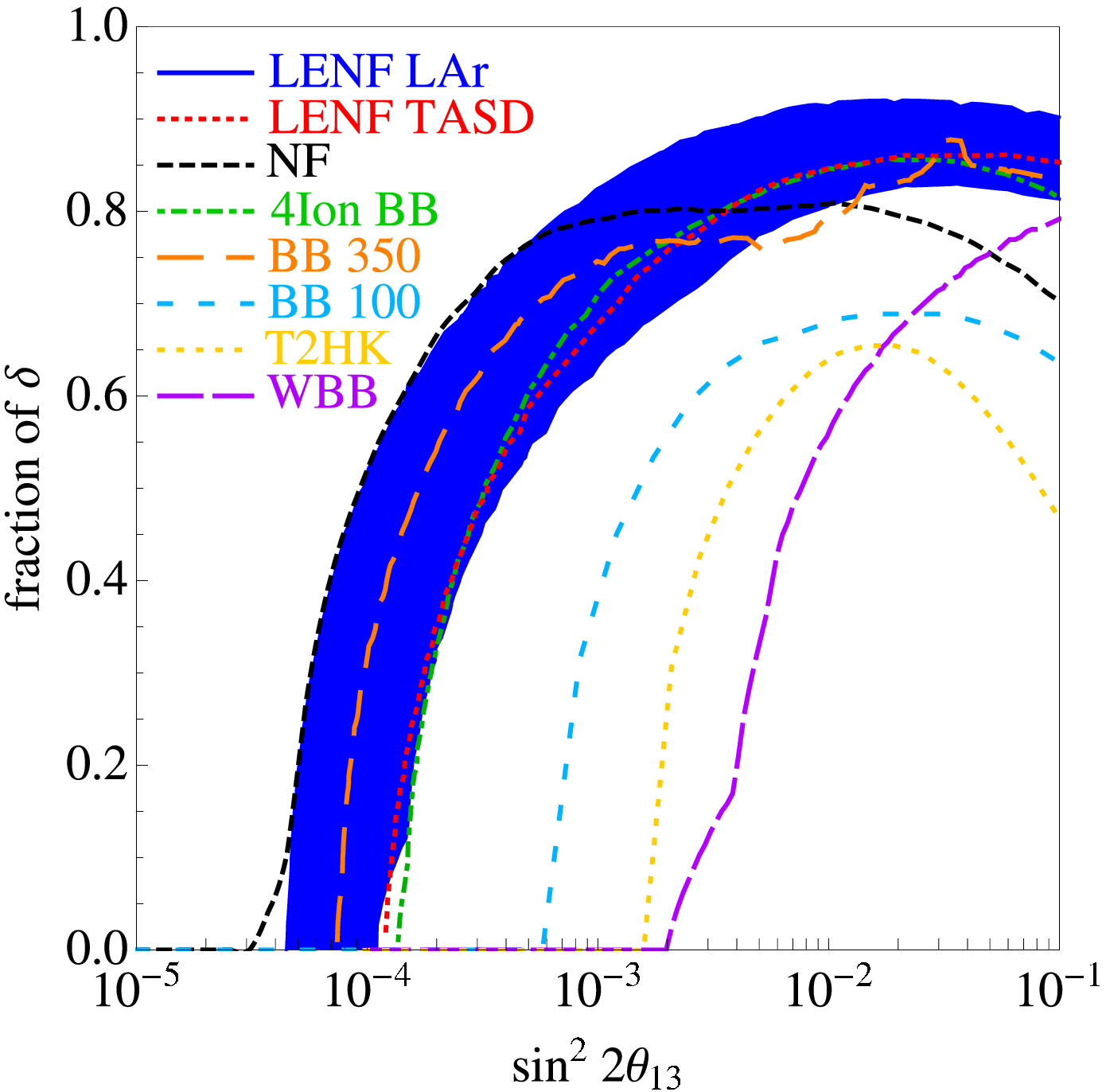}}
     \hspace{.3in}
     \subfigure[~Hierarchy sensitivity]{
          \includegraphics[scale=0.4]{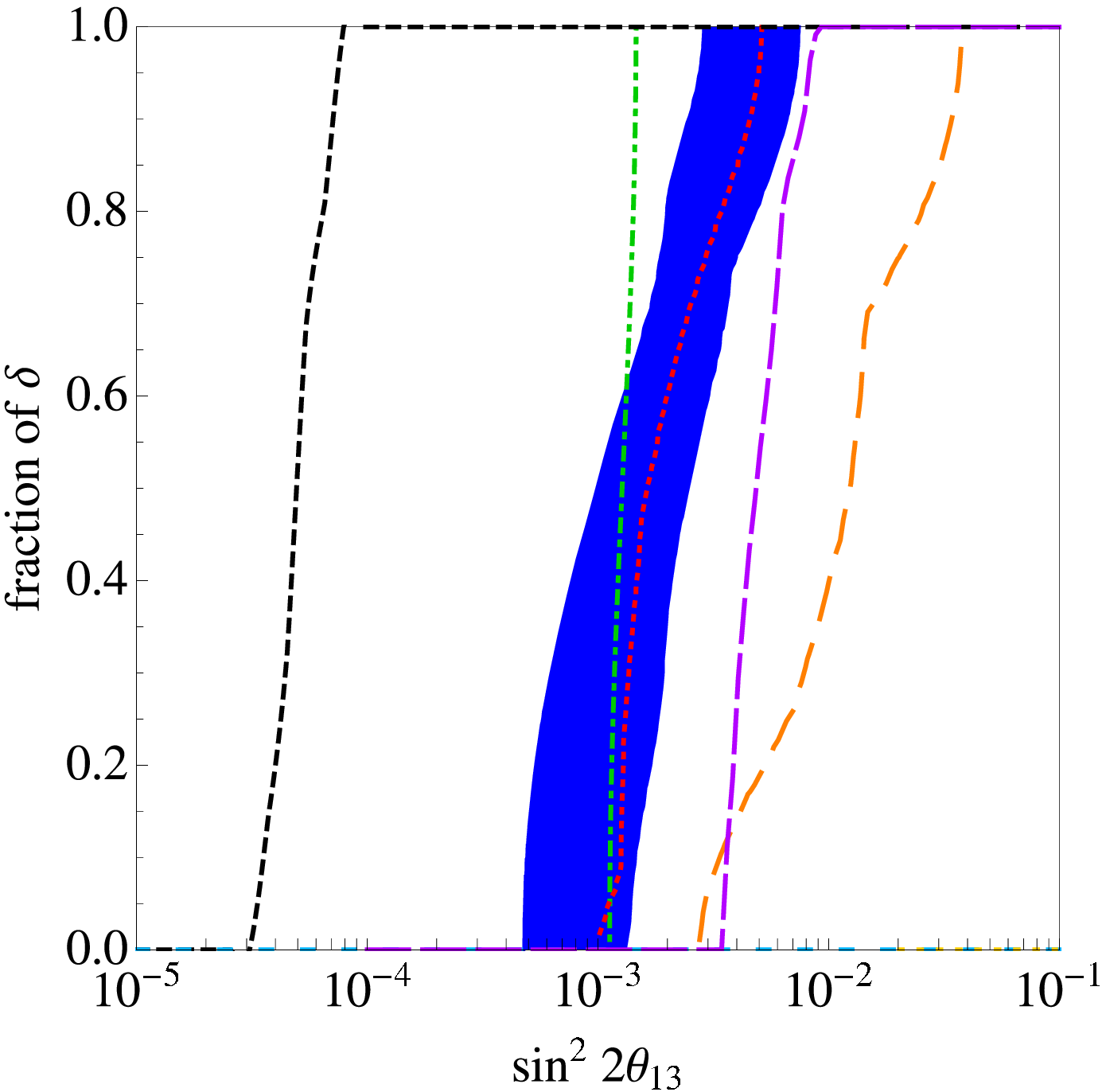}}
     \caption{~$3\sigma$ a) CP discovery potential and b) hierarchy sensitivity of the LENF with a 20 kton TASD and 100 kton LAr detector, the HENF, $\beta$-beams, T2HK, and the wide-band beam (WBB).}
\label{fig:osc}
\end{figure}

The $3\sigma$ CP discovery potential and sensitivity to the mass hierarchy of the LENF with either a LAr detector or a TASD is shown in Fig.~\ref{fig:osc} 
(taken from Ref.~\cite{FernandezMartinez:2010zza}). We also show the sensitivities of the high energy neutrino factory (HENF) \cite{ISS}, various $\beta$-beams \cite{Zucchelli:2002sa,Burguet-Castell:2003vv,Burguet-Castell:2005pa,Choubey:2009ks}, T2HK \cite{ISS} and the wide-band beam (WBB) \cite{WBB}. We see that a LENF with an optimistic LAr detector (left-hand edge of the blue band) has sensitivity to CP violation comparable to that of the HENF, for all values of $\theta_{13}$. A TASD also performs competitively for $\sin^{2}(2\theta_{13})\gtrsim10^{-3}$. The sensitivity to the mass hierarchy is an order of magnitude better than that of the wide-band beam which uses the same 1300 km baseline. 

\subsubsection{Sensitivity to non-standard interactions}

\begin{figure}[htp]
     \centering
     \subfigure[~$\varepsilon_{e\mu}=0$]{
          \includegraphics[scale=0.45]{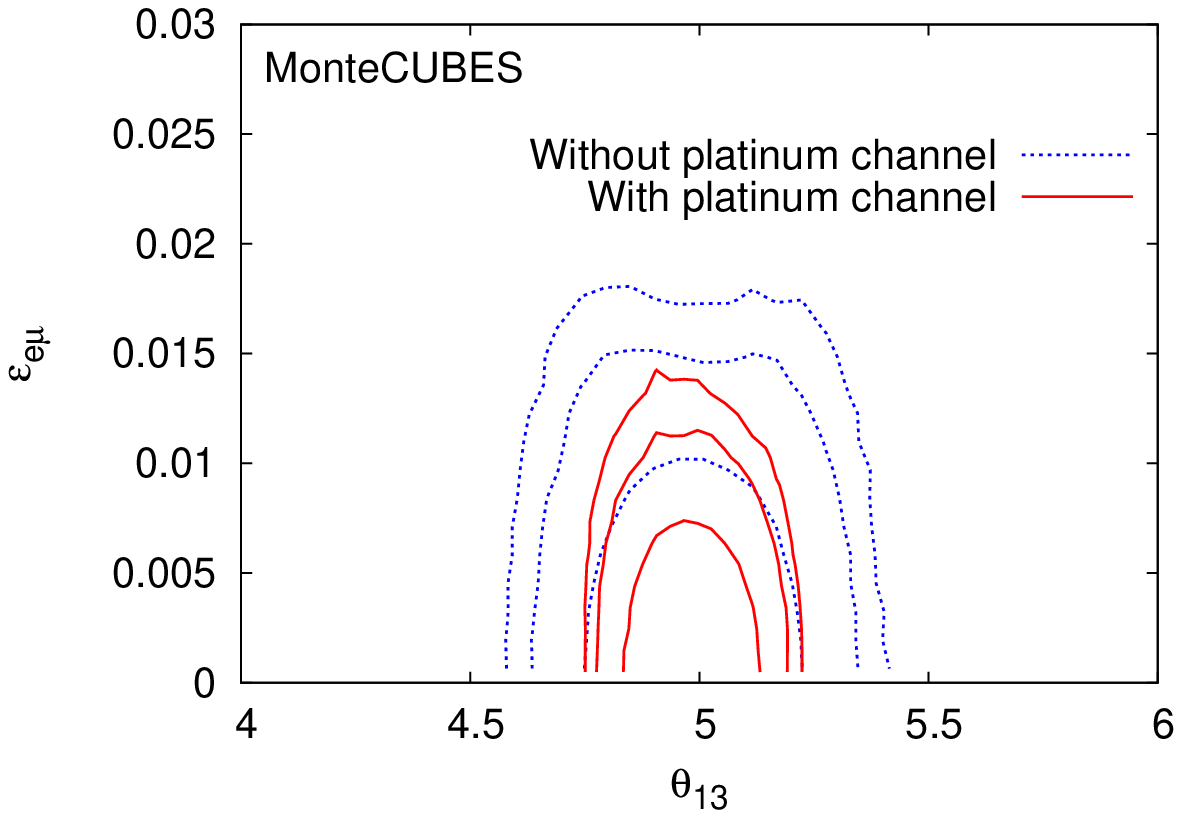}}
     \hspace{.3in}
     \subfigure[~$\varepsilon_{e\mu}=0.01$]{
          \includegraphics[scale=0.45]{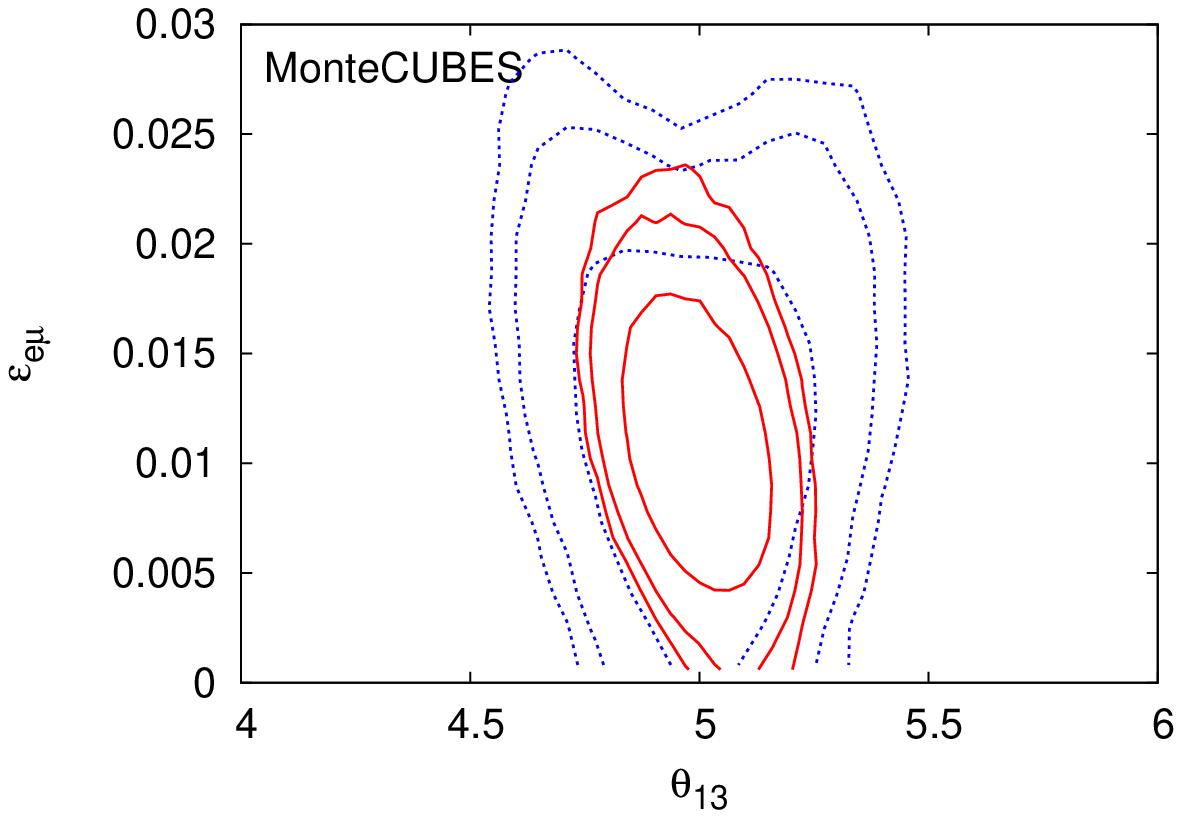}}
     \caption{$68\%$, $90\%$ and $95\%$ allowed regions in the $\theta_{13}-\varepsilon_{e\mu}$ plane, with and without the platinum channel, for true values of a) $\varepsilon_{e\mu}=0$ and b) $\varepsilon_{e\mu}=0.01$.}
\label{fig:NSI}
\end{figure}

Non-standard interactions (NSI's) can be parameterised by the parameters $\varepsilon_{\alpha\beta}$ \cite{Kopp:2008ds,Gago:2009ij}, which describe the rate of the transition $\nu_{\alpha}\rightarrow\nu_{\beta}$. The golden and platinum channels have leading order sensitivity to the parameters $\varepsilon_{e\mu}$ and $\varepsilon_{e\tau}$ \cite{Kopp:2008ds,Gago:2009ij}. We show the sensitivity of the LENF with a 20 kton TASD to the NSI parameter $\varepsilon_{e\mu}$ \cite{NSIpaper}, simulated by MonteCUBES \cite{Blennow:2009pk}. We illustrate how the platinum channel enhances the sensitivity of the experiment, by showing the $68\%$, $90\%$ and $95\%$ allowed regions in the $\theta_{13}-\varepsilon_{e\mu}$ plane, both with the platinum channel (red solid lines) and without (blue dashed lines), for the case of $\varepsilon_{e\mu}=0$ (Fig.~\ref{fig:NSI}a) and $\varepsilon_{e\mu}=0.01$ (Fig.~\ref{fig:NSI}b). The current bound on $\varepsilon_{e\mu}$ is $O(1)$ \cite{Biggio:2009nt}. Fig.~\ref{fig:NSI}a shows that the LENF could improve upon this and Fig.~\ref{fig:NSI}b shows that $\varepsilon_{e\mu}=0$ could be excluded at $\sim90\%$ confidence if $\varepsilon_{e\mu}\sim0.01$. We find that this sensitivity is not affected much by statistics, unlike for the case of standard oscillations. This is an indication that if NSI's are present, then the experimental sensitivity is limited by the degeneracies between the oscillation and NSI parameters, not by statistics. Unless this degeneracy can be resolved, for instance by including the platinum channels, then the sensitivity to both standard oscillation and NSI parameters will be severely limited.

%% file: CONTENT/11taus.tex
\subsection{The $\tau$-contamination of the golden muon sample at the Neutrino Factory \cite{Donini:2010xk}}
\label{sec:taucont}

The problem of $\tau$-contamination was first studied in Ref.~\cite{Indumathi:2009hg} in the context of precision measurements of the atmospheric parameters
$(\Delta m^2_{32}, \theta_{23})$ using the $\nu_\mu \to \nu_\mu$ channel at a Neutrino Factory. As it was shown in Ref.~\cite{Huber:2006wb}, it is useful to add all muons
in the final state without muon charge identification, as the overall efficiency and threshold are much better and the backgrounds are still under control. The improved efficiency
in the low-energy part of the neutrino spectrum, however, has the drawback that a previously irrelevant background becomes now potentially harmful. Oscillations into 
$\nu_\tau$'s, otherwise suppressed by low efficiency at low energy,
enhance both the right- and wrong-sign muon samples\footnote{They actually contaminate the electron sample, also \cite{Dutta:2011mc}.}. 
Oscillations of $\nu_e,\nu_\mu \to \nu_\tau$ will produce $\tau$'s through
$\nu_\tau N$ CC interactions within the detector that will, eventually, decay into muons (approximately 17\% of them). These muons from taus will, therefore, "contaminate" the "direct" muon samples (coming from $\nu_e,\nu_\mu \to \nu_\mu$ oscillations). Notice that muons from taus are not background but as good signal as the direct muons.

It is very hard to remove the muons from taus by applying kinematical cuts at an iron detector. Any cuts that attempt to do so drastically reduce the direct muon events as well
and hence worsen the sensitivity to the oscillation parameters. They escape essentially all filters designed to kill the dominant backgrounds and directly add to the direct muon sample,
see Ref.~\cite{Indumathi:2009hg}. On the other hand, neglect of the tau contribution will lead to an incorrect
conclusion about the precision achievable at the Neutrino Factory on a given observable.The
Ó$\tau$ -contaminationÓ must be added to the signal and it must be studied together with it.

The problem of $\tau$ -contamination of $\nu_e \to \nu_\mu$ was studied in detail in Ref.~\cite{Donini:2010xk}. A good
signal-to-background ratio is crucial to simultaneously determine with good precision $\theta_{13}$ and
$\delta$, since in this channel we have a statistics of tens of events at most. To separate highÐ-energy
charged currents from the lowÐ-energy dominant neutral current background. is important a 
good reconstruction of the neutrino energy. For this reason, in the standard MIND analysis
at the Neutrino Factory, the neutrino energy is reconstructed by adding the energy of the
muon and that of the hadronic jet. This operation, however, yields a wrong result when
the muon comes from a tau decay and it is detected at an iron calorimeter such as MIND,
since no additional information regarding the neutrino missing energy in the $\tau \to \nu_\tau \bar \nu_\mu \mu^-$
decay can be provided\footnote{This would not be the case at ECC \cite{Guler:2000bd} or Liquid Argon detectors \cite{Rubbia_LAr}, capable of separating the
$\nu_e \to \nu_\tau$ signal from the $\nu_e \to \nu_\mu$ one and to measure precisely the kinematic of the process under study (so
that the neutrino energy can be reconstructed precisely even $\nu_e \to \nu_\tau \to \tau \to \mu$ transitions).}. 
The sample of wrong-sign muons from the decay of wrong--sign taus
will be distributed erroneously in neutrino energy bins, thus contaminating the wrong-sign
muon sample by events whose parent neutrino energy is reconstructed wrongly.

\begin{figure}
    \includegraphics[width=0.45\textwidth]%
      {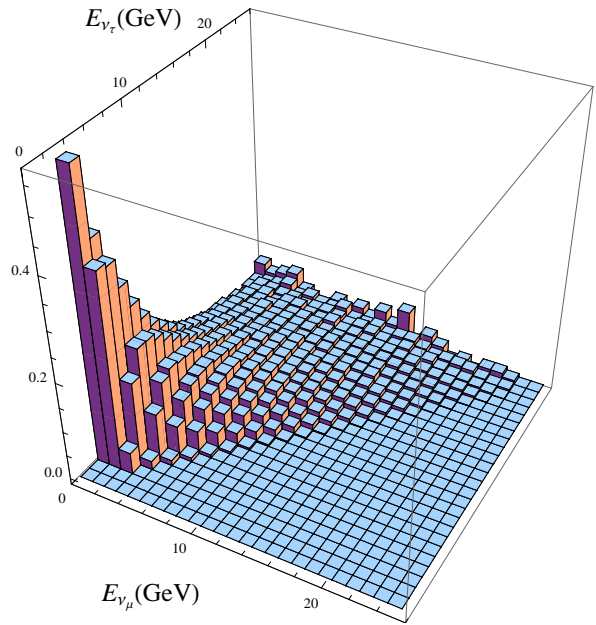} 
  \caption{ 
   The migration matrix $M_{ij}$. $M_{ij}$ has been statistically computed through a MonteCarlo
simulation of the events, using the GENIE neutrino generator \cite{Andreopoulos:2009rq} with $10^6$ simulated $\nu_\tau$Õs per neutrino
energy bin and 25 bins in the range $E_{\nu_\tau} \in [0, 25]$ GeV. From Ref.~\cite{Donini:2010xk}. 
  }
  \label{fig:tau1}
\end{figure}

Consider a $\nu_\tau$ of energy $E_{\nu_\tau}$, interacting in MIND and producing a wrongÐsign $\tau$ of energy
$E_\tau$ together with a hadronic jet of energy $E_h$. After $\tau$ -decay, $E_{\nu_\tau} = E_\tau + E_h = (E_\mu+E_{miss})+
E_h$, where $E_{miss}$ is the missing energy carried away by the two neutrinos in the $\tau$ -decay.
Experimentally, we observe the secondary muon and a hadronic jet, a signal essentially
indistinguishable from that of a wrongÐsign muon from CC $\nu_\mu$ interactions. However, in
this latter case, the addition of the (primary) muon energy $E_\mu$ and of the hadronic jet energy
$E_h$ results in the correct parent $\nu_\mu$ energy, $E_{\nu_\mu} = E_\mu+E_h$. On the other hand, in the former
case the addition of the (secondary) muon energy $E_\mu$ and of the hadronic jet energy $E_h$
results in the wrongly reconstructed fake neutrino energy $E_{fake} = E_\mu+E_h = E_{\nu_\tau} + E_{miss}$. 
If we divide the $\tau$ three-body decay energy distribution in discrete fake neutrino energy bins,
we find that for a monochromatic $\nu_\tau$ beam of energy $E_{\nu_\tau}$, the final muon will be assigned to a
given fake neutrino energy bin of energy $E^\mu_j$ with probability $V_j(E_{\nu_\tau})$, where $j = 1, . . . ,N^\mu_{bin}$.
We can compute the distribution of $\nu_\tau$ of a given energy $E_{\nu_\tau}$ and divide them into $\nu_\tau$ energy bins of energy 
$E^\tau_i$, where $i = 1, . . . , N^\tau_{bin}$. The ensemble of the probability vectors $V_j(E^\tau_i )$,
for $i$ and $j$ running over all the $\nu_\mu$ and $\nu_\tau$ energy bins, is represented by the migration matrix
$M_{ij}$ , Fig.~\ref{fig:tau1}.

After having computed $M_{ij}$ , the number of total wrong-sign muons in a given neutrino energy bin is given by
\begin{equation}
N_i(\theta_{13},\delta) = \sum_{i=1, N_{bin}} \left [ 
N^\mu_i (\theta_{13},\delta) + \sum_{j=1, N_{bin}} M_{ij} N^\tau_j (\theta_{13},\delta ) \right ] \, .
\end{equation}

\begin{figure}[!htb]
    \includegraphics[width=0.65\textwidth]%
      {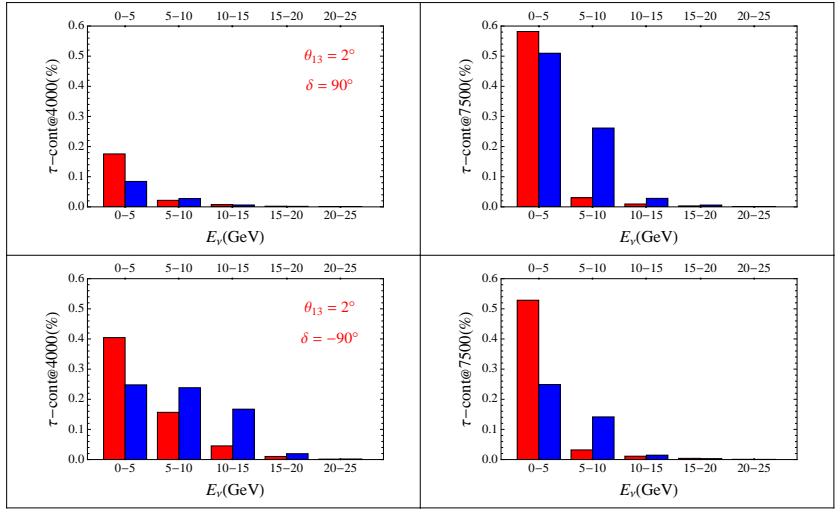}
  \caption{ 
 The fraction of $\tau$-contamination of the golden muon sample as a function of the reconstructed
neutrino energy, for $\theta_{13} = 2^\circ$. Left: L = 4000 Km; Right: L = 7500 Km. Top: $\delta= 90^\circ$; Bottom: $\delta = ?90^\circ$.
In red, we present the results obtained with normal hierarchy; in blue, with inverted hierarchy.
  }
  \label{fig:tau2}
\end{figure}

In Fig.~\ref{fig:tau2} it is shown the fraction of muons coming from $\tau$-decay that can be found in the
wrong-sign muon sample\footnote{The setup is a 25 GeV Neutrino Factory with $5 \times 10^{20}$ useful muon decays per year aiming at two 50
Kton MINDs located at L = 4000, 7500 Km, with 5 year of running time per muon polarity. An overall 2\%
systematic error has been considered.} after binning in the reconstructed neutrino energy, using the
MIND efficiency \cite{Abe:2007bi} at L = 4000 Km (left) and L = 7500 Km (right). The data are shown
for $\theta_{13} = 2^\circ, \delta = \pm 90^\circ$ (top and bottom, respectively) and the two possible neutrino mass
hierarchies (red: normal hierarchy; blue: inverted hierarchy). In all cases, a significant
$\tau$ -contamination can be observed below 5 GeV, ranging from a minimum of 10\% (inverted hierarchy, L = 4000 Km, $\delta = 90^\circ$)
to a maximum of 60\% (normal hierarchy, L = 7500 Km, $\delta = 90^\circ$). For normal hierarchy, the contamination in the energy range
$E_\nu \in [5, 10]$ GeV drops to the percent level (with the only exception of L = 4000 Km, $\delta = 90^\circ$). On
the other hand, for inverted hierarchy the decrease of the $\tau$ -contamination with the neutrino energy is softer, 
being as large as 30\% for L = 7500 Km, $\delta = 90^\circ$. Above 10 GeV, however,
the contamination is at the percent level for both hierarchies (again, with the only exception
of L = 4000 Km, $\delta = 90^\circ$ for inverted hierarchy).

\begin{figure}
  \begin{tabular}{cc}
    \includegraphics[width=0.45\textwidth]{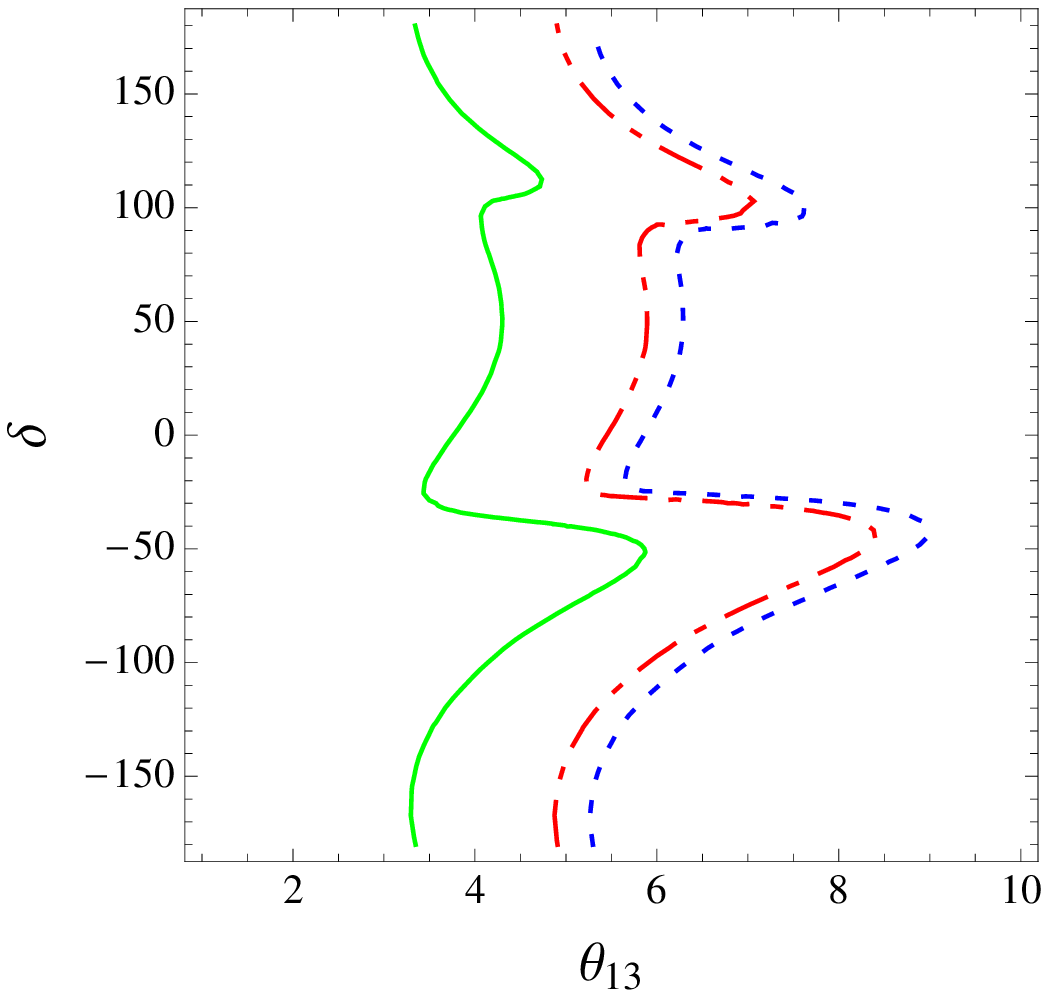} & 
    \includegraphics[width=0.45\textwidth] {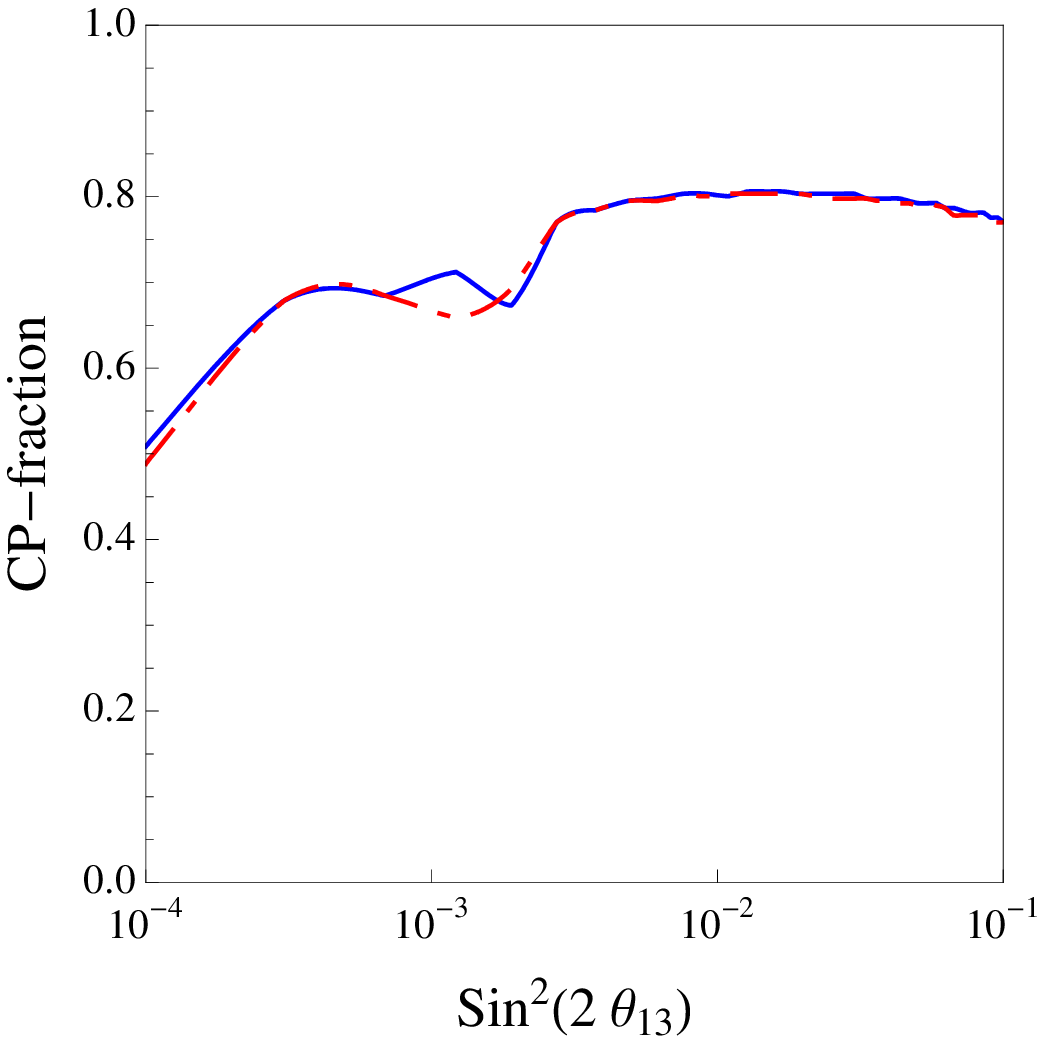} 
  \end{tabular}
  \caption{ 
    Left: Test of the hypothesis that a
    simulation of the data that includes the effect of the
    $\tau$-contamination at $L = 4\,000$~km can be fitted with the
    golden muon theoretical distribution.  
    In the regions to the right of the
    contour lines, the hypothesis can be rejected at 1, 2 or 3$\sigma$
    (from left to right), assuming the goodness-of-fit statistics
    follows the $\chi^2$ distribution with $n =8$ dof. From
    reference \cite{Donini:2010xk}.  
    Right: Comparison of the CP-fraction
    achievable at the IDS-NF baseline Neutrino Factory when
    the $\tau$-contamination is correctly taken into account (dashed
    red line) with respect to the ideal one in which no muons from
    taus are considered (solid blue line).  
  }
  \label{fig:tau3}
\end{figure}

The $\tau$ -contamination introduces, if not properly treated, an intolerable systematic error,
in particular for large $\theta_{13}$. Fig. ~\ref{fig:tau3} (left), from Ref.~\cite{Donini:2010xk}, shows that the test of the hypothesis
that simulated data including the $\tau$ -contamination, $N_i(\theta_{13}, \delta)$, can be fitted using the direct
wrong--sign muon distribution, $N^\mu_i (\theta_{13}, \delta)$, fails at more than $3 \sigma$ for $\theta_{13} \geq 5^\circ$. For $\theta_{13} \in [1^\circ, 5^\circ]$, 
even if $N^\mu_i (\theta_{13}, \delta)$ can fit the $\tau$ -contaminated data (with a relatively poor $\chi^2$, though),
the error in the joint measurement of $\theta_{13}$ and $\delta$ can be so large that it could actually
prevent the use of the Neutrino Factory as a precision facility (see, again, Ref.~\cite{Donini:2010xk} for a
detailed analysis of the errors introduced by a wrong treatment of the $\tau$ -contamination).
On the other hand, once $M_{ij}$ has been statistically computed, experimental data distributed
in reconstructed neutrino energy bins, can be fitted using the complete wrong-sign muons
distribution $N_i(\theta_{13}, \delta)$, properly taking into account the $\tau$ -contamination of the golden muon
sample. Using this procedure, the systematic error introduced by the muons from taus is
completely removed. The remaining error is the statistical error of the migration matrix
elements, that is under reasonable control.

It is worth noting that the $\tau$ -contamination of the wrong-sign muon sample, once properly
treated, does not worsen the measurement of $\theta_{13}$ and $\delta$, as it was the case for the atmospheric
parameters measurement. Fig.~\ref{fig:tau3}(right) shows the comparison of the CP-fraction
when only golden muons are considered (solid blue line) or when the total wrong-sign muon
signal is taken into account (dashed red line), as a function of $\sin^2 2 \theta_{13}$. It can be seen
that the only difference between the two lines is a slight displacement of the wiggles at
$\sin^2 2\theta_{13} \sim 10^{-3} (\theta_{13} \sim 1^\circ)$. The wiggles are a consequence of the loss of sensitivity to CP
violation introduced by the so-called Ósign clonesÓ for negative $\delta$ (a phenomenon known
as Ó$\pi$-transitÓ, \cite{Huber:2002mx}). Since the location of the clones in the two samples differs, a small
difference in the location of the wiggles is found when the two lines are compared. We can
see, however, that once the $\tau$ -contamination is properly treated, no (significant) loss in the
CP-fraction is found everywhere else. This is a consequence of the extremely low statistical
weight of the $\tau$ -contamination in the golden channel. Since the signal is represented by tens
of events, the $\tau$ -contamination amounts to a few events, at most. Once the problem of the
wrong assignment of muons from taus into reconstructed energy bins is solved by means of
the migration matrix approach, the residual statistical impact of the $\tau$ -contamination for
$\theta_{13} \leq 10^\circ$ is marginal.

%% file: CONTENT/12Neutrinoless.tex
\subsection{Neutrinoless double beta decay in seesaw models}

Neutrino oscillations, implying the massive nature of neutrinos, constitute an evidence for physics beyond the Standard Model (SM).
Thus, models accommodating  this neutrino masses become an important component in the search for new physics.
In this context, one of the most promising processes is the neutrinoless double beta decay (\znbb\ decay), whose detection is the purpose of several ongoing and upcoming experiments~\cite{Avignone:2007fu}.
Since this process is lepton number violating, its observation would imply that neutrinos are Majorana fermions~\cite{Schechter:1981bd}.

Among the most popular models for neutrino masses, we find the different types of seesaw mechanisms:
type-I~\cite{Minkowski:1977sc,Yanagida:1979as,Mohapatra:1979ia,GellMann:1980vs},
type-II~\cite{Magg:1980ut,Schechter:1980gr,Wetterich:1981bx,Lazarides:1980nt,Mohapatra:1980yp},
and type-III~\cite{Foot:1988aq,Ma:1998dn,Ma:2002pf,Hambye:2003rt}.
These are the extensions of the SM particle content that lead to the Weinberg $d=5$ effective 
operator~\cite{Weinberg:1979sa} after the extra mediators have been integrated out.

The extra degrees of freedom associated to the just mentioned models, and required to induce the 
Majorana nature of the SM neutrinos, can also contribute to the $0\nu\beta\beta$ decay. 
The effects of the SM neutrinos and these extra states, introduced to generate the neutrino masses, 
are usually analyzed as if they were independent (see, e.g., 
Refs.~\cite{Benes:2005hn,Atre:2009rg,Belanger:1995nh,Simkovic:1999re,delAguila:2008cj}).
However, if the light neutrino masses are generated in the context
of the model, important constraints between the light neutrino contribution and the one 
associated to the
extra states arise. Here we will study the phenomenological implications of these correlations.

In order to take into account the extra neutrino contributions to the \znbb\ decay rate in the 
most general type-I seesaw scenarios, we have computed the nuclear matrix element (NME) without any 
assumption on the mass of the neutrinos mediating the process. They are related to the observable 
\znbb\ decay rate as follows:
\begin{equation}
\frac{\Gamma_{0\nu\beta\beta}}{\ln2}=
G_{01}\left|\sum_{j}U_{ej}^{2}\frac{m_j}{m_e}M^{0\nu\beta\beta}(m_j)\right|^{2},
\label{eq:t-1_rep}
\end{equation}
\noindent
where $G_{01}$ is a well-known kinematical factor, $m_e$ the electron mass, $m_j$ the mass of the $j$ neutrino, 
$M^{0\nu\beta\beta}(m_j)$ the NME, which contain the dependence on the neutrino propagator,
and $U_{ej}$ are elements of the neutrino mixing matrix.

The calculation was performed using Interacting Shell Model (ISM) nuclear wavefunctions, one of the most popular methods employed to obtain \znbb\ decay NMEs~\cite{Menendez:2009dis}.
The results show two distinct regions for the behaviour of the NME as a function of the virtual neutrino mass:
almost constant up to $m_i \simeq 100$~MeV and then decreasing quadratically as the neutrino mass increases beyond $100$~MeV.
The transition mass $100$~MeV is the typical momentum exchange of the decay, which is the momentum of the virtual neutrino $|\mathbf{p}|$.
This scale is fixed by the typical distance between the two decaying nucleons, $r\simeq$~1 fm, and is sometimes referred as the nuclear scale.

The results of this computation are available at Ref.~\cite{Blennow:2010dt}.
Different approximations are required in order to perform the calculation, both obtaining the wavefunctions and in the treatment of the two-body transition operator,
and consequently some uncertainties are induced into the NMEs.
These can be estimated in $\sim ^{+25~\%}_{-35~\%}$ for light neutrino exchange ($m_j\le 100$~MeV) and $\sim ^{+35~\%}_{-40~\%}$ for heavy neutrinos ($m_j\ge 100$~MeV),
more sensitive to the short range part of the transition operator.

Within the type-I seesaw, depending on whether the extra mass eigenstates fall in the light or heavy neutrino mass regimes we can split their respective contributions to the amplitude:
\begin{equation}
 A \propto \sum_i^{\rm light} m_i U_{ei}^2 M^{0\nu\beta\beta}(m_i) + \sum_I^{\rm light} m_I U_{eI}^2 M^{0\nu\beta\beta}(m_I) + \sum_I^{\rm heavy} m_I U_{eI}^2 M^{0\nu\beta\beta}(m_I),
\label{eq:typeIampl}
\end{equation}
\noindent
where capital letters denote the mass index of the mostly sterile states and lowercase letters the  mostly active SM states.
Moreover, the diagonalization of the neutrino mass matrix provides an important constraint between 
the light and extra parameters:
\begin{equation}
\label{eq:constraint}
\sum_I^{\rm light} m_I U_{eI}^2 +\sum_I^{\rm light} m_I U_{eI}^2=0.
\end{equation}
\noindent
This simple relation, stemming from the fact that a Majorana mass coupling for the active neutrinos is forbidden
by the gauge symmetry of the SM, provides a useful constraint that should be always satisfied.
We can now distinguish three cases exhibiting very different phenomenologies depending on the mass 
regime of the extra mass eigenstates:

 {\bf All extra mass states are light.}
In this case Eq.~(\ref{eq:constraint}) implies
\begin{eqnarray}
 A &\approx& -\sum_I^{\rm light} m_I U_{eI}^2 \left(M^{0\nu\beta\beta}(0) - M^{0\nu\beta\beta}(m_I)  \right).
\label{eq:zero}
\end{eqnarray}
Since in this regime the NMEs are basically independent of the neutrino mass,
$M^{0\nu\beta\beta}(m_i) =M^{0\nu\beta\beta}(0) \simeq M^{0\nu\beta\beta}(m_I)$ the rate of 
\znbb\ decay is very suppressed. Indeed, only the different neutrino masses in the NME prevent a full 
cancellation leading to a suppression driven by $\Delta m^2/p^2$ with $|p^2| \simeq (100\ {\rm MeV})^2$.
Therefore, in this regime, the \znbb\ decay becomes experimentally inaccessible even being the neutrinos
Majorana particles. 

\begin{figure}
\begin{center}
\includegraphics[width=0.7\textwidth]{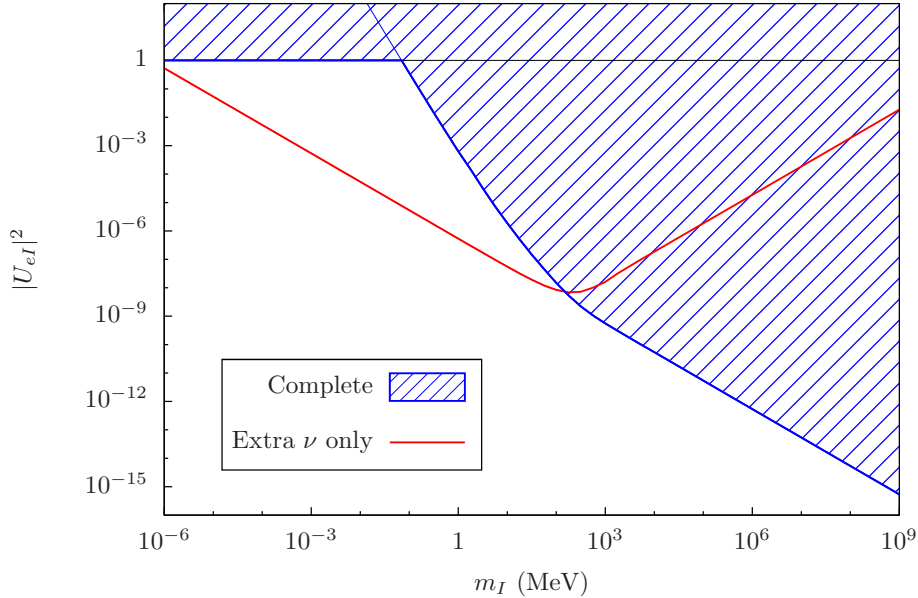}
 \bigskip 
 \caption{Bounds from CUORICINO on the extra neutrino mixing from \znbb\ decay in $^{130}$Te, with a 90~\%
CL half-life \cite{Arnaboldi:2008ds}. We have assumed the extra neutrinos are non hierarchical and show the bounds as a function of their 
common mass. We compare the case in which the contribution from the SM neutrinos is properly taken into account (striped area) to that in which only the extra contribution is considered (above the red line). 
\label{fig:U_bounds}}
\end{center}
\end{figure}

{\bf All extra mass states in the heavy regime.}
Now the NMEs for this extra states are very suppressed compared to the SM ones.
Furthermore, Eq.~(\ref{eq:constraint}) implies that \begin{equation}
 A \approx -\sum_I^{\rm heavy} m_I U_{eI}^2 \left(M^{0\nu\beta\beta}(0) - M^{0\nu\beta\beta}(m_I)  \right) \approx \sum_i^{\rm light} m_i U_{ei}^2 M^{0\nu\beta\beta}(0).
\label{eq:heavyfromlight}
\end{equation}
Thus, the contribution from the light active neutrinos (first term) dominates the transition rate. 
However, a much stronger bound on the mixing with the heavy neutrinos than the one usually 
shown in the literature can be extracted from it. This is shown in 
Fig.~\ref{fig:U_bounds} using our calculation for the NME and the results from CUORICINO. 

{\bf Extra mass states in the light and heavy regimes.}
In this scenario the leading terms stem from the light states:
\begin{equation}
 A \propto  \sum_i^{\rm light} m_i U_{ei}^2 M^{0\nu\beta\beta}(m_i) + \sum_I^{\rm light} m_I U_{eI}^2 M^{0\nu\beta\beta}(m_I).
\label{eq:tuning}
\end{equation}
Now it is possible to satisfy Eq.~(\ref{eq:constraint}) even in a situation where $m_i U_{ei}^2 \ll m_I U_{eI}^2$,
by canceling the contribution of the extra heavy states against that of the extra light ones while keeping the light neutrino masses small. 
In such a situation, the contribution of the light extra states could dominate over that of the active. As an example, if we consider the controversial Heidelberg-Moscow claim for a positive \znbb\ decay signal \cite{KlapdorKleingrothaus:2006ff},
the accommodation of this signal through only SM neutrinos would require, using our ISM NMEs, $0.24~\mathrm{eV} < m_{\beta \beta} < 0.89~\mathrm{eV}$ at $2\sigma$.
The interpretation of this claim as light active SM neutrinos is very disfavoured by the constraints from cosmology and neutrino oscillation data~\cite{Fogli:2008ig,Hannestad:2010yi,Komatsu:2010fb}.
However, this signal could be accommodated in a model with heavier neutrinos (which are not bounded by cosmology) mediating the process.
Indeed, we could reinterpret the result as $0.24~\mathrm{eV} < \left| \sum_I^{\rm heavy} m_I U_{eI}^2 \right| < 0.89~\mathrm{eV}$.

As for the type-II and type-III seesaws, current bounds from accelerator experiments place the extra degrees of freedom in the heavy regime~\cite{Abazov:2008ab}.
This effectively reduces the situation to that which appears for the type-I seesaw with only heavy extra states.
In the same manner, mixed type-I and type-II/type-III seesaw models (with extra neutrinos below 
$100$ MeV) resemble the situation of type-I seesaw with both light and heavy extra states.

%% file: CONTENT/13GiuntiLaveder.tex
\subsection{Short-Baseline $\bar\nu_{\mu}\to\bar\nu_{e}$ Oscillations \cite{Giunti:2010jt}}

\begin{table*}[t!]
\footnotesize
\begin{center}
\begin{tabular}{ccccccccc}
&
&
MB
&
LS
&
MB+LS
&
KA
&
MB+LS+KA
&
Re
&
(MB+LS+KA)+Re
\\
\hline
 No Osc. & $\chi^{2}$ & $ 17.8 $ & $ 15.0 $ & $ $ & $ 6.8 $ & $ $ & $ 51.0 $ & $ $ \\
 & NDF & $ 16 $ & $ 5 $ & $ $ & $ 8 $ & $ $ & $ 56 $ & $ $ \\
 & GoF & $ 0.33 $ & $ 0.010 $ & $ $ & $ 0.55 $ & $ $ & $ 0.66 $ & $ $ \\
\hline Osc. & $\chi^{2}_{\text{min}}$ & $ 10.6 $ & $ 1.4 $ & $ 13.3 $ & $ 6.4 $ & $ 24.1 $ & $ 48.5 $ & $ 75.6 $ \\
 & NDF & $ 14 $ & $ 2 $ & $ 18 $ & $ 6 $ & $ 26 $ & $ 54 $ & $ 82 $ \\
 & GoF & $ 0.71 $ & $ 0.51 $ & $ 0.77 $ & $ 0.38 $ & $ 0.57 $ & $ 0.69 $ & $ 0.68 $ \\
 & $\sin^22\vartheta_{\text{bf}}$ & $ 0.83 $ & $ 0.0058 $ & $ 0.0059 $ & $ 0.0010 $ & $ 0.89 $ & $ 0.042 $ & $ 0.013 $ \\
 & $\Delta{m}^2_{\text{bf}}$ & $ 0.069 $ & $ 8.13 $ & $ 4.57 $ & $ 6.76 $ & $ 0.055 $ & $ 1.86 $ & $ 0.46 $ \\
\hline PG & $\Delta\chi^{2}_{\text{min}}$ & $ $ & $ $ & $ 1.30 $ & $ $ & $ 5.80 $ & $ $ & $ 2.94 $ \\
 & NDF & $ $ & $ $ & $ 2 $ & $ $ & $ 4 $ & $ $ & $ 2 $ \\
 & GoF & $ $ & $ $ & $ 0.52 $ & $ $ & $ 0.21 $ & $ $ & $ 0.23 $ \\
\hline
\end{tabular}
\caption{ \label{003}
Values of
$\chi^{2}$,
number of degrees of freedom (NDF),
goodness-of-fit (GoF)
and
best-fit values
$\sin^22\vartheta_{\text{bf}}$, $\Delta{m}^2_{\text{bf}}$
of the oscillation parameters
obtained from
the fit of various combinations of
MiniBooNE (MB),
LSND (LS),
KARMEN (KA)
and
reactor Bugey and Chooz (Re)
antineutrino data.
The first three lines correspond to the case of no oscillations (No Osc.).
The following five lines correspond to the case $\bar\nu_{\mu}\to\bar\nu_{e}$ oscillations (Osc.).
The last three lines give the parameter goodness-of-fit (PG) \protect\cite{Maltoni:2003cu}.
The variations of
$\sin^22\vartheta_{\text{bf}}$ and $\Delta{m}^2_{\text{bf}}$
depending on the fitted data sets
are due to the oscillating character of
$P_{\bar\nu_{\mu}\to\bar\nu_{e}}$
in Eq.~(\ref{002}).
See Ref.~\cite{Giunti:2010jt} for details.
}
\end{center}
\end{table*}

The MiniBooNE collaboration
\cite{AguilarArevalo:2010wv}
recently reported the observation of a signal
of short-baseline $\bar\nu_{\mu}\to\bar\nu_{e}$ transitions
compatible with that observed in the LSND experiment
\cite{Aguilar:2001ty}.
The agreement of the MiniBooNE and LSND signals in favor of neutrino oscillations
is remarkable,
because the two experiments observed the signal
of $\bar\nu_{\mu}\to\bar\nu_{e}$ transitions
at different source-detector distances and different neutrino energy ranges.
Since only the ratio of distance and energy is similar in the two experiments
and neutrino oscillations depend just on this ratio
(see Refs.~\cite{Bilenky:1998dt,Bilenky:2002aw,Giunti:2003qt,Maltoni:2004ei,Fogli:2005cq,Strumia:2006db,GonzalezGarcia:2007ib,Giunti-Kim-2007}),
the neutrino oscillation explanation of the two signals is strongly favored.
On the other hand,
the MiniBooNE collaboration did not observe any signal of short-baseline
$\nu_{\mu}\to\nu_{e}$ transitions
\cite{AguilarArevalo:2008rc}
compatible with the MiniBooNE and LSND signals of
$\bar\nu_{\mu}\to\bar\nu_{e}$ transitions.
Therefore,
it is possible that the effective parameters which govern neutrino and antineutrino oscillations
are different, maybe because of a violation of the CPT symmetry
\cite{Murayama:2000hm,Barenboim:2001ac,Bilenky:2001ka,Barenboim:2002rv,Strumia:2002fw,Bahcall:2002ia,Murayama:2003zw,Barger:2003xm,Minakata:2005jy,GonzalezGarcia:2003jq,Laveder:2007zz,Giunti:2007xv,Antusch:2008zj,Giunti:2009zz,Dolgov:2009yk,Giunti:2009en,Barenboim:2009ts,Giunti:2010wz,Giunti:2010zs}.
From a phenomenological point of view it is interesting to
consider the neutrino and antineutrino sectors independently,
especially in view of possible experimental checks of the
short-baseline $\bar\nu_{\mu}\to\bar\nu_{e}$ signal
\cite{Baibussinov:2009tx,Stancu:2009vq,Agarwalla:2010zu,AndreRubbia:NEU2012}.
Here we adopt this point of view
\cite{Giunti:2010jt}
and we present the results of a combined fit of the
MiniBooNE and LSND antineutrino data
in favor of
short-baseline $\bar\nu_{\mu}\to\bar\nu_{e}$ transitions,
together with the constraints imposed by the data of the
KARMEN experiment
\cite{Armbruster:2002mp}
in which the transitions have not been observed.
We also take into account the
constraints imposed by the absence of short-baseline
$\bar\nu_{e}$ disappearance observed in reactor antineutrino experiments.

In the analysis of the data of
$\bar\nu_{\mu}\to\bar\nu_{e}$
oscillation experiments
we consider the simplest case of an effective two-neutrino-like
short-baseline oscillation probability,
similar to that obtained in the case of four-neutrino mixing
(see Refs.~\cite{Bilenky:1998dt,Maltoni:2004ei,Strumia:2006db,GonzalezGarcia:2007ib}),
\begin{equation}
P_{\bar\nu_{\mu}\to\bar\nu_{e}}(L/E)
=
\sin^22\vartheta
\sin^2\left(\dfrac{\Delta{m}^2 L}{4 E}\right)
\,,
\label{002}
\end{equation}
where $\Delta{m}^2$ is the relevant neutrino squared-mass difference
and
$\vartheta$ is the effective mixing angle for $\bar\nu_{\mu}\to\bar\nu_{e}$ transitions.

Tab.~\ref{003} shows the results of the fit of various combinations of
MiniBooNE (MB),
LSND (LS),
KARMEN (KA)
and
reactor Bugey \cite{Declais:1995su} and Chooz \cite{Apollonio:2002gd} (Re)
antineutrino data.

\begin{figure}[t!]
\begin{center}
\includegraphics*[bb=7 14 563 569, width=0.49\linewidth]{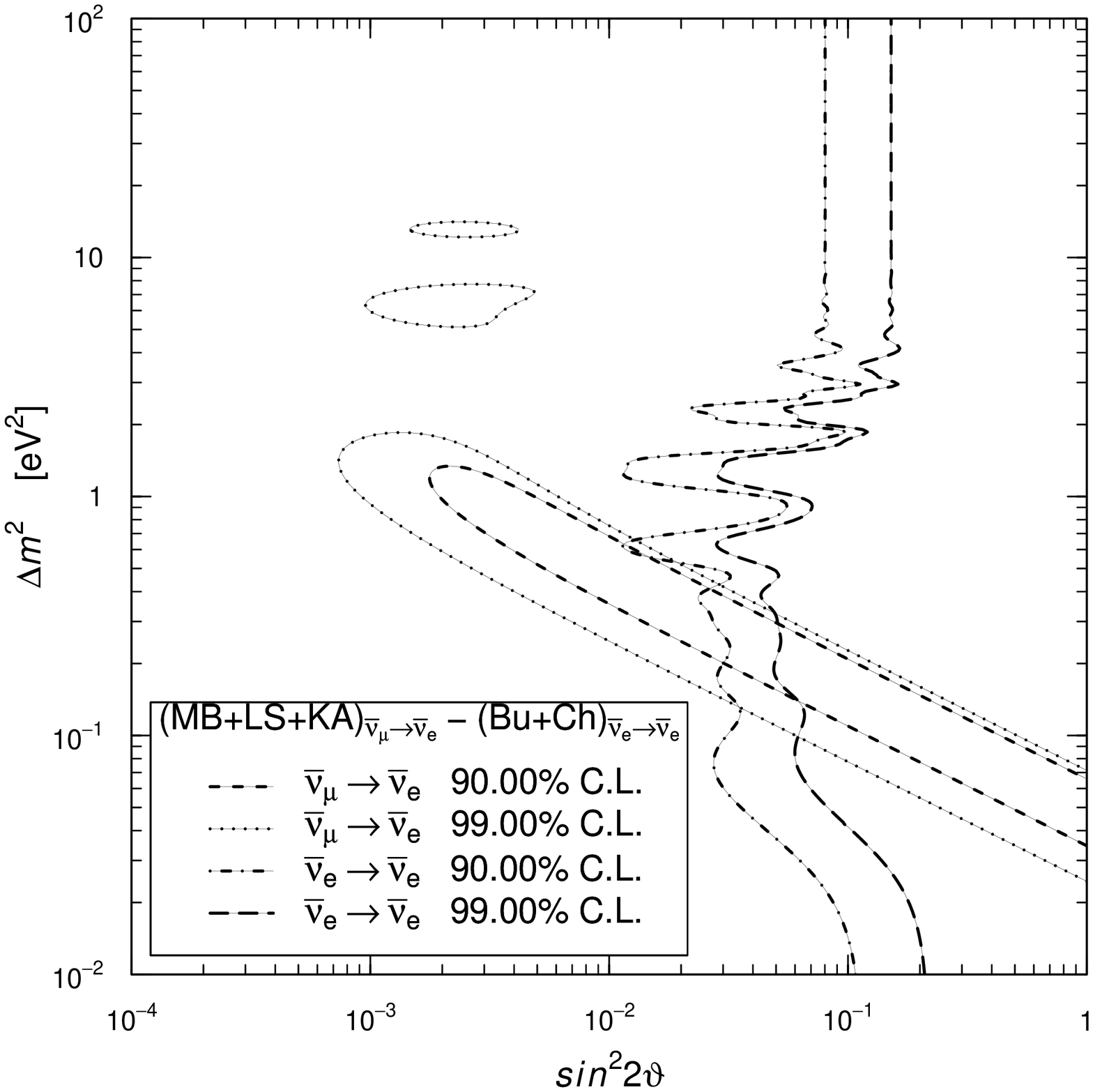}
\hfill
\includegraphics*[bb=5 11 571 571, width=0.49\linewidth]{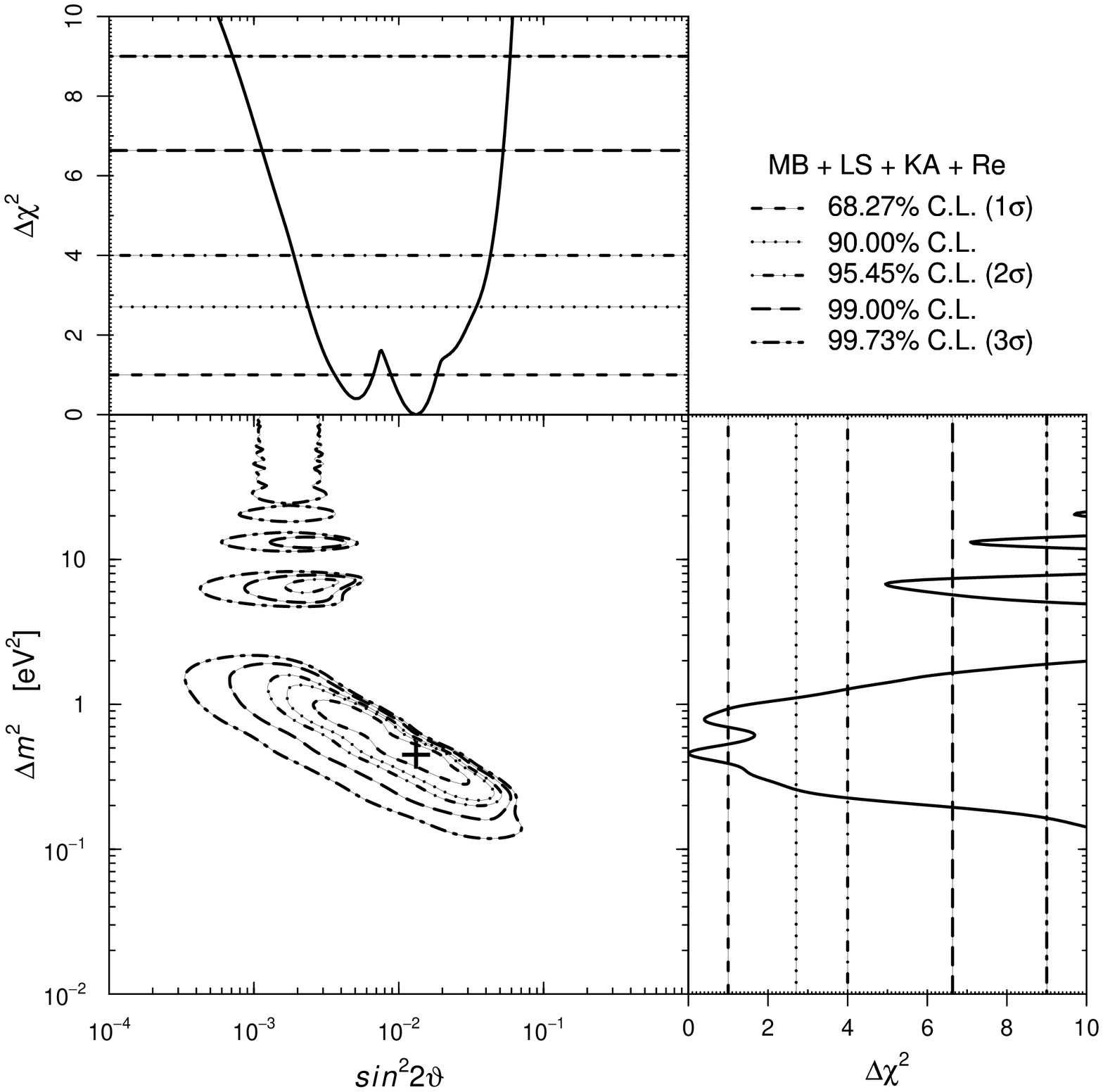}
\end{center}
\caption{ \label{005}
Left:
superposition of the allowed regions in the
$\sin^{2}2\vartheta$--$\Delta{m}^{2}$ plane
obtained from the combined fit of MiniBooNE (MB), LSND (LS) and KARMEN (KA)
$\bar\nu_{\mu}\to\bar\nu_{e}$ data and
the exclusion curves obtained from the fit of reactor Bugey (Bu) and Chooz (Ch)
$\bar\nu_{e}\to\bar\nu_{e}$ data.
Right:
allowed regions in the
$\sin^{2}2\vartheta$--$\Delta{m}^{2}$ plane
and
marginal $\Delta\chi^{2}=\chi^2-\chi^2_{\text{min}}$'s
for
$\sin^{2}2\vartheta$ and $\Delta{m}^{2}$
obtained
from the combined fit of MiniBooNE (MB), LSND (LS) and KARMEN (KA)
$\bar\nu_{\mu}\to\bar\nu_{e}$ data and
the exclusion curves obtained from the fit of reactor Bugey and Chooz (Re)
$\bar\nu_{e}\to\bar\nu_{e}$ data.
The best-fit point is indicated by a cross.
See Ref.~\cite{Giunti:2010jt} for details.
}
\end{figure}

The left panel in Fig.~\ref{005}
shows a superposition of the 90\% and 99\% C.L.
allowed regions in the
$\sin^{2}2\vartheta$--$\Delta{m}^{2}$ plane
obtained from the combined fit of MiniBooNE, LSND and KARMEN
$\bar\nu_{\mu}\to\bar\nu_{e}$ data and
the exclusion curves obtained in Ref.~\cite{Acero:2007su} from the fit of reactor
Bugey and Chooz
$\bar\nu_{e}\to\bar\nu_{e}$ data,
which currently provide the most stringent constraints on short-baseline reactor $\bar\nu_{e}$ disappearance.
The model-independent inequality
\begin{equation}
P_{\bar\nu_{\mu}\to\bar\nu_{e}}
\leq
1 - P_{\bar\nu_{e}\to\bar\nu_{e}}
\label{034}
\end{equation}
implies that in the left panel in Fig.~\ref{005}
the large-$\sin^22\vartheta$ part of
the straight region below $\Delta{m}^2\approx2\,\text{eV}^2$
allowed by the combined fit of MiniBooNE, LSND and KARMEN
$\bar\nu_{\mu}\to\bar\nu_{e}$ data
is excluded by
the results of reactor antineutrino experiments.
Quantitatively,
only the parts with
$\sin^22\vartheta \lesssim 3\times10^{-2}$
and
$\sin^22\vartheta \lesssim 5\times10^{-2}$
are allowed at
90\% and 99\% C.L., respectively.

The right panel in Fig.~\ref{005} and the last column of Tab.~\ref{003}
give the results of the combined fit of
accelerator MiniBooNE, LSND and KARMEN
$\bar\nu_{\mu}\to\bar\nu_{e}$ data and
reactor Bugey and Chooz
$\bar\nu_{e}\to\bar\nu_{e}$ data
assuming an equality in Eq.~(\ref{034}).
The value of the parameter goodness-of-fit in Tab.~\ref{003}
shows that the accelerator and reactor data are compatible
under the hypothesis of $\bar\nu_{\mu}\to\bar\nu_{e}$ oscillations.

From the right panel in Fig.~\ref{005} one can see that
there is a favorite region at about 95\% C.L. around the best-fit point for
$2\times10^{-3} \lesssim \sin^22\vartheta \lesssim 5\times10^{-2}$
and
$0.2 \lesssim \Delta{m}^2 \lesssim 2 \, \text{eV}^2$.
Larger values of
$\Delta{m}^2$
are allowed only at more than about 95\% C.L.
for
$5\times10^{-4} \lesssim \sin^22\vartheta \lesssim 5\times10^{-3}$.

This region is interesting for a study of the possibilities to check the
LSND and MiniBooNE indication of
short-baseline
$\bar\nu_{\mu}\to\bar\nu_{e}$
oscillations
with future experiments
\cite{Baibussinov:2009tx,Stancu:2009vq,Agarwalla:2010zu,AndreRubbia:NEU2012}.

%% file: CONTENT/14AkhmedovSchwetz.tex
\subsection{MiniBooNE/LSND data: NSI's in a (3+1)-scheme vs. (3+2)-oscillations \cite{Akhmedov:2010vy}}

Recent MiniBooNE anti-neutrino data \cite{AguilarArevalo:2010wv} indicate an
excess of $\bar\nu_e$ events, in agreement with the LSND evidence
\cite{Aguilar:2001ty} for $\bar\nu_\mu\to\bar\nu_e$ transitions. It is known
that oscillations with one or more sterile neutrino(s) at the eV scale are
not sufficient to explain the global data \cite{Maltoni:2007zf,
Karagiorgi:2009nb}. The preprint EURONU-WP6-10-24,
Ref.~\cite{Akhmedov:2010vy}, investigated the possibility that in addition
to a sterile neutrino there are some non-standard neutrino interactions
(NSI), beyond the Standard Model weak interactions. Since matter effects are
tiny for the short baselines relevant here, charged-current (CC) type NSI in
the neutrino source and detector are considered. Thanks to the interference
between NSI effects and oscillations with $\Delta m^2_{41} \sim 1$~eV$^2$ CP
violation is obtained, even in the presence of only one mass scale. This
effect is used to reconcile the indication for $\bar\nu_\mu\to\bar\nu_e$ in
anti-neutrino experiments (LSND and MiniBooNE) with the absence of a signal
in MiniBooNE neutrino data. 

A general parameterisation of the relevant transition and survival
probabilities in the presence of oscillations (within the one-mass scale
approximation) and NSI is presented, and particular combinations of mixing
matrix elements $U_{\alpha 4}$ and NSI parameters
$\varepsilon_{\alpha\beta}$ entering in the probabilities are identified.
This drastically reduces the number of independent parameters and makes a
general fit to global short-baseline data feasible.

Two versions of the (3+1) NSI model are considered. In the general case
(denoted NSI$^g$) one makes use of the fact that the neutrino production
mechanism in LSND (and in KARMEN) is muon decay (purely leptonic), whereas
in all other experiments neutrino production and detection are
semi-leptonic, involving transitions between $u$ and $d$ quarks. Therefore,
in the presence of suitable NSI parameters one can decouple the transition
probabilities in LSND and KARMEN from the rest of the data. In this case one
obtains an excellent fit to the global data and the tension between
appearance and disappearance experiments is resolved. In this case MiniBooNE
does not provide a direct test of LSND, since different combinations of
parameters are relevant for them. Also, in the global fit the excess
observed in MiniBooNE anti-neutrino data is not reproduced. 

For the second version of the (3+1) NSI model the assumption is adopted that
NSI involving the charged muon can be neglected. In this case exactly the
same NSI parameters are relevant for LSND and KARMEN as for all other
experiments. In this constrained model (NSI$^c$) one makes use of the CP
violation due to NSI--oscillation interference to reconcile neutrino and
anti-neutrino data. It is shown that in the NSI$^c$ model there is a
factorisation between appearance and disappearance amplitudes, similar to
that in the (3+1) oscillation scheme. Therefore, it is more difficult to
satisfy constraints from disappearance experiments and some tension is left
in the fit. However, also this model provides significant improvement of the
global fit compared to the pure oscillation case. 

The results of the fits are presented in terms of effective parameters,
representing the specific combinations of NSI parameters entering in the
transition probabilities. However, for both cases, NSI$^c$ and NSI$^g$, 
examples are provided of how to realise the required parameters in terms of
the fundamental mixing and NSI parameters. It is shown that values in safe
agreement with bounds on the various $\varepsilon$'s can be found to realise
the fits which require $\varepsilon$'s of order a few$\,\times 10^{-2}$.

The quality of the (3+1) NSI fits are compared to an updated fit in the
(3+2) oscillation scheme, which also allows for CP violation due to the
presence of two relevant mass scales. Similarly to (3+1) NSI, in (3+2) the
appearance experiments can be described very well. However, as 
previously \cite{Maltoni:2007zf}, for (3+2) oscillations significant tension
remains in the global fit between appearance and disappearance experiments.
The improvement of (3+2) compared to (3+1) is not significant, in terms of
$\chi^2$ gain per new parameter. 
None of the scenarios considered in \cite{Akhmedov:2010vy} can explain the
MiniBooNE low energy excess of events when disappearance data are taken into
account. Therefore, the data below 475~MeV is excluded from the analysis,
relying on a separate explanation for this anomaly.

The predictions of the model for future experiments depend on the detailed
realization in terms of mixing and NSI parameters. In general one may expect
some signals in searches for deviations from the standard three-flavour
oscillation picture in both respects, sterile neutrino oscillations as well
as NSI. Several proposals to search for sterile neutrinos at the eV scale
have been presented recently, see for example~\cite{Donini:2008wz,
Giunti:2009en, Baibussinov:2009tx, Agarwalla:2009em, Meloni:2010zr,
Agarwalla:2010zu}. In~\cite{Hamann:2010bk} implications of sterile neutrinos
for latest cosmological data have been investigated. Recent studies on NSI
in the context of upcoming and far future experiments can be found, e.g.,
in~\cite{Kopp:2007ne, Ohlsson:2008gx, Kopp:2008ds, Meloni:2009cg}.
A specific prediction of the scenario are zero-distance effects in
appearance searches~\cite{Langacker:1988up, FernandezMartinez:2007ms,
Meloni:2009cg}. Hence, the observation of an energy independent appearance
probability at very short distances is a characteristic signature from this
kind of models. 

The model may also provide a signature at the LHC. Typically, realising
CC-like interactions as the ones considered here require a charged particle
as mediator. The NSI parameters $\varepsilon$ measure the strength of the
new interactions relative to the standard weak interaction strength set by
$G_F$. Therefore, from the fit results, $\varepsilon \sim 0.01$, one expects
that the mass of a mediator for a dimension-6 operator should be roughly one
order of magnitude larger than the $W$ boson mass. Hence, one might expect
charged particles to show up at the TeV scale, with good prospects to be
observed at LHC. However, the results of~\cite{Antusch:2008tz,Gavela:2008ra}
suggest that NSI at the level of 0.01 are difficult to obtain from
dimension-6 operators without being in conflict with bounds on
charged-lepton processes. As discussed there, a possibility to obtain such
large NSI would be to go to dimension-8 operators and allow for some fine
tuning.

%% file: CONTENT/17MeloniTangWinter.tex
\subsection{Sterile neutrinos beyond LSND at the Neutrino Factory \cite{Meloni:2010zr}}

\begin{figure}[!t]
\label{fig:theta-mass}
\includegraphics[width=0.32\textwidth]{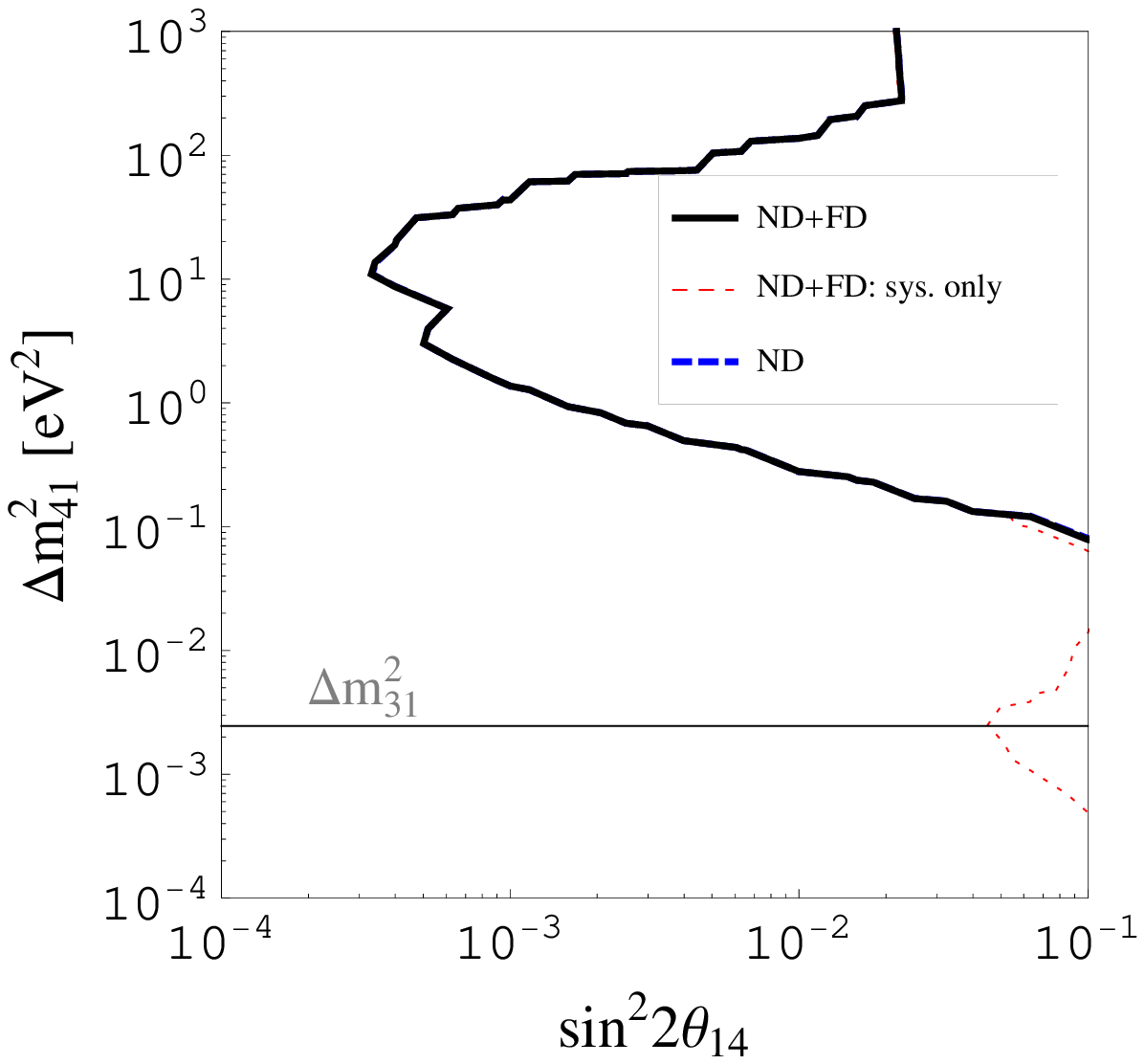} 
\includegraphics[width=0.32\textwidth]{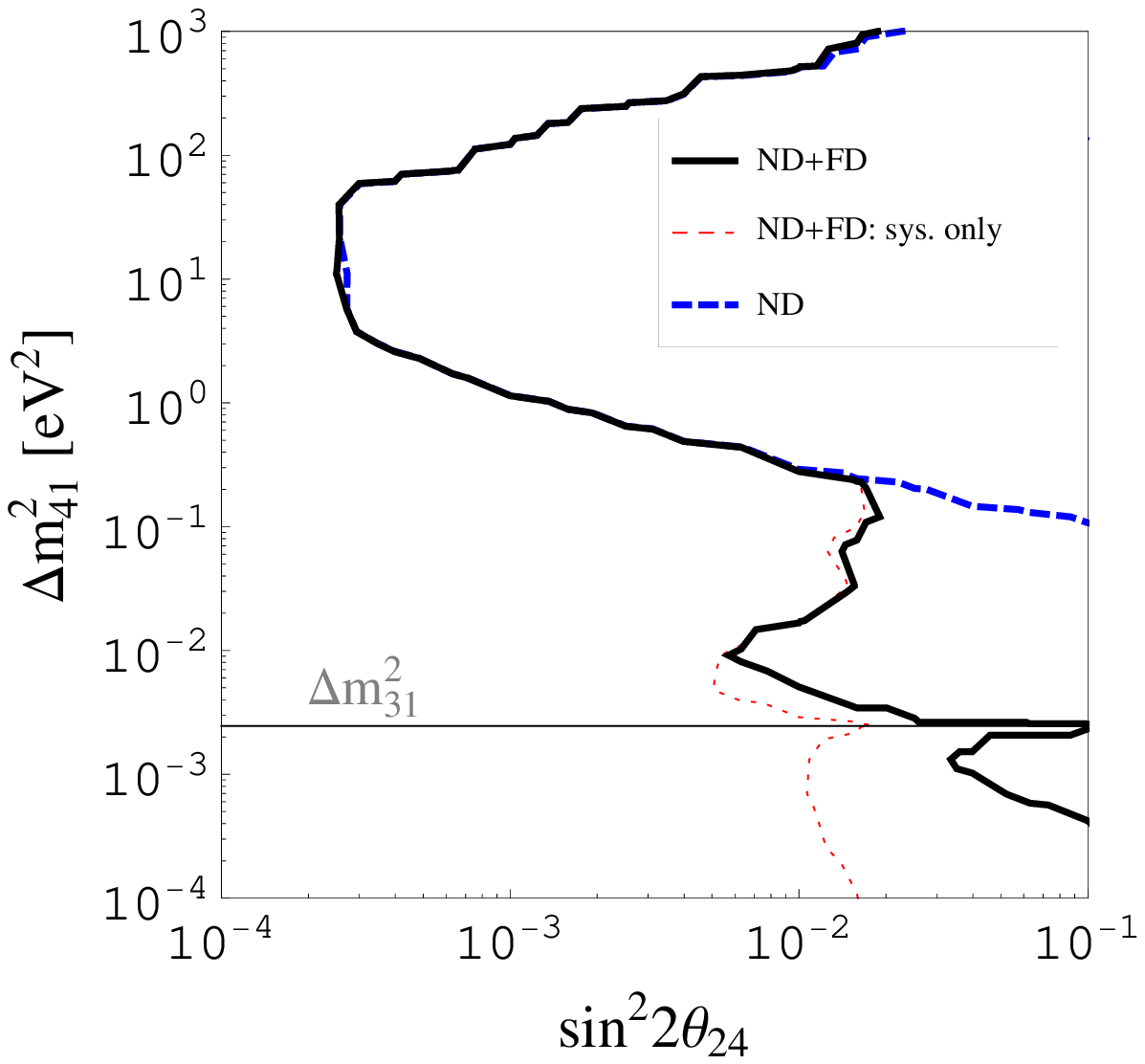} 
\includegraphics[width=0.32\textwidth]{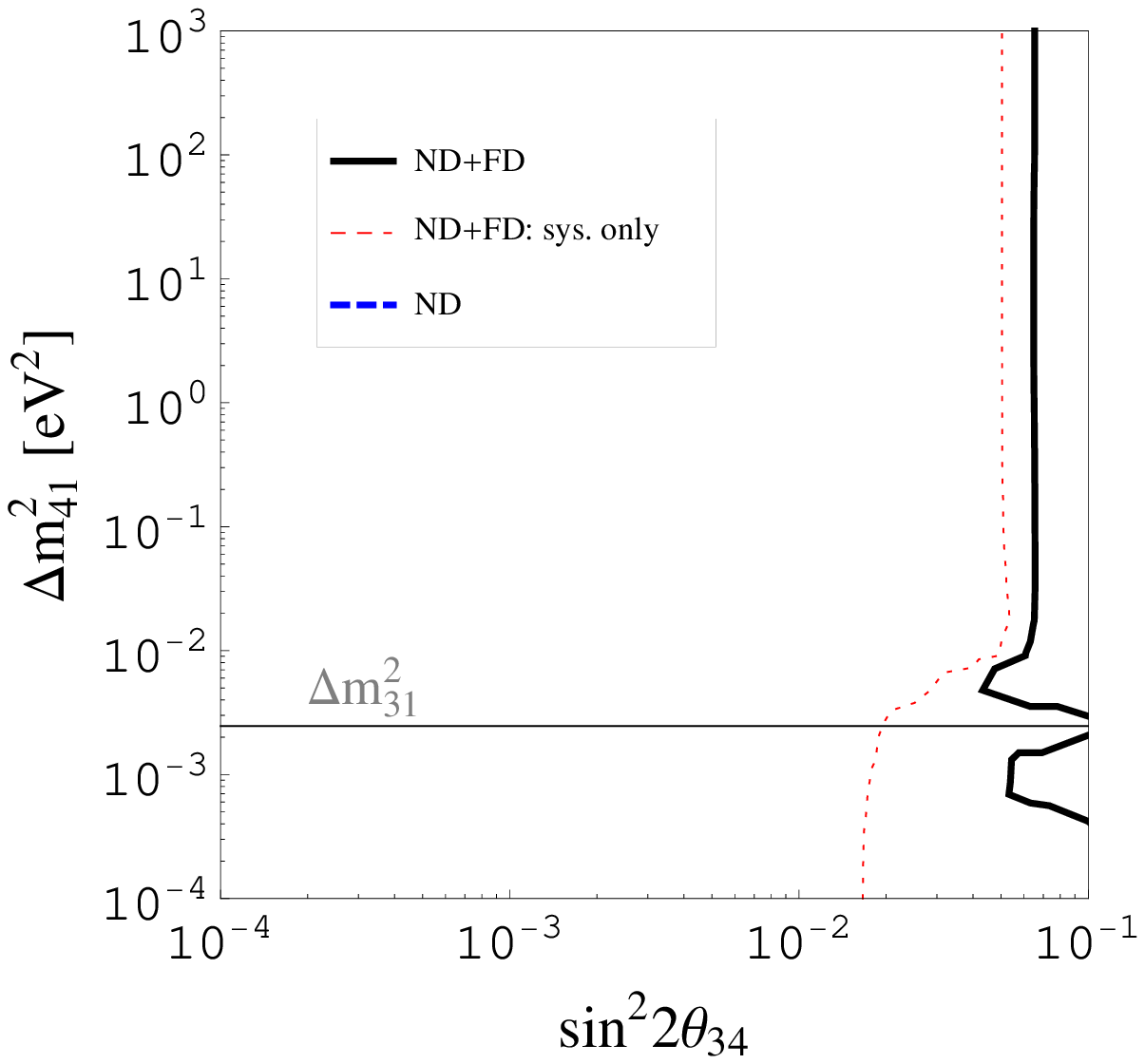}
\caption{  \it The exclusion limit at 90\% CL (2 d.o.f) for $\sin^22\theta_{i4}$--$\Delta m_{41}^2 (i=1,2,3)$ (region on r.h.s. of curves excluded). Here gives results of the standard IDS-NF ($4000$ km and $7500$ km) setup with near detectors (thick solid curves), and the impact of the near detectors is also shown separately (thick dashed curves). The thin dashed curves only consider systematics.}
\end{figure}

In this section we will discuss the discovery reach for sterile neutrinos at the standard IDS-NF baseline.
We consider the simplest scenario in which only one sterile neutrino is added to the three active  
neutrinos of the Standard Model in the so-called 3+1 scheme which recovers the standard picture in the case of small active-sterile mixings
\cite{Meloni:2010zr,Donini:1999jc,Kalliomaki:1999ii,Donini:2001xp,Donini:2001xy,Dighe:2007uf,Donini:2008wz,Goswami:2008mi}. 
For the sake of simplicity, we just focus on the scheme where the fourth state is the heaviest 
and normal hierarchy is assumed in the standard sector.
The numerical results on the discovery reach can be understood from the analytical expressions of the transition probabilities; we found particularly illuminating the following 
parameterization of the $4 \times 4$ unitary mixing matrix $U$: 
\begin{equation}
    \label{equ:3+1param1}
    U =
    R_{34}(\theta_{34} ,\, 0) \; R_{24}(\theta_{24} ,\, 0) \;
    R_{14}(\theta_{14} ,\, 0) \;
    R_{23}(\theta_{23} ,\, \delta_3) \;
    R_{13}(\theta_{13} ,\, \delta_2) \; 
    R_{12}(\theta_{12} ,\, \delta_1) \,,
\end{equation}
where $R_{ij}(\theta_{ij},\ \delta_l)$ are the complex rotation matrices in the $ij$-plane.
In the short-baseline limit $|\Delta_{41}|= \Delta m^2_{41}L/4 E \sim \mathcal{O}(1) \gg |\Delta_{31}|$,  the matter effects can be safely ignored, and the relevant probabilities read:
\begin{align}
&\mathcal{P}_{e\mu}= \mathcal{P}_{\mu e} = 4 c_{14}^2 s_{14}^2 s_{24}^2 \sin ^2 \Delta_{41} \label{equ:pem2}  \\
&\mathcal{P}_{ee}=1-\sin ^2\left(2 \theta _{14}\right)\sin ^2 \Delta_{41}  \label{equ:pee2}\\
&\mathcal{P}_{\mu \tau}= 4 c_{14}^4 c_{24}^2 s_{24}^2 s_{34}^2 \sin ^2 \Delta_{41} \\
&\mathcal{P}_{\mu\mu}=1-c_{14}^2 s_{24}^2  \left[3+2 c_{14}^2 \cos \left(2 \theta _{24}\right)-\cos \left(2 \theta _{14}\right)\right]\sin ^2 \Delta_{41}  \label{equ:pmm2}
\end{align}
where we used the short-hand notation $\Delta_{ij}= \Delta m^2_{ij} L/4 E$.
From these probabilities we can see that $\theta_{24}$ can be measured by $\mathcal{P}_{\mu \mu}$  and  $\theta_{14}$  by   $\mathcal{P}_{e e}$. On the other hand, $\theta_{34}$ only shows up in combination with the other small mixing angles. 
For long baselines,  some of the relevant features of the probability transitions can be well understood using simple 
perturbative expansions: for $\Delta_{31} = \mathcal{O}(1) \ll \Delta_{41}$ and
up to the second order in 
$
s_{13},  s_{14},  s_{24},  s_{34}, \hat s_{23}=\sin\theta_{23}-\frac{1}{\sqrt{2}}
$,
and considering
$
\Delta_{21}
$
as small as $s_{ij}^2$,
we obtain: 
\begin{eqnarray}
\mathcal P_{\mu\mu} &=& \cos^2 (\Delta_{31}) (1 - 2 s_{24}^2) + 8 \hat s_{23}^2\sin^2(\Delta_{31}) + c_{12}^2 \Delta_{12}   \sin(2\Delta_{31}) +  \nn \\
&&   2 s_{24} s_{34} \cos \delta_3 \Delta_n  \sin(2\Delta_{31}) - \label{pmumulo}\\ 
&& 2 s_{13}^2 \Delta_{31}\cos( \Delta_{31})\,\frac{(\Delta_{31}-\Delta_e) \Delta_e \sin (\Delta_{31})-\Delta_{31}\sin\left(\Delta_{31}-\Delta_e \right)\sin (\Delta_e )}{(\Delta_{31}-\Delta_e)^2} \, ,
 \nn \\
\mathcal P_{\mu\tau} &=&\sin^2 (\Delta_{31})(1-8 \hat s_{23}^2-s_{24}^2-s_{34}^2)-c_{12}^2 \Delta_{12} \sin (2\Delta_{31})-\nn \\
&& s_{24}s_{34}\sin (2\Delta_{31})\left[2 \Delta_n \cos \delta_3-\sin\delta_3\right] - \label{pmutaulo}  \\ 
&& s_{13}^2 \Delta_{31}\sin \Delta_{31} \,\frac{\Delta_{31}
\left\{\sin (\Delta_{31}-\Delta_e) +\sin (\Delta_e )\right\}-2 (\Delta_{31}-\Delta_e) 
\Delta_e \cos (\Delta_{31})}{(\Delta_{31}-\Delta_e)^2} \, .\nn 
\end{eqnarray}
from which we learn that at the long baselines, $\theta_{24}$ is best accessible by $\mathcal{P}_{\mu \mu}$ with the first term proportional to $\cos^2 (\Delta_{13})$. 
The leading sensitivity to $\theta_{34}$ can be expected from 
$\mathcal{P}_{\mu \tau}$  (the {\it discovery channel} as claimed by \cite{Donini:2008wz}). 
Notice also the dependence on the phase $\delta_3$ in both probabilities which makes them useful to check whether other sources of CP violation beside the standard one ($\delta_2$ in our parameterization) can be 
tested at the Neutrino Factory. 
Let us first discuss general constraints to the new mixing angles $\theta_{14}$, $\theta_{24}$, and $\theta_{34}$  and the additional mass squared difference $\vldm$ without any additional assumptions \cite{Meloni:2010zr}. 
The considered oscillation channels, simulated with a properly modified version of the GLoBES software, 
are electron to muon neutrino (appearance channels) and muon to muon neutrino (disappearance channels) oscillations. 
Beside the standard IDS-NF setup, we also considered near detectors 
with a fiducial mass of $32 \, \mathrm{t}$ and a distance of $d=2 \, \mathrm{km}$ 
from the end of the decay straight, which corresponds to the effective 
baseline of 2.28~km. We show three different exclusion planes $\theta_{i4}$-$\vldm$ in Fig.\ref{fig:theta-mass}. 
The main sensitivity is obtained at about $\Delta m_{41}^2 \simeq 10 \, \mathrm{eV}^2$, which comes from the distance chosen for the near detectors. Since the efficiencies for muon neutrino detection are typically better, the sensitivity to $\theta_{24}$ is slightly better than that to $\theta_{14}$ for our assumptions. As expected, there is no sensitivity to $\theta_{34}$ coming from the near detectors, because the $\nu_\tau$ disappearance channel does not exist. 
For the effect at the long baselines, it is first of all useful to consider the thin dashed curves with systematics only. In all three panels, the sensitivity changes as a function of $\vldm$ in the region where $\vldm \sim \ldm$. It comes from the fact that the Neutrino Factory is sensitive to the atmospheric oscillation frequency, whereas for $\vldm \sim \sdm$, no particular additional effects from the solar frequency can be found. As expected (see Eq.~(\ref{pmumulo})), 
the main sensitivity is found for $\theta_{24}$.
However, there is also some sensitivity to $\theta_{14}$, which vanishes after the marginalization, and some sensitivity to $\theta_{34}$, which is even present for $\vldm=0$ for systematics only. After marginalization (thick solid curves), only the sensitivities to $\theta_{24}$ and $\theta_{34}$ remain in the $\vldm$ regions close to the atmospheric $\ldm$ and above, where the effects of $\vldm$ average out. Very interestingly, note that mixing angle correlations destroy the sensitivities for $\vldm=\ldm$, where $m_4=m_3$ and no additional $\vldm$ is observable, leading to small gaps (see horizontal lines). We have tested that the sensitivity to $\theta_{34}$ is a matter potential-driven, statistic limited higher order effects present in the muon neutrino disappearance channels.  
In view of the three panels, it is not easy to disentangle the parameters for arbitrary massive sterile neutrinos. Parameter correlations lead to a pollution of the exclusion limit of a particular mixing angle with $\Delta m_{41}^2$. In addition, there is a competition between $\Delta m_{41}^2$ and $\Delta m_{31}^2$ at the long baseline. Near detectors, on the other hand, have very good sensitivities to $\theta_{14}$ and $\theta_{24}$ but cannot measure $\theta_{34}$. Nevertheless, the absolute values of the sensitivities are quite impressive. Especially, $\theta_{24}$ can be very well constrained close to the atmospheric mass squared difference range. This indicates that sterile neutrino bounds in that range should be also obtainable from current atmospheric neutrino oscillation experiments.

%% file: CONTENT/15MeloniOhlssonWinterZhang.tex
\subsection{NSI's vs. non-unitary lepton flavor mixing at a neutrino factory \cite{Meloni:2009cg}}

Apart from the study of neutrino oscillation parameters, one of the
most important tasks in future neutrino oscillation experiments,
e.g., a neutrino factory (NF), is to search for non-standard effects
stemming from the exchange of heavy fields presented in high-energy
theories. In the language of the effective theory, the impacts of
heavy particles can be parametrized in terms of a tower of higher
dimensional non-renormalizable operators, e.g., the dimension-five
operator $\mcl{O}^5_W = (\overline{L^{c}} {\rm i} \tau^{2} \phi)\,
(\phi {\rm i} \tau^2 L)$ responsible for the generation of neutrino
masses. According to the nature of the heavy fields, the
dimension-six operators can be classified into two categories, the
scalar mediated operators and the fermion mediated operators (See
discussions and references in, e.g.,
Refs.~\cite{Malinsky:2008qn,Malinsky:2009gw,Ohlsson:2009vk} on the
possible theories behind these operators). These dimension-six
operators typically suffer from severe experimental constraints on
the lepton flavor violating processes, since $\ell\to\ell\ell\ell$
decays may be induced at tree level. However, there are two leptonic
dimension-six operators which do not lead to charged lepton flavor
violation,
\begin{equation}
{\cal O^S} = \left(\overline{L^c} \cdot L \right)\left(\overline{L}
\cdot L^c \right) \, , \ \ \ \ \ {\cal O^F} =
\left(\overline{L}_\alpha \tilde\phi\right) {\rm i} \slc\partial
\left(\tilde\phi^\dagger L_\beta\right) \, , \label{equ:exampleop}
\end{equation}
where $L$ and $\phi$ stand for the Standard Model lepton doublets
and the Higgs field. The scalar mediated operator ${\cal O^S}$ leads
to non-standard neutrino interactions (NSIs), whereas the fermion
mediated operator $ {\cal O^F} $ results in non-unitarity (NU)
effects. Note that, the NU effects are similar to these from NSIs,
and therefore, both NSIs and NU can be described by using the same
parametrization, i.e., the conventional $\varepsilon^A$ parameters,
where $A=s,m,d$ correspond to non-standard effects in the neutrino
production, propagation, and detection processes, respectively.

The generic bounds on above two dimension-six operators come from
rare lepton decays $\ell_\alpha \to \ell_\beta \gamma$, the
invisible decay width of the $Z$-boson and the universality test of
weak interactions, and can be found in
Refs.~\cite{Antusch:2008tz,Biggio:2009nt}. An important question to
be answered is if we can distinguish these two operators at future
neutrino oscillation experiments so as to find hints on the origin
of non-standard effects. To this end, we remark that the NU effects
are fundamental and process-independent, whereas the NSI effects are
experiment-dependent. For example, ${\cal O^S}$ does not affect the
neutrino production process at a superbeam experiment since it is
essentially a hadron decay process. We summarize in
Tab.~\ref{tab:allowedNSIs} which non-standard effects are allowed
for a neutrino factory and a superbeam if the origin are the
discussed leptonic dimension-six operators.
\begin{table}[h]\vspace{.3cm}
\begin{minipage}[t]{20pc}
\begin{tabular}{|c|c|c|c|c|c|c|c|c|c|c|c|c|c|c|c|c|c|c|c|}
\hline  & \multicolumn{2}{|c|}{$\nu$-factory} &
\multicolumn{2}{|c|}{SB} && \multicolumn{2}{|c|}{$\nu$-factory} &
\multicolumn{2}{|c|}{SB} && \multicolumn{2}{|c|}{$\nu$-factory} &
\multicolumn{2}{|c|}{SB} \\
\cline{2-5} \cline{7-10} \cline{12-15} & ${\cal O^S}$ & ${\cal O^F}$
& ${\cal O^S}$& ${\cal O^F}$ & & ${\cal O^S}$ & ${\cal O^F}$ &
${\cal O^S}$ & ${\cal O^F}$ & & ${\cal O^S}$ & ${\cal O^F}$ & ${\cal
O^S}$ & ${\cal O^F}$
\\
\hline  $\varepsilon^m_{ee}$ &   & \ding {52} &   & \ding {52} & $\varepsilon^s_{ee}$ & \ding {52} & \ding {52} & n/a & n/a &&&&& \\
$\varepsilon^m_{e\mu}$ &   &  &  & & $\varepsilon^s_{e\mu}$ & & \ding {52} & n/a & n/a &&&&&\\
 $\varepsilon^m_{e\tau}$ &   &  &   & & $\varepsilon^s_{e\tau}$ & \ding {52} & \ding {52} & n/a & n/a & $\varepsilon^d_{\alpha\beta}$ & & \ding {52} & & \ding {52} \\
 $\varepsilon^m_{\mu\mu}$ & \ding {52}  & \ding {52} &  \ding {52}  & \ding {52} & $\varepsilon^s_{\mu e}$ &  & \ding {52} &  &  \ding {52}& & & & & \\
 $\varepsilon^m_{\mu\tau}$ &  \ding {52} & \ding {52} &  \ding {52}  & \ding {52} & $\varepsilon^s_{\mu\mu}$ & \ding {52} & \ding {52} &  & \ding {52} & & & & & \\
$\varepsilon^m_{\tau\tau}$ &  \ding {52} & \ding {52} &  \ding {52}  & \ding {52} & $\varepsilon^s_{\mu\tau}$ & \ding {52} & \ding {52} & & \ding {52} & & & & & \\
\hline
\end{tabular}
\end{minipage}
\hspace{4pc}%
\begin{minipage}[b]{15pc}
\caption{\it Allowed parameters from the discussed dimension-six
effective operator classes in a neutrino factory ($\nu$-factory) and
a superbeam experiment (SB). \label{tab:allowedNSIs}}
\end{minipage}
\vspace{0.3cm}
\end{table}
In addition, there exist non-trivial correlations between the source
and matter non-standard effects for both ${\cal O^S}$ and ${\cal
O^F}$~\cite{Gavela:2008ra}, i.e.,
\begin{eqnarray}
\varepsilon^m_{\mu \tau}  =     - (\varepsilon^{\mathrm{NF}}_{\mu
\tau})^*  \quad \text{(NSIs)} \, , \ \ \ \ \ \  \ \
\varepsilon^m_{\mu \tau}  = -\varepsilon^s_{\mu\tau} \quad
\text{(NU)} \, , \label{equ:corr2}
\end{eqnarray}
which motivate us to investigate the possibility of distinguishing
the origin of non-standard effects at a neutrino factory
alone~\cite{Meloni:2009cg}.

\begin{figure}[t]
\begin{center}\vspace{-0.3cm}
\includegraphics[width=5.5cm,bb=40 0 760 720]{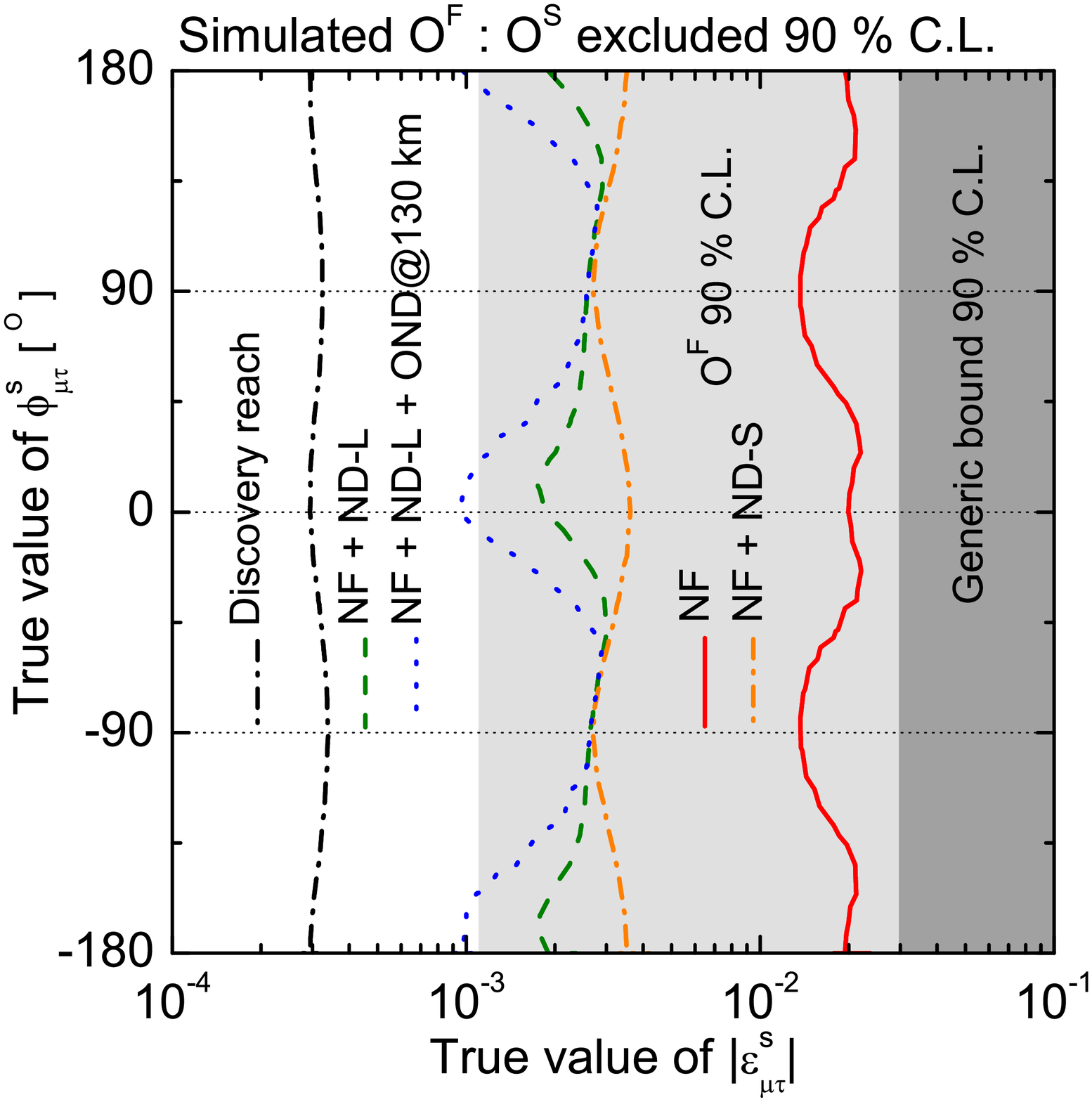}
\includegraphics[width=5.5cm,bb=40 0 760 720]{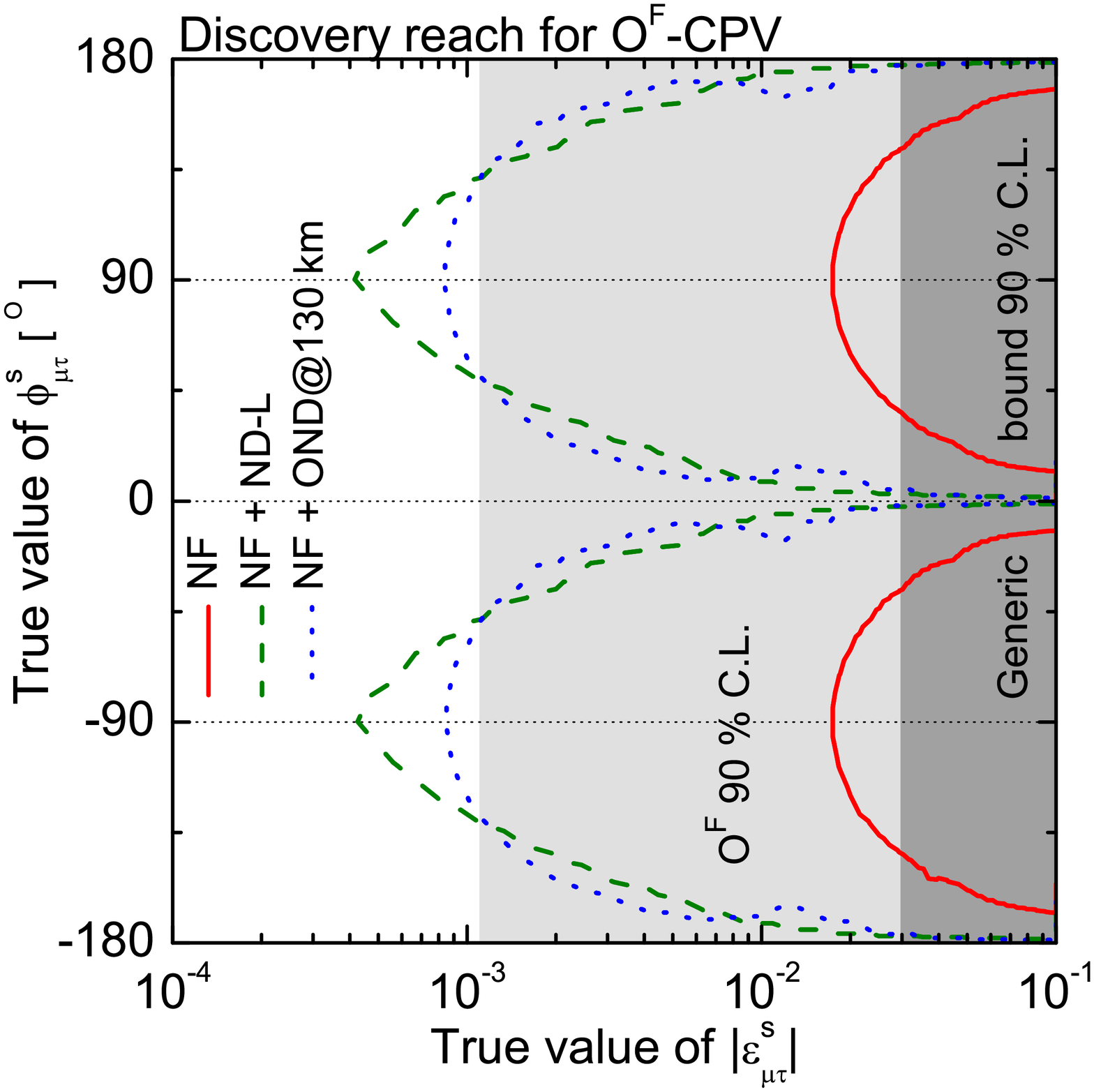}
\includegraphics[width=5.5cm,bb=40 0 760 720]{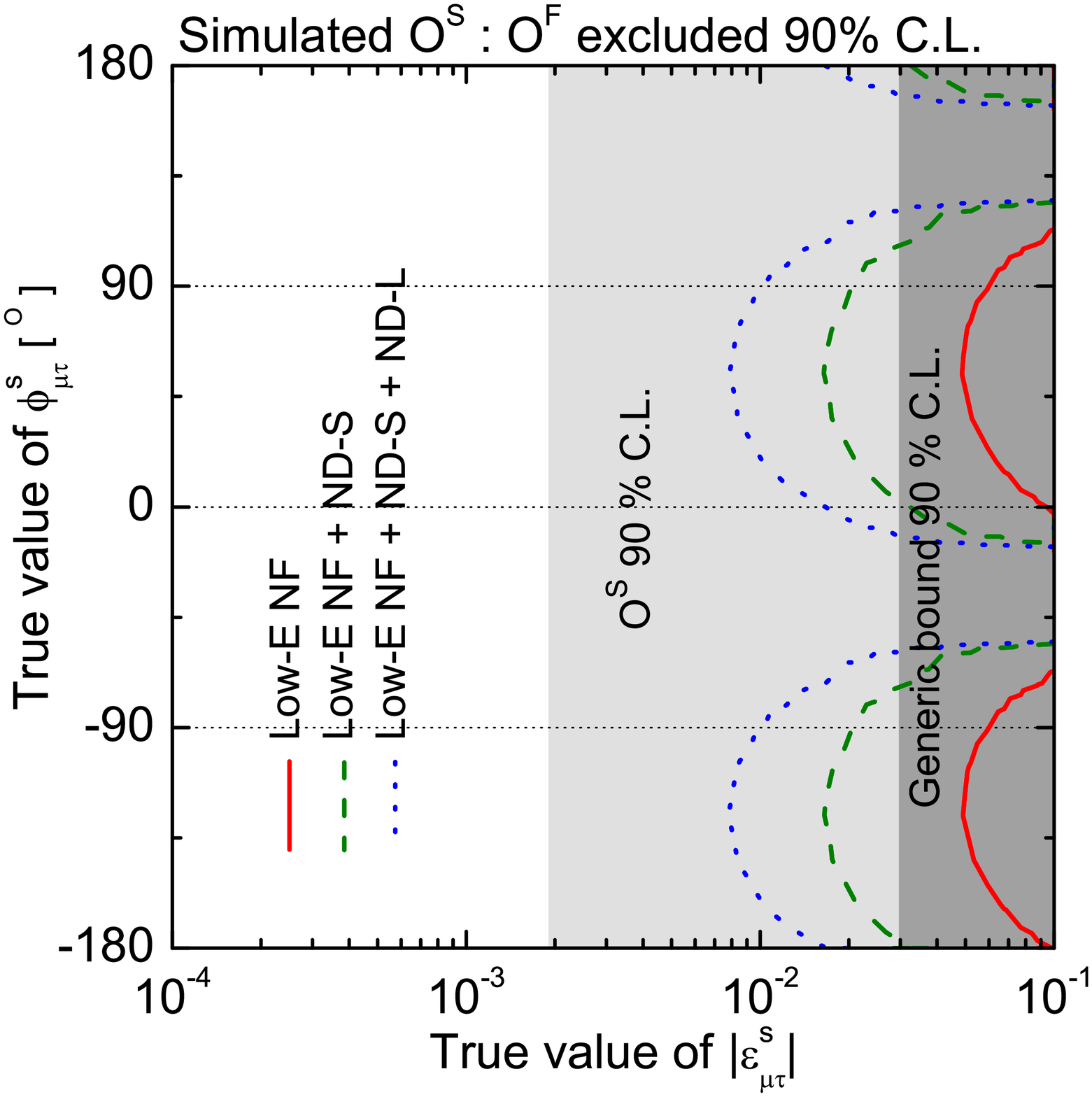}
\includegraphics[width=5.5cm,bb=400 0 1120 720]{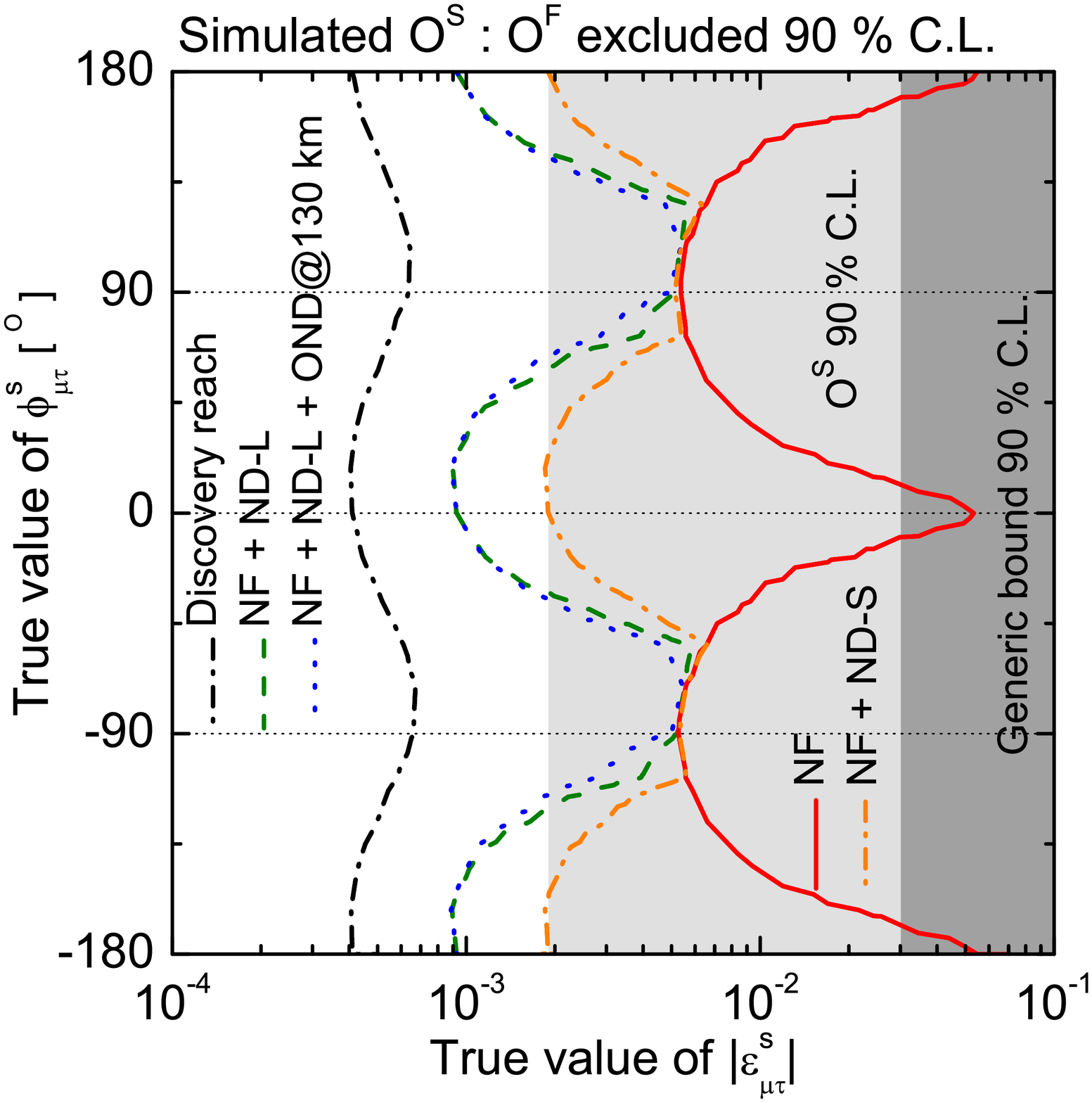}
\includegraphics[width=5.5cm,bb=400 0 1120 720]{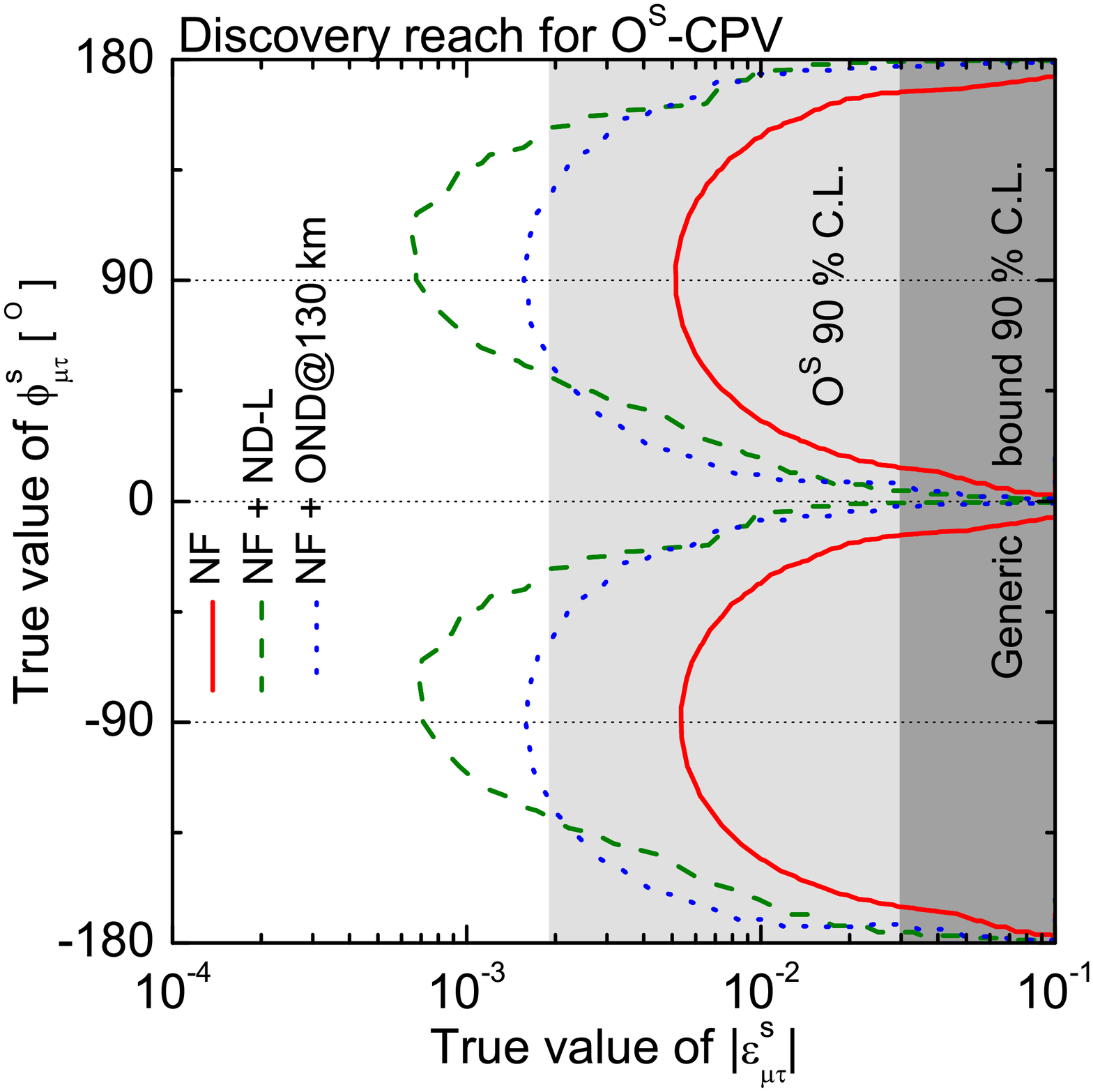}
\put(-80,135){{\bf ND-L}: Large size (OPERA-like)}%
\put(-48,125){fiducial mass $2~{\rm kt}$, $d=1~{\rm
km}$}%
\put(-80,105){{\bf ND-M}: Medium size (e.g., SciBar-sized)}%
\put(-48,95){fiducial mass  $25~{\rm t}$, $d=80~{\rm
m}$}%
\put(-80,75){{\bf ND-S}: Small size (e.g., silicon vertex-sized)}%
\put(-48,65){fiducial mass $100~{\rm kg}$, $d=80~{\rm
m}$}%
\put(-80,45){{\bf OND@130km}: Large size (OPERA-like)}%
\put(-48,35){fiducial mass $2~{\rm kt}$, $L=130~{\rm km}$}%
\caption{\label{fig:fig1mowz} \it Left column: regions in the
($|\varepsilon^s_{\mu\tau}|$-$\phi^s_{\mu\tau}$)-plane where the
simulated $\epsilon^s_{\mu\tau}$ induced by one type of operator can
be uniquely established, \ie, the other type of operator is excluded
at the 90~\% C.L. (regions on the right-hand side of the curves) at
an IDS-NF ($E_\mu=25~{\rm GeV}$). The discovery reach is also
displayed in each plot. Middle column: 90~\% C.L. of CP discovery
potentials in the NU framework (upper plot) and NSI framework (lower
panel) at an IDS-NF. Right column: same plot as the lower panel of
the left column, but for a low-energy neutrino factory ($E_\mu = 4.5
~{\rm GeV}$) alone and in combination with the ND-S and ND-S+ND-L.
Taken from~\cite{Meloni:2009cg}.}
\end{center}\vspace{-.6cm}
\end{figure}

In the left column of Fig.~\ref{fig:fig1mowz}, we illustrate the
potential of discriminating ${\cal O^S}$ from ${\cal O^F}$ at an
IDS-NF (International Design Study) with parent muon energy $E_\mu
=25~{\rm GeV}$~\cite{ids} (See also a detailed description of the
experiment setups in Ref.~\cite{Meloni:2009cg}).
In our numerical analysis, we make use of the ``true'' parameters in
the case of ${\cal O^F}$ (${\cal O^S}$), and then fit the data with
only ${\cal O^S}$ (${\cal O^F}$). For the IDS-NF neutrino factory
combined with several different near detectors, the curves in the
left column show that there is just a very small region beyond the
bound at the 90~\% C.L. on the ${\cal O^F}$ operators, where the
data generated with ${\cal O^F}$ can be distinguished from the
${\cal O^S}$ even if the OPERA-like near detector at the longer
baseline is used. If ${\cal O^S}$ is simulated, however, it can be
distinguished from ${\cal O^F}$ for a part of the parameter space
beyond the current bound with ND-L. In the middle column of
Fig.~\ref{fig:fig1}, we plot the CP discovery potential for both
${\cal O^F}$ (upper panel) and ${\cal O^S}$ (lower plot) induced CP
violations. This is defined as the ensemble of true values of
$\phi^s_{\mu\tau}$ (phase of $\varepsilon^s_{\mu\tau}$), which
cannot be fitted with the CP-conserving values
$\phi^s_{\mu\tau}=0,\pm \pi$ at 90~\% C.L. The combination of the
standard IDS-NF neutrino factory with different large enough near
detectors may discover CP violation, somewhat beyond the current
bounds, especially for $\phi^s_{\mu\tau}\sim \pm \pi/2$. There are
no qualitative differences between ${\cal O^F}$ and ${\cal O^S}$.
Finally, we show, in the right upper panel of Fig.~\ref{fig:fig1},
that a low-energy neutrino factory ($E_\mu = 4.5 ~{\rm
GeV}$)~\cite{Bross:2007ts,Bross:2009gk} is not able to exclude the
${\cal O^F}$ operator at 90~\% C.L., since the exclusion regions are
excluded by current limits already.

In conclusion, differentiating between NSIs and NU should be one of
the key priorities of searches for new physics effects, since the
nature of non-standard effect points towards the nature of the heavy
mediator. The components necessary for this search are $\nu_\tau$
detection at least in near detectors, both at high-intensity
superbeams and a neutrino factory. For the neutrino factory, a high
enough muon energy is mandatory for the discussed non-standard
effects searches, which means that the high-energy neutrino factory
should at least be an upgrade option even for large $\theta_{13}$.
In addition, for non-standard effect searches, the size of the near
detector is very important, which means that for all applications,
large enough detectors are needed.

%% file: CONTENT/16OhlssonSchwetzZhang.tex
\subsection{Non-standard neutrino interactions in the Zee--Babu model \cite{Ohlsson:2009vk}}

While the origin of neutrino masses emerges as one of the unsolved
and fundamental problems in particle physics, the mechanism of
radiative mass generation provides a natural method to obtain small
neutrino masses. In such a framework, neutrino masses are exactly
vanishing at tree level, and are induced as finite radiative
corrections. The simplest version of this kind of models, i.e., the
Zee model~\cite{Zee:1980ai}, cannot accommodate current experimental
data, since the predicted leptonic mixing angle $\theta_{12}$ is too
large. Alternatively, in the Zee--Babu
model~\cite{Zee:1985rj,Zee:1985id,Babu:1988ki}, two $SU(2)_{\rm L}$
singlet scalars (one singly charged $h^+$ and one doubly charged
$h^++$) are introduced besides the Standard Model (SM) particle
content, and neutrino masses can be generated at two-loop level. In
addition to the generation of neutrino masses, the exchange of heavy
scalars also results in lepton flavor violating processes such as
$\mu \to 3e$ and $\mu \to e\gamma$, which can be dramatically
enhanced compared to those in the SM. Most interestingly, if
kinematically accessible, the new scalars could be directly produced
at the Large Hadron Collider or the future International Linear
Collider, and the decay of the doubly charged Higgs may induce very
clean like-sign bi-lepton events, which are basically free from 
SM background, and opening the possibility of collider tests of the
neutrino mass generation mechanism.

Besides colliders, the next generation neutrino oscillation experiments,
e.g., neutrino factories, will also help us to unveil the underlying physics
behind neutrino masses, in particular, through exploring non-standard
neutrino interactions (NSIs). Unfortunately, in most of the low-scale
models~\cite{Malinsky:2008qn}, NSIs are typically linked to lepton flavor
violating processes of charged leptons, yielding too tight bounds, see,
e.g., Refs.~\cite{Davidson:2003ha, Antusch:2008tz, Biggio:2009nt}. However,
in the case of the Zee--Babu model, the situation is more involved, since
the masses of singly and doubly charged Higgs in principle can be well
separated and a different set of Yukawa couplings controls charged lepton
and neutrino interactions with the scalars. In the following we summarize
the results of Ref.~\cite{Ohlsson:2009vk}, where a detailed investigation of
NSIs in the Zee--Babu model has been performed.

The Lagrangian of the minimal Zee--Babu model reads
\begin{eqnarray}\label{eq:L}
{\cal L} & = & {\cal L}_{\rm SM} + f_{\alpha\beta} L^{T}_{L\alpha} C
i\sigma_2 L_{L\beta} h^{+} + g_{\alpha\beta} \overline{e^c_\alpha}
e_{\beta} k^{++} - \mu h^- h^- k^{++} + {\rm h.c.} + V_H \ ,
\end{eqnarray}
where $L_L$ denote left-handed lepton doublets, $e$ are the
right-handed charged leptons, and the scalar potential $V_H$
contains the couplings among scalar fields. $f$ and $g$ are
antisymmetric and symmetric Yukawa couplings, respectively. Neutrino
masses are generated at two-loop level as
\begin{eqnarray}\label{eq:mnu}
m_{\nu} \simeq \frac{1}{48\pi^2} \frac{\mu}{M^2} \, \tilde I \, f
D_e g^\dagger D_e f^T \, ,
\end{eqnarray}
where $D_e = {\rm diag}(m_e,m_\mu,m_\tau)$ contains the
charged-lepton masses, $M={\rm max}(m_k, m_h)$, and $\tilde I(r)$ is
a dimensionless function of order unity~\cite{Nebot:2007bc}.
Furthermore, the tree-level exchange of $h^+$ induces non-standard
lepton interactions via dimension-six operators
\begin{eqnarray}
{\cal L}^{\rm NSI}_{d=6} =  4 \frac{f_{\alpha\beta}
f^*_{\rho\sigma}}{m^2_h} \left( \overline{\ell^c_\alpha} P_L
\nu_\beta \right) \left( \overline{\nu_\sigma} P_L \ell^c_\rho
\right)  = 2\sqrt{2}{G_F}
\varepsilon^{\rho\sigma}_{\alpha\beta} \left(
\overline{\nu_\alpha}\gamma^\mu P_{L} \nu_\beta \right) \left(
\overline{\ell_\rho}\gamma_\mu P_{L} \ell_\sigma \right)\ ,
\end{eqnarray}
where $\varepsilon^{\rho\sigma}_{\alpha\beta} \simeq 0.06 \,
f_{\sigma\beta} f^*_{\rho\alpha} \left(\frac {m_h}{\rm
TeV}\right)^{-2}$ are the canonical NSI parameters. Using the
conventional notation in the literature, e.g.,
Ref.~\cite{Ohlsson:2008gx}, we find that the following NSI
parameters are induced
\begin{eqnarray}
\varepsilon^m_{\alpha\beta} = \varepsilon^{ee}_{\alpha\beta} =
\frac{f_{e\beta} f^*_{e\alpha}}{\sqrt{2} G_F m^2_h} \, , \ \ \ \ \
  \varepsilon^s_{\mu\tau} =
  \varepsilon^{e\mu}_{\tau e} =
  \frac{f_{\mu e} f^*_{e\tau}}{\sqrt{2} G_F m^2_h} \ , \ \ \ \ \
  \varepsilon^s_{e\tau} =
  \varepsilon^{e\mu}_{\mu\tau} =
  \frac{f_{\mu\tau} f^*_{e\mu}}{\sqrt{2} G_F m^2_h} \ , \label{eq:es}
\end{eqnarray}
where the superscript $m$ and $s$ correspond to NSI effects in matter and in
the neutrino source at a Neutrino Factory, respectively.

Equations~\eqref{eq:mnu} and \eqref{eq:es} indicate that there
exists a firm connection between neutrino parameters and NSI
parameters. For example, in the of normal mass hierarchy ($m_1\ll
m_2\ll m_3$, NH), the relation $f_{e\mu} \simeq f_{e\tau}\simeq
f_{\mu\tau}/2$ approximately holds, while in the case of inverted
mass hierarchy ($m_2>m_1 \gg m_3$, IH), $|f_{e\mu}| \sim
|f_{e\tau}|$ and $|f_{\mu\tau}| \sim |f_{e\tau}| s_{13}/s_{23} $ can
be obtained. Taking into account the experimental constraints from
the universality in $\ell_a \to \ell_b \overline{\nu} \nu$
decays~\cite{Nebot:2007bc}, we can roughly estimate that $|f_{e\mu}|
\sim |f_{e\tau}| \lesssim 0.05~(m_h/{\rm TeV})$ and $|f_{\mu\tau}|
\lesssim 0.1~(m_h/{\rm TeV})$. Therefore, compared with
Eq.~\eqref{eq:es}, the important NSI parameters in the NH case is
$\epsilon^s_{e\tau}$, while in the IH case,
$\varepsilon^m_{\mu\tau}$, $\varepsilon^m_{\mu\mu}$,
$\varepsilon^m_{\tau\tau}$, and $\varepsilon^s_{\mu \tau}$ can be
sizable.

\begin{figure*}[t]
\begin{center}\vspace{-0.7cm}
\includegraphics[width=5.8cm,bb=0 0 750 750]{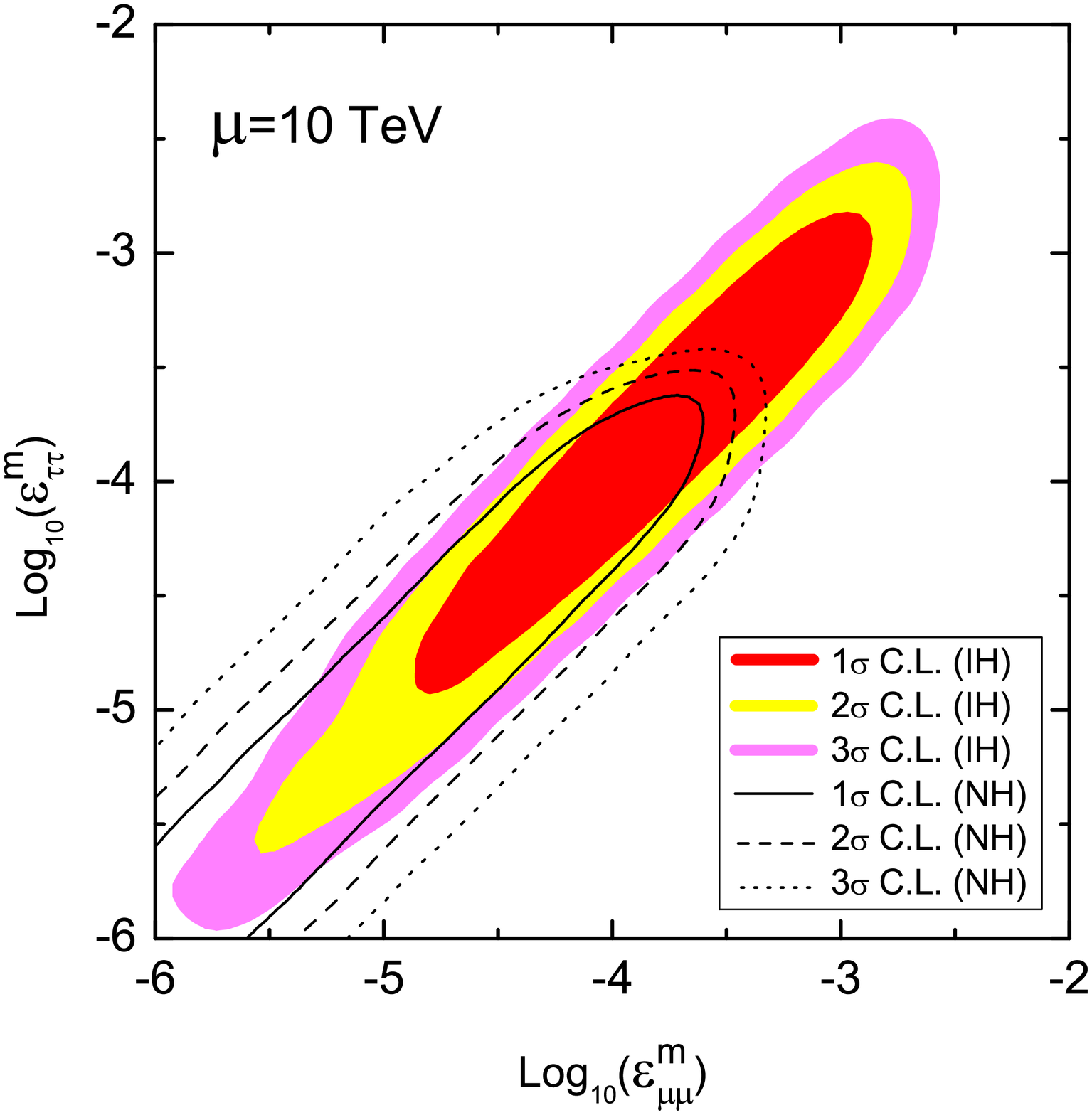}
\includegraphics[width=5.8cm,bb=0 0 750 750]{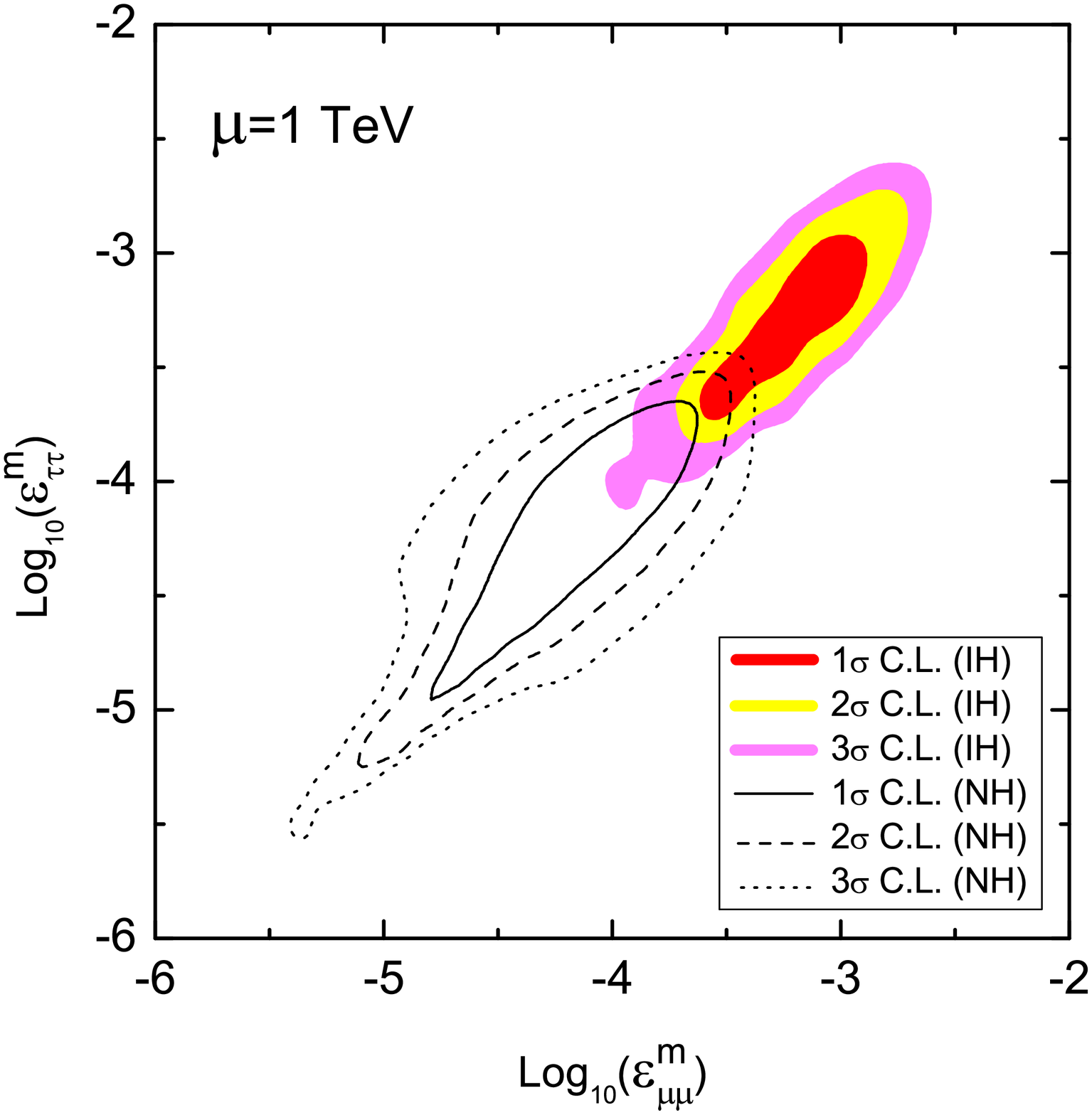}
\includegraphics[width=5.8cm,bb=0 0 750 750]{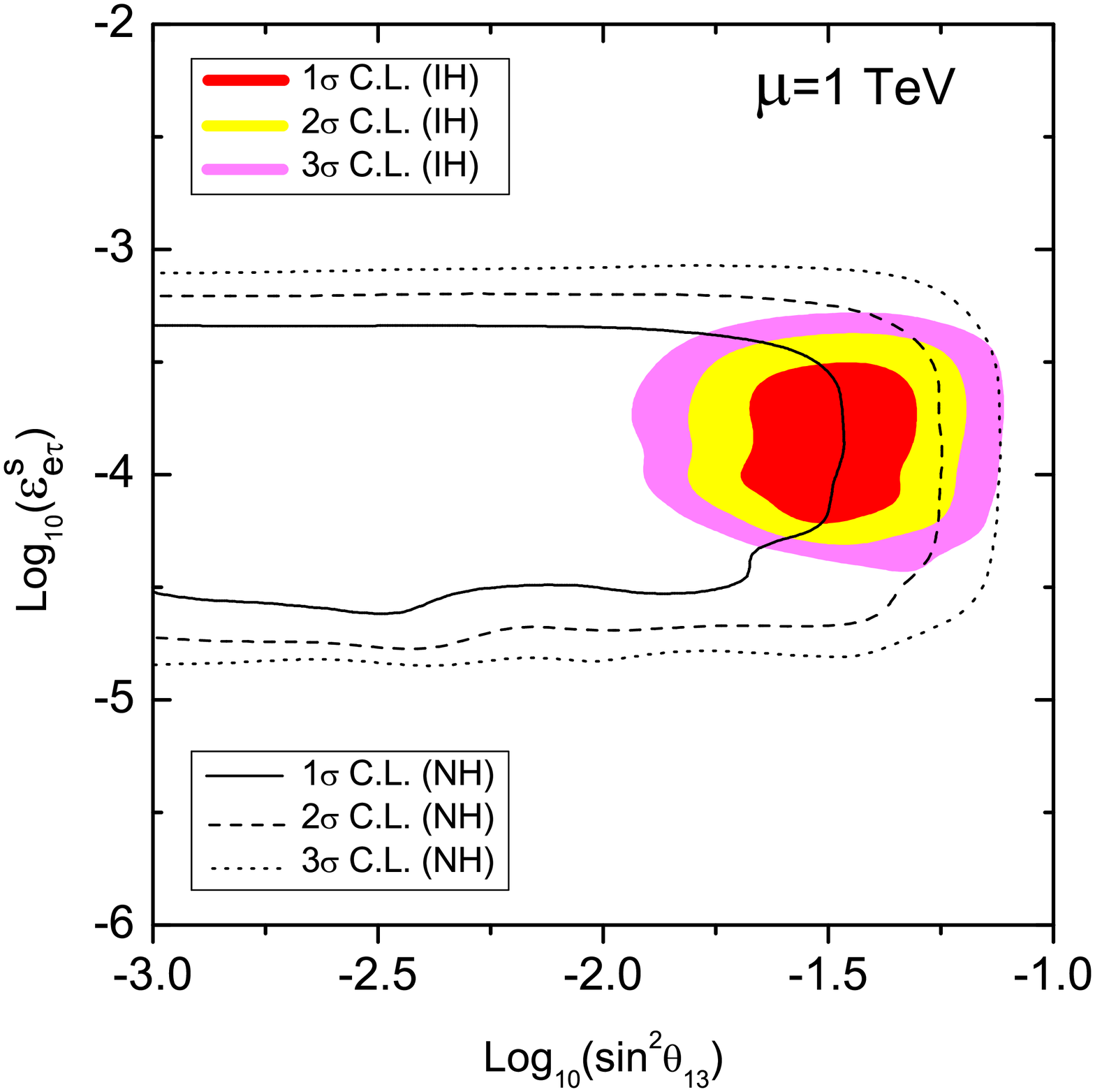}
\includegraphics[width=5.8cm,bb=0 -90 750 660]{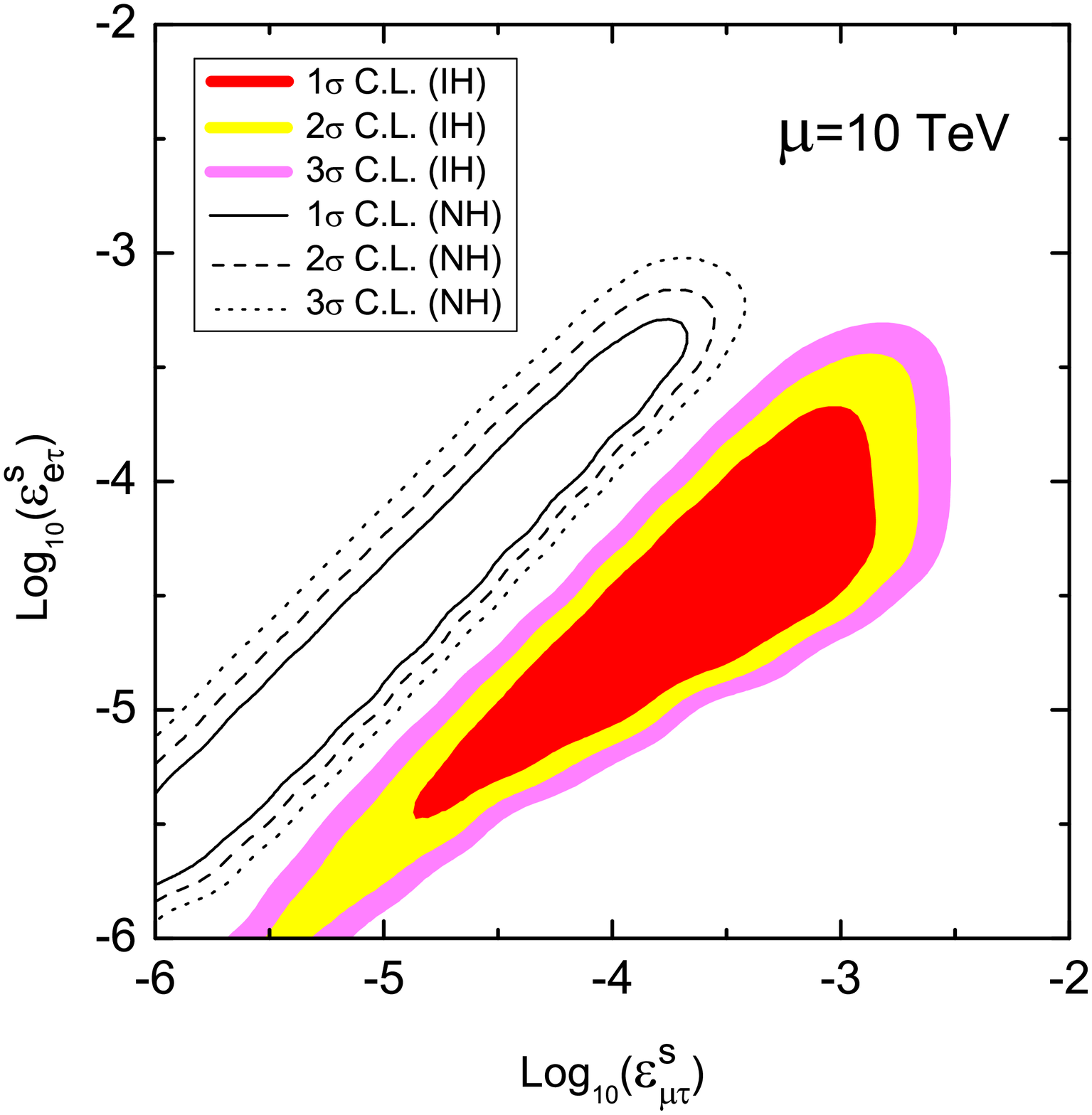}
\includegraphics[width=5.8cm,bb=0 -90 750 660]{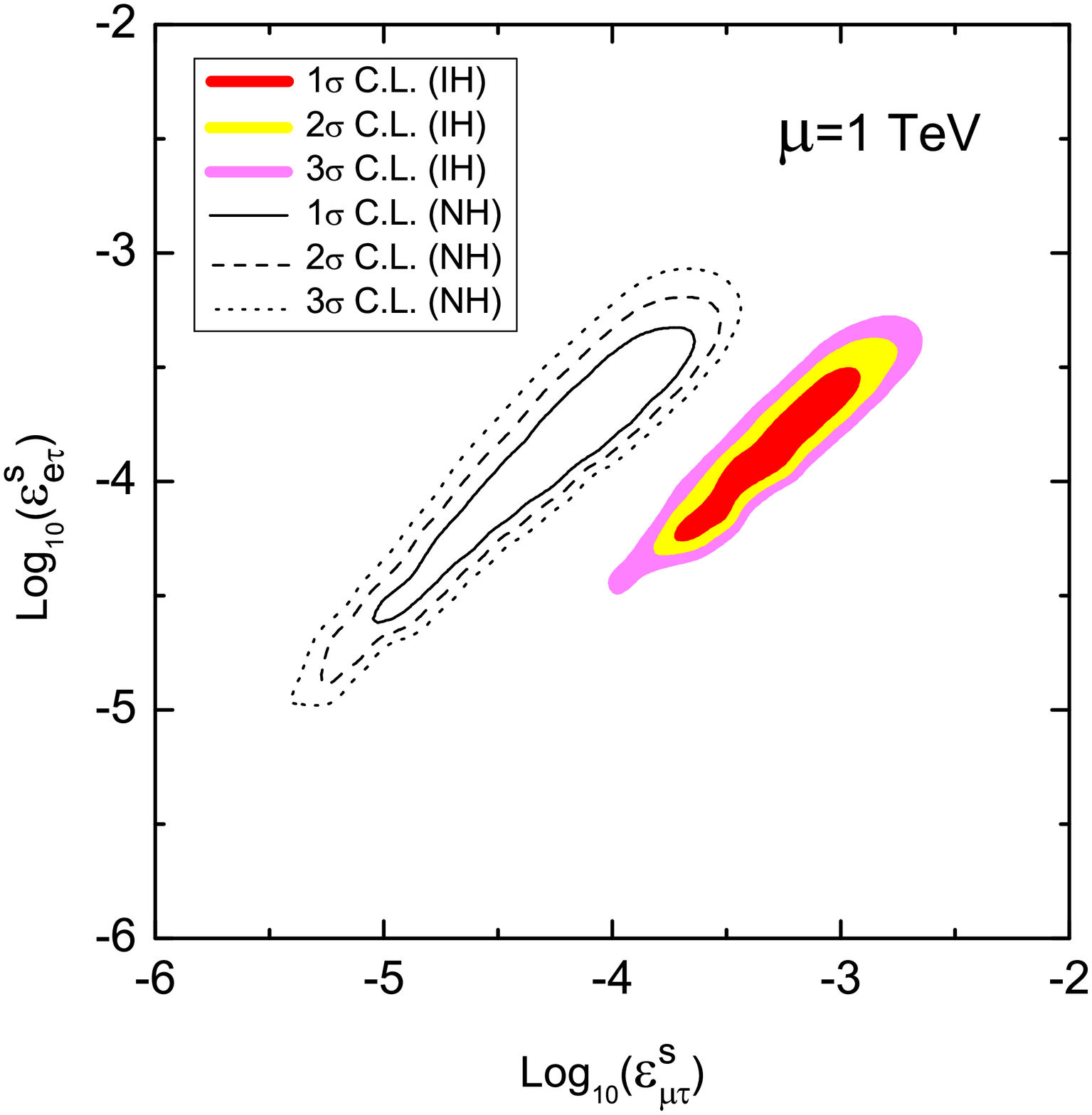}
\includegraphics[width=5.8cm,bb=0 -90 750 660]{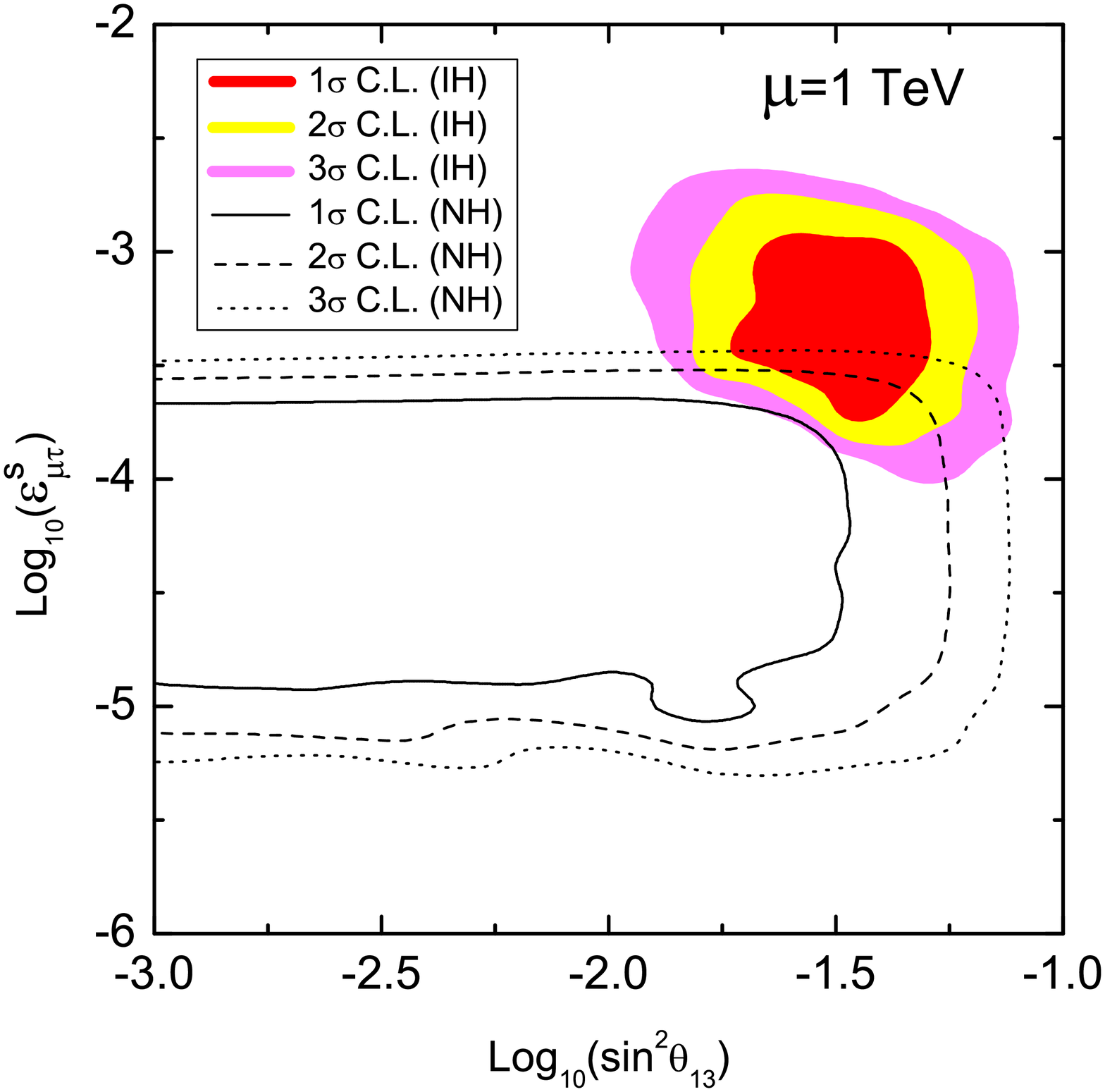}
  \vspace{-0.7cm} \caption{\label{fig:fig1} Allowed regions of NSI parameters and $\sin\theta_{13}$ at 1$\sigma$,
  2$\sigma$, and 3$\sigma$ C.L. We take $m_h=m_k=\mu=10~{\rm TeV}$
  for the figures in the left column and $m_h=m_k=\mu=1~{\rm TeV}$ for
  the figures in the middle and right column. Taken from~\cite{Ohlsson:2009vk}.} \vspace{-0.5cm}
\end{center}
\end{figure*}

Taking into account the experimental bounds summarized in
Ref.~\cite{Nebot:2007bc}, we present, in Fig.~\ref{fig:fig1}, the
allowed regions of NSI parameters and $\sin\theta_{13}$ at
1$\sigma$, 2$\sigma$, and 3$\sigma$ C.L.
We find, from the left column, that $\varepsilon^m_{\mu\mu}$ and
$\varepsilon^m_{\tau\tau}$ can only be sizable in the IH case, while
the present long-baseline experiments are not very sensitive to
these parameters. As for the source related NSI parameters,
$\varepsilon^s_{e\tau}$ may reach values up to $10^{-3}$ for both
hierarchies, whereas $\varepsilon^s_{\mu\tau}$ can be as large as
few $\times 10^{-3}$ only in the IH case. In particular, for a
scalar mass scale of $1~{\rm TeV}$, a non-trivial lower bound on the
NSI parameters of order $10^{-4}$ is found, indicating that the
model is rather constrained from the requirement of a correct
neutrino mass matrix. The right column in Fig.~\ref{fig:fig1} shows
correlations between NSI parameters and the mixing angle
$\theta_{13}$. Especially, in the IH case, one obtains a quite
strong prediction for $\theta_{13}$, i.e., a lower bound
$\sin^2\theta_{13} \gtrsim 10^{-2}$. Such a sizable lower bound is
of particular interest, since it would guarantee a discovery at the
forthcoming reactor or long-baseline experiments in the near future
\cite{Huber:2009cw}.

In conclusion, sizable NSIs can be accommodated in the Zee--Babu
model in a natural way, and they depend on both the mass scale of
the charged scalars and on the type of neutrino mass hierarchy, NH
or IH. We further remark that, in addition to the NSIs, the singly
and doubly charged scalars could be directly produced through the
$s$-channel processes at the Tevatron and the LHC. Therefore, the
interplay of the phenomenology at colliders, search for lepton
flavor violation, and NSI effects at neutrino oscillation
experiments could play a complementary role towards the goal of
identifying the true mechanism of neutrino mass generation.

%% file: CONTENT/18Comparison.tex

\begin{figure}[htbp]
\begin{center}
\includegraphics[width=12cm]{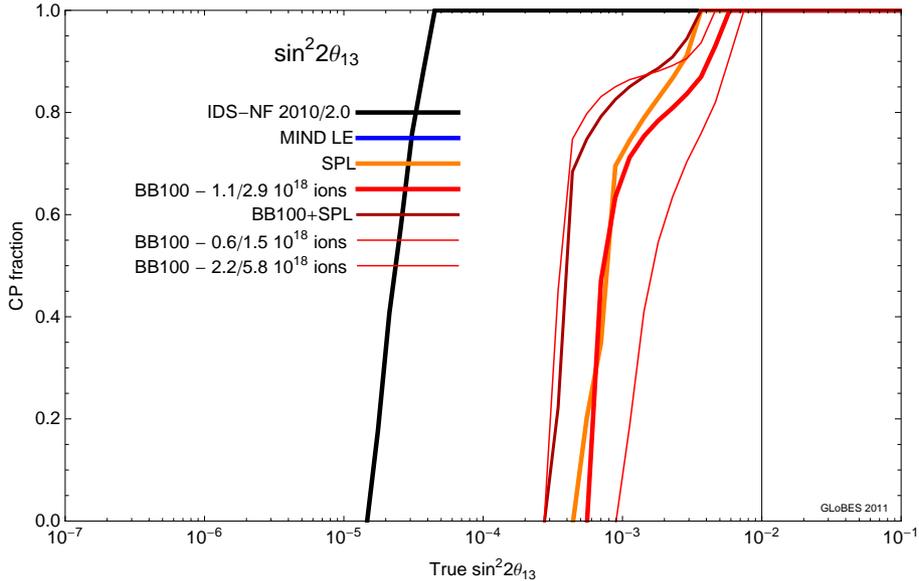}\\
\caption{
Comparison of the sensitivity to $\theta_{13}$ of different future facilities as a function of  $\sin^2 2 \theta_{13}$. 
Prepared by P.~ÊHuber for the EUROnu Technical Review, April 13th, 2011, 
using the GLoBES package \cite{Huber:2004ka,Huber:2007ji}. Curves are taken from [a] \cite{Campagne:2006yx}, [b] \cite{Barger:2007jq} , [c] \cite{Choubey:2009ks},  
[d] \cite{Huber:2009cw} and [e] \cite{Bross:2007ts}.
}
\label{fig:euronu2010th}
\end{center}
\end{figure}

\begin{figure}[htbp]
\begin{center}
\includegraphics[width=12cm]{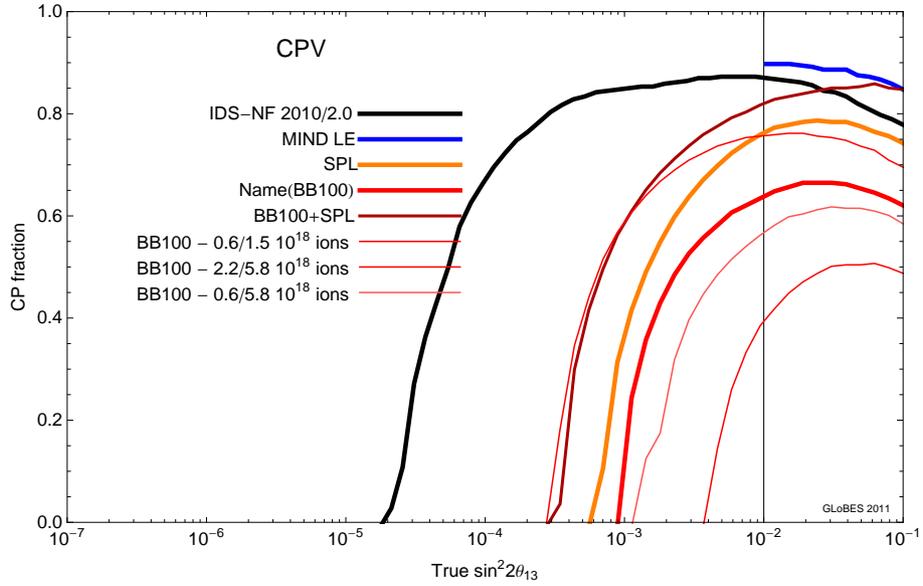}\\
\caption{
Comparison of the CP discovery potential of different future facilities as a function of  $\sin^2 2 \theta_{13}$. 
Prepared by P.~ÊHuber for the EUROnu Technical Review, April 13th, 2011, 
using the GLoBES package \cite{Huber:2004ka,Huber:2007ji}. Curves are taken from [a] \cite{Campagne:2006yx}, [b] \cite{Barger:2007jq} , [c] \cite{Choubey:2009ks},  
[d] \cite{Huber:2009cw} and [e] \cite{Bross:2007ts}.
}
\label{fig:euronu2010cp}
\end{center}
\end{figure}

\begin{figure}[htbp]
\begin{center}
\includegraphics[width=12cm]{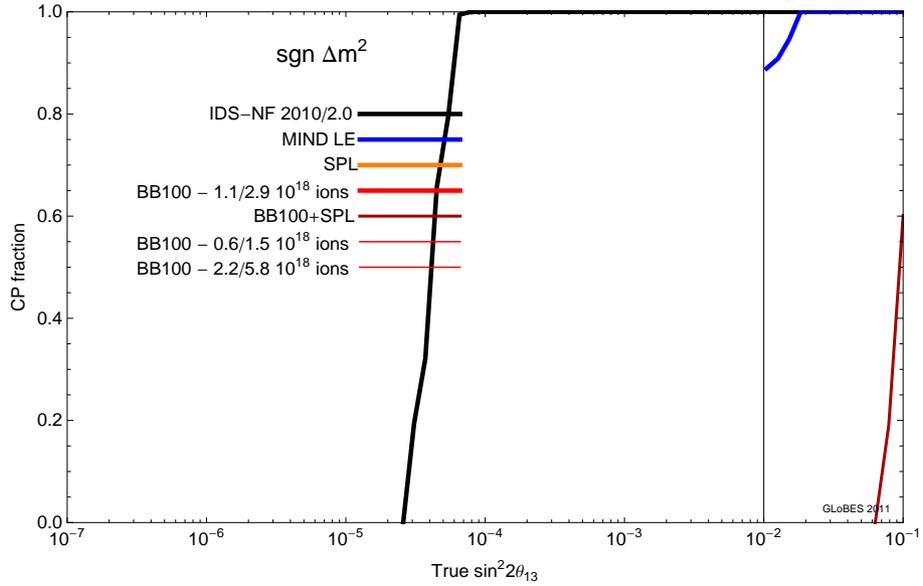}
\caption{
Comparison of the sensitivity to the neutrino mass hierarchy of different future facilities as a function of  $\sin^2 2 \theta_{13}$. 
Prepared by P.~ÊHuber for the EUROnu Technical Review, April 13th, 2011, 
using the GLoBES package \cite{Huber:2004ka,Huber:2007ji}. Curves are taken from [a] \cite{Campagne:2006yx}, [b] \cite{Barger:2007jq} , [c] \cite{Choubey:2009ks},  
[d] \cite{Huber:2009cw} and [e] \cite{Bross:2007ts}.
}
\label{fig:euronu2010sgn}
\end{center}
\end{figure}

In Figs.~\ref{fig:euronu2010th}-\ref{fig:euronu2010sgn} we present the comparison of the physics potential of the three EUROnu facilities (the 25 GeV Neutrino Factory, 
the $\gamma = 100$ Beta-Beam and the SPL Super-Beam) as shown at the Technical Review Meeting in Geneva, April 13th, 2011. Results for the $\gamma = 100$ beta-beam 
are shown for several values of the neutrino fluxes, as well as in combination with the SPL. In addition to the 25 GeV Neutrino Factory, results for the 10 GeV Low-Energy
Neutrino Factory are also shown in case $\sin^2 2 \theta_{13}$ were found to be larger than $10^{-2}$ (as it is indeed the case according to the recent results 
of T2K, MINOS, Daya Bay, RENO and Double Chooz experiments).

The three figures correspond to: (1) sensitivity to $\theta_{13}$, Fig.~\ref{fig:euronu2010th}; (2) CP discovery potential, Fig.~\ref{fig:euronu2010cp}; 
and, (3) sensitivity to the neutrino mass hierarchy, Fig.~\ref{fig:euronu2010sgn}. The curves correspond to $3\sigma$ CL  (1 dof). 
The other parameters have been fixed to $\Delta m^2_{31}= 0.0024$eV$^2$, $\Delta m^2_{21}= 8 \times 10^{-5}$ eV$^2$ and $\theta_{23} =45^\circ$.

\vspace{1cm}

%% file: CONTENT/19NuFlavour.tex

In 8--10 June 2009, the workshop "Flavour physics 
in the era of precision neutrino experiments" at 
 Coseners House, Abingdon, UK, focussed 
on a critical review of the physics case for neutrino physics and long baseline neutrino oscillations from a theoretical perspective. 
This workshop was organised with partial support from
the European Community under the European Commission Framework
Programme 7 Design Studies: EUROnu (Project Number 212372).
The topics which were discussed included:
i) lepton flavor violation (LFV) from grand unified theory (GUT)
 see-saw models and from TeV see-saw models, 
ii) Leptogenesis in the context of neutrino mass models,
iii) A theoretical perspective on lepton flavor physics at the TeV scale 
within supersymmetric (SUSY) models and extra-dimension models,
iv) Interplay between neutrino masses and other phenomenological signatures,
v) Neutrino physics and cosmology/astroparticle physics, and finally
vi) Performance indicators in long baseline neutrino oscillation experiments.

\subsection{LFV from GUT see-saw models and from TeV see-saw models}

See-saw models provide a natural solution
to the smallness of neutrino masses 
and they can be embedded
in Grand Unified Theories (talk by G. Ross) or emerge at the TeV scale, being testable at the LHC
in the forecoming future (talk by E. Ma). All these models
need also to account for the low energy flavor structure (talk by M. Schmidt).
A series of theoretical questions emerged at the meeting:
 Is it possible to discriminate between GUT and TeV-scale see-saw models?
How do these models inscribe into a more general theory?
Where does the smallness of neutrino masses come from, and does it require fine-tuning anywhere?
How can flavor mixing be implemented? Is it possible to define some "benchmark" scenarios? If so, which ones?
Is it possible to have arguments for the precisions required for 
$\theta_{13}$, $\delta$, and the Majorana phases?
What information on the origin of neutrino masses can be obtained from the mass hierarchy measurement?
These questions were addressed in the discussion session lead by M. Frigerio. 
It was mainly emphasized that there are two aspects which need to be studied:
the origin of neutrino masses and the rationale for the leptonic flavor structure. 
Three contributed talks were also given by A. Adulpravitchai
on Non-abelian Discrete Flavor Symmetries from $T^2/Z_N$ Orbifolds, by 
D.  Hernandez on the  Minimal Flavor Seesaw Models
and by M. Satriawan on Possible mechanism for generating a very small Dirac neutrino mass.

\subsection{Neutrino physics and the cosmology/astroparticle physics complementarity}

The bath of relic cosmological neutrinos, which decoupled at $T\sim 1$~MeV, 
affects the evolution
of the Universe, in particular Big Bang Nucleosynthesis, large scale structure formation and
the Cosmic Microwave Background
and allows to constrain neutrino masses, the number of neutrinos and
the type of neutrino interactions (talk by S. Hannestad).
Sterile neutrinos are the leading candidate for warm dark matter and can 
be tested in x-ray searches from dark matter overdensities,
as the center of the galaxy or nearby dwarf galaxies (talk by A. Kusenko).
In the discussion session lead by P. Di Bari, various issues 
were discussed, in particular those related to the comparison between
cosmological bounds and terrestrial ones, 
the estimation of systematic errors in the measurement 
of neutrino parameters from cosmological observations, 
and the impact of a non-standard evolution of the Early Universe.

\subsection{A theoretical perspective on lepton flavor physics at the TeV scale within i) SUSY models ii) extra-dimension models}

The origin of neutrino masses requires
new particles and interactions which need to be embedded
in a wider scheme to solve the hierarchy problem
and explain the pattern of masses and mixing. 
Possible scenarios are supersymmetric (talks by S. Lavignac, S. Khalil and A. Texeira), 
extra dimensions, and little Higgs models. 
One needs to consider 
how these models fit into a wider particle physics theory 
and if there are any model-independent signatures. 
These issues were discussed by A. Santamaria.
Generically, the embedding of neutrino mass models
in a wider scheme induces new signatures 
as LFV processes at low energy such as $\mu\rightarrow e \gamma$,
$\mu-e$ conversion, and $\tau \rightarrow e \gamma$.

\subsection{Leptogenesis in the context of neutrino mass models: model dependent versus model independent considerations}

Leptogenesis (talk by A. Abada) is one 
of the favored mechanisms for the generation of the baryon
asymmetry of the Universe and naturally takes place
in see-saw models.
In the presence of CP-violation
the decays of heavy Majorana neutrinos in the Early Universe
 produce a lepton asymmetry which is partially washed-out and 
converted into a baryon asymmetry by sphaleron processes. 
Compelling questions arise (discussion session lead by E. Nardi):
Under what conditions can a connection between low and high energy CP violation be established?
Is it obvious to believe in leptogenesis if a low energy CP violation and lepton number violation is observed?
In see-saw type I models, 
in general, there is not a direct connection between 
low energy parameters and high energy ones. 
This link arises if a theory of flavor is present,
which explains the observed mixing pattern
and reduces the number of free-parameters. 
For masses of the heaviest neutrino smaller than $10^{12}~\mathrm{GeV}$,
 flavor effects play an important role. In this case, 
 generically
 a connection between leptogenesis and low energy CP-violating phases
 can be obtained in the sense that if CP-violation is found in 
 future neutrino experiments a lepton asymmetry needs to be generated 
 in the Early Universe. 
 Leptogenesis can also take place in other see-saw models,
for example see-saw type II models and in other scenarios
of neutrino mass generation. 

\subsection{Interplay between neutrino masses and other phenomenological signatures}

Small neutrino masses require new physics beyond the Standard Model
which can induce also
other signatures at low energy.
Additional sources of lepton flavor violation induce
processes at low energy such as $\mu\rightarrow e \gamma$,
$\mu-e$ conversion, $\tau \rightarrow e \gamma$ (talk by A. de Gouvea)
at rates which could be close to present bounds. 
The LHC experiment
will provide information on the physics at the TeV scale which 
might also be at the origin of neutrino masses (talk by T. Schwetz).
Additional effects can arise in neutrino oscillations due to 
non-standard interactions of neutrinos with matter
which can be tested in future long-baseline neutrino experiments
(talk by E. Fernandez-Martinez). 
Neutrinoless double beta decay can be mediated not only
by light Majorana neutrino masses but also by other mechanisms
as heavy sterile neutrinos, supersymmetric R-parity violation.
Once a positive signal is found, it will be critical to discriminate
between the various mechanisms. 
S. Kom gave a contributed talk on LHC probes of SUSY neutrinoless double beta 
decay mechanism and C. Jackson on the issues related to understanding the mechanism of neutrinoless double beta decay with the SuperNEMO experiment.
In the discussion session T. Ohlsson addressed the synergy
and complementarity of these different experimental signatures in 
pinning down the mechanism at the origin of neutrino masses.

\subsection{Discussion on performance indicators in long baseline experiments}

Future long baseline neutrino oscillation experiments play a crucial role in determining the value 
of the unknown mixing angle $\theta_{13}$, the type of neutrino mass hierarchy
and the existence of CP-violation in the leptonic sector.
A rich experimental program is under
consideration for the near future: conventional and superbeam experiments (MINOS, OPERA, T2K, NOvA, LBNE)
are already taking data, under construction 
or at the R\&D phase while even more ambitious projects
such as neutrino factories and beta-beams are studied.
The sensitivity of the various setups is studied in detail.
T. Li discussed the physics reach of the low energy neutrino factory
and J. Lopez Pavon the search for 3+1 sterile neutrinos
in contributed talks.
It is necessary to compare the reach of different experiments in an
objective way and with comparable assumptions. 
These issues where discussed in the round table lead by K. Long. 

\subsection{Conclusions}

With the discovery of neutrino oscillations, 
neutrino physics has opened a new window on the physics BSM.
This workshop has given the opportunity to review 
the status of the studies in neutrino phenomenology, theory and astroparticle
physics with particular focus on the theoretical motivations
for future precise neutrino oscillation experiments.